\documentclass[useAMS,usenatbib]{mn2e}
\usepackage{placeins} 
\usepackage{lineno}
\usepackage{amssymb}
\usepackage{graphicx}
\usepackage{color}
\usepackage[normalem]{ulem}

\title[Chemical evolution of the magnetised ISM]{The impact of magnetic fields on the chemical evolution of the supernova-driven ISM}
\author[A. Pardi, et al.]{A. Pardi$^{1}$\thanks{email: pardia@MPA-Garching.MPG.DE} , P. Girichidis$^{1}$, T. Naab$^{1}$, S. Walch$^{2}$, T. Peters$^{1}$, F. Heitsch$^{3}$, \\
\newauthor S.C.O. Glover$^{4}$, R.S. Klessen$^{4,5}$, R. W\"{u}nsch$^{6}$, A. Gatto$^{1}$\\
\\
$^{1}$Max-Plank-Institut f\"{u}r Astrophysik, Karl-Schwarzschild-Str. 1, 85741 Garching, Germany\\
$^{2}$Physikalisches Institut, Universit\"{a}t zu K\"{o}ln, Z\"{u}lpicher Str. 77, 50937 K\"{o}ln, Germany\\
$^{3}$University of North Carolina Chapel Hill, CB 3255, Phillips Hall, Chapel Hill, NC 27599-3255, USA \\
$^{4}$Universit\"{a}t Heidelberg, Zentrum f\"{u}r Astronomie, Institut f\"{u}r Theoretische Astrophysik, Albert-Ueberle-Str. 2, 69120 Heidelberg, Germany\\
$^{5}$Member of the Interdisciplinary Center for Scientific Computing at Heidelberg
University\\
$^{6}$ Astronomical Institute, Academy of Science of the Czech Republic, Bocni II 1401, 141 31 Prague, Czech Republic}
\begin{document}

\date{}


\maketitle


\begin{abstract}

We present three-dimensional magneto-hydrodynamical simulations of the
self-gravitating interstellar medium (ISM) in a periodic (256 pc)$^3$ box with a
mean number density of 0.5 cm$^{-3}$. At a fixed supernova rate we investigate the
multi-phase ISM structure, H$_{2}$ molecule formation and density--magnetic field
scaling for varying initial magnetic field strengths (0, $6\times 10^{-3}$, 0.3, 3 $\mu$G).
All magnetic runs saturate at mass weighted field strengths of $\sim$ 1 $-$ 3 $\mu$G
but the ISM structure is notably different. With increasing initial field
strengths (from $6\times 10^{-3}$ to 3 $\mu$G) the simulations develop an ISM with a more
homogeneous density and temperature structure, with increasing mass (from 5\% to
85\%) and volume filling fractions (from 4\% to 85\%) of warm (300 K $<$ T $<$ 8000 K)
gas, with decreasing volume filling fractions (VFF) from $\sim$ 35\% to $\sim$ 12\% of hot
gas (T $> 10^5$ K) and with a decreasing H$_{2}$ mass fraction (from 70\% to $<$ 1\%).
Meanwhile the mass fraction of gas in which the magnetic pressure dominates over
the thermal pressure increases by a factor of 10, from 0.07 for an initial field
of $6\times 10^{-3}$ $\mu$G to 0.7 for a 3 $\mu$G initial field. In all but the simulations with the
highest initial field strength self-gravity promotes the formation of dense gas
and H$_{2}$, but does not change any other trends. We conclude that magnetic fields
have a significant impact on the multi-phase, chemical and thermal structure of
the ISM and discuss potential implications and limitations of the model.

\end{abstract}


%
\section{Introduction}

Magnetic fields are ubiquitously observed in the interstellar medium
(ISM; e.g. \citealp{Han09,Cr12,Beck13}) and they are expected to
impact the dynamical evolution of the gas on galactic scales
(e.g. \citealp{Han03,Frau12,Alv14}). Therefore a good understanding of the
connection between magnetic fields, ISM structure, and star formation
is crucial for a complete theory of the dynamical and chemical evolution
of the ISM (see \citealp{McK07,Cr12}).

Magnetic fields have been suggested to affect dynamics of the ISM and star
formation in various ways. Most popular is a distinction between weak and strong
field models. Weak field models \citep{Elm,ML04} are
based on the idea that the formation of molecular clouds (MCs) is controlled by
turbulent flows that compress the gas and form MCs, with the magnetic field being
too weak to stop the collapse. Still, even in that case, the fields can modify the
gas dynamics (and in consequence the star formation activity) of molecular clouds
\citep[e.g.][]{Vaz05, PB08}. Strong field models \citep{Mou75, KB07} rely on the magnetic fields being the dominant
mechanism controlling formation and evolution of molecular clouds and star
formation. In this case, the magnetic field is thought to be strong enough initially
to prevent wholesale gravitational collapse of the cloud \citep[see, however,][]{BalHar07}. Gravitational fragmentation and collapse in
this scenario may be enabled via turbulent compression in combination with ambipolar
diffusion \citep{KB08}, or accumulation of additional matter along field
lines \citep{Chen14, HeiHar14}.

Supernovae (SNe) play an important role in the evolution of the ISM,
possibly regulating the formation of a structured multi-phase ISM
\citep[e.g.][]{Kim14, Tan14, Gatto, Walch15, Gir15, Gatto16, Thom1, Thom2} and also the evolution and structure of the
magnetic fields \citep{Gent13}. SN feedback leads to a
multitude of complex mechanisms like the formation of the hot phase of
the ISM  \citep{Cox74, McK77}, the driving of turbulence \citep{ML04,Kle16},
local dispersal of gas \citep{McK77,ML05,Avill07,Jou09} or the
formation of colliding flows that regulate MC formation and the
formation of stars \citep{Hei205, Hen08, Bane09}. The impact of SNe location
and spatial as well as temporal clustering has been extensively studied in
both stratified boxes \citep{Hill12,Gent13,Kim13,Walch15,Gir15} and homogenous
boxes with open, mixed and periodic boundary
conditions \citep{Bal,ML05,Slyz,Gatto,Li15}. In particular \citet{Gatto} show that the
positioning and clustering of SN have significant impact on the structure of the ISM.
Explosions predominantly in intermediate and low density
environments favour the formation of a three-phase ISM with properties
similar to the observed ones in the solar neighbourhood. 


Observational information on the magnetic field of the Milky Way, on nearby and
distant galaxies and cosmic magnetic fields has increased recently
due to the improvement of observational methods \citep{Beck13}. Some
of these methods are: Zeeman splitting \citep{Cr93}, dust polarisation
\citep{Hilde88, Laz03}, optical and far-IR polarisation \citep{Laz08}, 
spectral-line linear polarisation \citep{Gol81}, Faraday rotation
\citep{Burn66} or synchrotron emission \citep{Mao12,Yoas13}.  Optical,
far-IR and sub-mm dust polarisation studies have shown that the
magnetic field is aligned with the Galactic plane
\citep{Fas02, Pav12}. Synchrotron emission analysis gives an average
field of $ 6 \pm 2$ $\mu$G in the local neighbourhood and $10 \pm 3$
$\mu$G at 3 kpc radius from the Galactic centre \citep{Beck01}. More
recent studies of radio continuum all-sky surveys combined with
thermal electron models and an assumed cosmic ray distribution reveal
an average field strength of $\sim 2$ $\mu$G for the regular field and
$\sim3$ $\mu$G for the random field, in the solar neighbourhood
\citep{Sun10,Jas12}. 

In the diffuse ISM of the Milky Way, the magnetic field is estimated to be around 10
$\mu$G along the line-of-sight \citep{Cr12}. 
In the dense regions of the ISM, such as MCs, there are examples 
supporting both the strong field and weak magnetic field theory. On the one hand 
it has been seen that the magnetic field can be regular and strong enough to resist turbulence.
In \citet{Cr03} two examples are discussed: L183 with
the magnetic field strength in the plane of sky of $\sim 80$  $\mu$G
and the line-of-sight field of $<$ 16 $\mu$G and DR 21(OH), a high
mass star forming region consisting of two compact cores \citep{Woody}
with an estimated total field strength of 1.1 mG. On the other hand, more recent work 
shows evidence for many MCs being supercritical, with magnetic fields that are too weak ($\sim  28 \mu$G in the line-of-sight) to
provide support against collapse \citep[see][]{Cr09,Cr10,CrHak10,Cr12}.


Recent magnetohydrodynamical (MHD) simulations that focused on the
impact of magnetic fields and turbulent flows in the ISM
indicate that magnetic fields
reduce the amount of dense and cold gas formed \citep{Hill12}, delay
H$_{2}$ formation \citep{Walch15}, reduce star formation and
fragmentation of massive cores significantly  \citep{Pet11, Hen11, Hen14} and
determine the filamentary structure orientation along the field lines
\citep{Nak, Avill05}, even though only the cold, dense regions 
are magnetically dominated. Recently, \citet{Pad16} have shown that MCs have a mean magnetic field enhanced by a factor of two relative to the
mean magnetic field of their simulation. Other studies have also shown that the
magnetic field has little impact on the scale height of magnetised
disks galaxies \citep{Hill12} and a weak effect on outflows \citep{Gir15}. 
However, those models did not follow the chemical evolution of the ISM
\citep[except for][]{Gir15}. Instead temperature and density cuts were
used to distinguish between the different ISM phases. \citet{Walch15}
showed that these simple cuts are insufficient to investigate the ISM
phases and dynamics, in particular for the cold molecular phase. As
this is the phase with the strongest magnetic fields as well as the
phase in which star formation occurs, it is important to follow the
evolution of the magnetised ISM together with its chemical evolution. 


In this study, which is part of the SILCC project (SImulating the Life-Cycle of molecular 
Clouds\footnote{see the SILCC website: \texttt{www.astro.uni-koeln.de/silcc}}), we investigate the interplay between the chemical evolution
of the SN-driven ISM and the magnetic field evolution. We use three-dimensional
magnetohydrodynamical simulations including a chemical network and shielding
from an interstellar radiation field, to explore how 
magnetic fields alter the formation of molecular
gas in the ISM. We present the details of the simulations, the
numerical methods and parameters, describing the chemistry network,
the SN implementation, the list of simulations and the initial
conditions in Section \ref{Discr}. We discuss the morphological
evolution of the simulations in Section \ref{Morp}. The chemical
evolution is described in Section \ref{Chem} together with 
the magnetic field evolution and the influence of self-gravity. In
this section we also evaluate the thermal, magnetic and  kinetic energy density
(Subsection \ref{Thmagpress}). In Subsection \ref{Scal} we show the  
magnetic field scaling and present our conclusions in Section \ref{Conc}.  

\section[]{Numerical method, parameters and simulations}
\label{Discr}

\begin{table*}
 \centering
 \begin{minipage}{140mm}
 
  \begin{tabular}{@{}llrrrrlrlr@{}}
  \hline

No. & Sim.\ name  & SN   &   $B_{\rm init}$   & Numerical & Physical & Self-gravity\\
    &                   & rate &        & resolution & resolution \\
    &          & [Myr$^{-1}$]  & [$\mu$G] & [cells]& [pc/cell] \\
\hline
1 & KS-lowB-L4& 1.2 & $6 \times 10^{-3}$   & $64^{3}$ & 4 &no\\

2 & KS-lowB& 1.2 & $6 \times 10^{-3}$  & $128^{3}$ & 2 &no\\

3 & KS-lowB-L6 & 1.2 & $6 \times 10^{-3}$  & $256^{3}$ & 1 &no\\

4 & KS-medB&  1.2 & 0.3  & $128^{3}$ & 2 &no\\

5 & KS-highB&  1.2 & 3  & $128^{3}$ & 2 &no\\

6 & KS&  1.2 & 0  & $128^{3}$& 2 &no\\ 
\hline
7 & KS-SG&  1.2 & 0  & $128^{3}$& 2 &yes \\
8 & KS-lowB-SG& 1.2 & $6 \times 10^{-3}$  & $128^{3}$& 2 & yes\\
9 & KS-medB-SG& 1.2 & $ 0.3 $  & $128^{3}$& 2 & yes\\
10 & KS-highB-SG&  1.2 &  3 & $128^{3}$& 2 & yes\\
11 & KS-medB-lateSG & 1.2 & $ 0.3 $  & $128^{3}$& 2 & from $t=150$ Myr
\end{tabular}
\caption{List of simulations and parameters. From left to right we
  list the name of the simulation, SN rate, initial magnetic field
  strength, numerical resolution, physical resolution, and whether or
  not the simulations were carried out with
  self-gravity. \label{table}} 
\end{minipage}
\end{table*}

\subsection{Numerical method}

We simulate the chemical, thermal and dynamical evolution of gas
driven by SN feedback in a periodic volume using the Eulerian,
adaptive mesh  refinement (AMR), three dimensional,
magnetohydrodynamic code FLASH 4 \citep{Fryx,DubeyEtAl2008}. To solve
the ideal MHD equations, we use the directionally-split 3-wave
MHD solver HLL3R \citep{Bou7,Waa9,Bou10,Waa11}. Our simulations
account for the effects of gas cooling and chemical evolution
(Section~\ref{meth:cool}) and SN feedback (Section~\ref{meth:SN})
using an approach very similar to the one adopted  by
\citet{Gatto}. Further details of our modifications to FLASH can also
be found in \citet{Walch15} and \citet{Gir15}. 
 
\subsection{Gas cooling and chemistry}
\label{meth:cool}

The simplified chemical network 
follows the evolution of seven chemical species (H$^+$, H, H$_{2}$,
C$^+$, O, CO and free electrons) and is based on the hydrogen
chemistry of \citet{GlMc07a,GlMc07b} and the CO chemistry of
\citet{NL97}. We model the effects of radiative cooling and
diffuse heating by cosmic rays, X-rays and the interstellar radiation
field \citep[see][]{Gl10,GlCl12,Walch15}. Cooling by metals in hot ionised gas
is modelled using rates from \citet{GF12}, as outlined in \citet{Walch15}.
Dust shielding and molecular self-shielding are
included using the TreeCol algorithm developed by 
\citet{Cl122}. Full details of the implementation of the chemical
network and the coupling of TreeCol with FLASH are described by
\citet{Walch15} and W\"{u}nsch et al. (in prep). 
In the simulations presented here, we consider
an interstellar radiation field of $G_{ \rm 0}=1.7$ \citep{Hb68,Dr78}
and a cosmic ray ionisation rate of $3\times10^{-17}$ s$^{-1}$. We
assume that the gas has solar metallicity and a constant gas-to-dust
mass ratio of $10^{2}$.

\subsection{SN driving}
\label{meth:SN}

We implement SN explosions at a fixed rate. To estimate it we
calculate the mean surface density of gas in the simulation volume,
$\Sigma_{\rm gas}=\rho_{i} L$, where $\rho_{i}$ is the initial mass
density of the gas and $L$ is the dimension of the simulation volume. 
We then determine the corresponding star formation rate surface
density $\Sigma_{\rm SFR}$ using the Kennicutt-Schmidt relation
\citep{Ken98}. 
Finally, we assume a Salpeter initial mass function
\citep[IMF, see][]{Sal}, with a low-mass cutoff at 0.01 M$_{\odot}$ and a
high-mass cutoff at  60 M$_{\odot}$, to find the number of massive
stars that will explode as Type II SN. This procedure yields 1 SN for
every $100 \: {\rm M_{\odot}}$ of stars formed, and so the final SN
rate in the volume is given by
\begin{equation}
\dot{N}_{\rm SN} = 2.5 \times 10^{-6} \left(\frac{L^{2}}{1 \: {\rm
    pc}^{2}} \right) \left(\frac{\Sigma_{\rm gas}}{\rm M_{\odot}
  pc^{-2}} \right)^{1.4} \: {\rm Myr}^{-1} .
\label{KSeq}
\end{equation} 
For each SN we inject thermal energy of $E_{\mathrm{SN}}=10^{51}$ erg into the
medium in a spherical region adjusted to encompass 800 $M_{\odot}$ of  
gas. The radius $R_{\rm inj}$ of this region is required to be
resolved with at least four cells.

In previous studies \citep{Gatto,Walch15}, we have explored the
influence of SN positioning on the evolution of the ISM. The four methods that we tested
were random SN positioning, peak positioning (the SN
are placed at the peaks in the density field), mixed 
positioning (a combination of random and peak driving), and clustered
positioning. With clustered positioning, 60\% of the SNe occur in clusters (the positions of 
which are random) and the remaining 40\% are individual SNe inserted at random
positions. We found in those studies that simulations
carried out with peak and mixed SN positioning lead to substantial
cooling of the SN remnants in dense regions,  and hence produce very
little hot gas. They also drive turbulence at a relatively low level
and just a small amount of  H$_{2}$ can form. On the other hand,
random or clustered positioning of the SNe results in H$_2$ mass fractions that are
in reasonable agreement with observational estimates \citep{Hon95,Kru09}.  
Simulations with random or clustered positioning also reproduce the large volumes
occupied by hot diffuse gas that are predicted by the classical
three-phase ISM model \citep{McK77} and result
in velocity dispersions for the individual phases that are in much better
agreement with observations \citep{Gir15}. As the clustering of the SNe was found to have only a minor effect on the mass distributions and volume filling fractions \citep{Walch15}, 
for the simulations presented in this paper, we
therefore adopt the random positioning model.  


\subsection{Simulations and initial conditions}

All simulations cover a box volume of (256 pc)$^{3}$. We start
with atomic gas of solar composition, at a uniform temperature of $T = 5000$~K and with a uniform number density of $n_{i} = 0.5 \: {\rm
  cm^{-3}}$. This yields a gas surface density of $\Sigma_{\rm gas} =
4.1 \: {\rm M_{\odot} \, pc^{-2}}$ for the box, and a total gas mass
of $2.9 \times 10^{5} \: {\rm M}_{\odot}$, comparable to the mass of
a giant molecular cloud \citep{Hug10}.

We adopt a fiducial, fixed-grid resolution of 128$^{3}$ cells across the domain, equivalent to a uniform cell size of 2 pc. To test the resolution effects, we ran additional 
simulations at a uniform resolution of 64$^{3}$ (labeled L4) and 256$^{3}$ (labeled L6)
cells (see Appendix \ref{A}, Fig. \ref{L4L5L6}).




To study the evolution of the magnetic fields, we carry out runs with
three different values for the initial magnetic field strength: a low
value of $6 \times 10^{-3}$ $\mu$G,  
an intermediate value of 0.3 $\mu$G and a high value of 3 $\mu$G
(comparable to observations in the local ISM). We note, however, that 
even our strong initial magnetic field is at the lower 
limit of what observations suggest for the atomic gas.

The role of self-gravity in shaping the ISM has been discussed by \citet{Walch15}, who show that without
it, it is impossible to obtain reliable estimates of dense and molecular gas. However, recently, \citet{Pad16} have seen little influence from self-gravity in a very similar
simulation setup to ours. In this study, we run all simulations with and without self-gravity (with the exception of the simulations run as part of our convergence testing,
described in Appendix~\ref{A}). In addition, we compare one simulation with self-gravity
switched on from $t=0$ to one simulation in which we  
let the ISM first evolve without self-gravity and then switch it on
after 150 Myr. As this procedure has been used in previous
studies \citep[see e.g.][]{Pad16,Huan16} we describe the morphological
and chemical differences between these two cases in Appendix
\ref{B}. A full list of the simulations is given in Table~\ref{table}.
The names of the simulations specify the supernova rate in SNe/Myr given by the Kennicutt-Schmidt relation via Eq.~\ref{KSeq}
(KS), the initial magnetic field strength (no label for runs with $B_{\rm init} =
0$, lowB for runs with a weak initial field, medB for the runs with $B_{\rm init} = 0.3 \, \mu$G
and highB for the runs with $B_{\rm init} = 3\,\mu\mathrm{G}$. To the
name, the level of refinement of the simulation (L4, L6) or a suffix for  
self-gravity (from $t=0$ as SG and from $t=150$ Myr as lateSG) is added where appropriate. 

\section{Morphological evolution}
\label{Morp}

\begin{figure*}
\centering
\begin{minipage}{175mm}
\includegraphics[width=175mm]{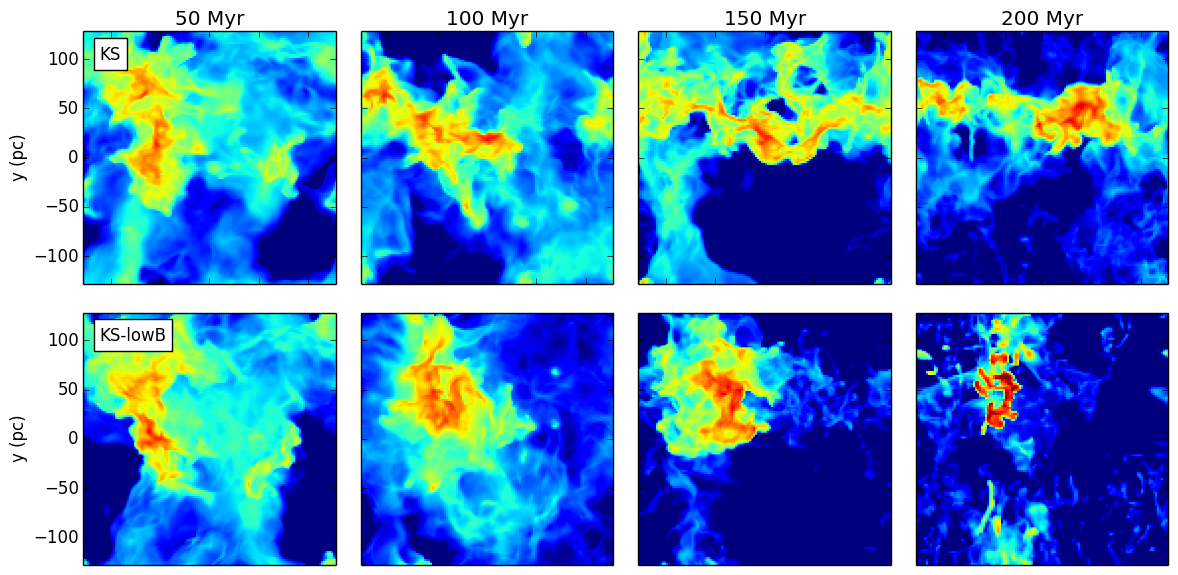}
\includegraphics[width=175mm]{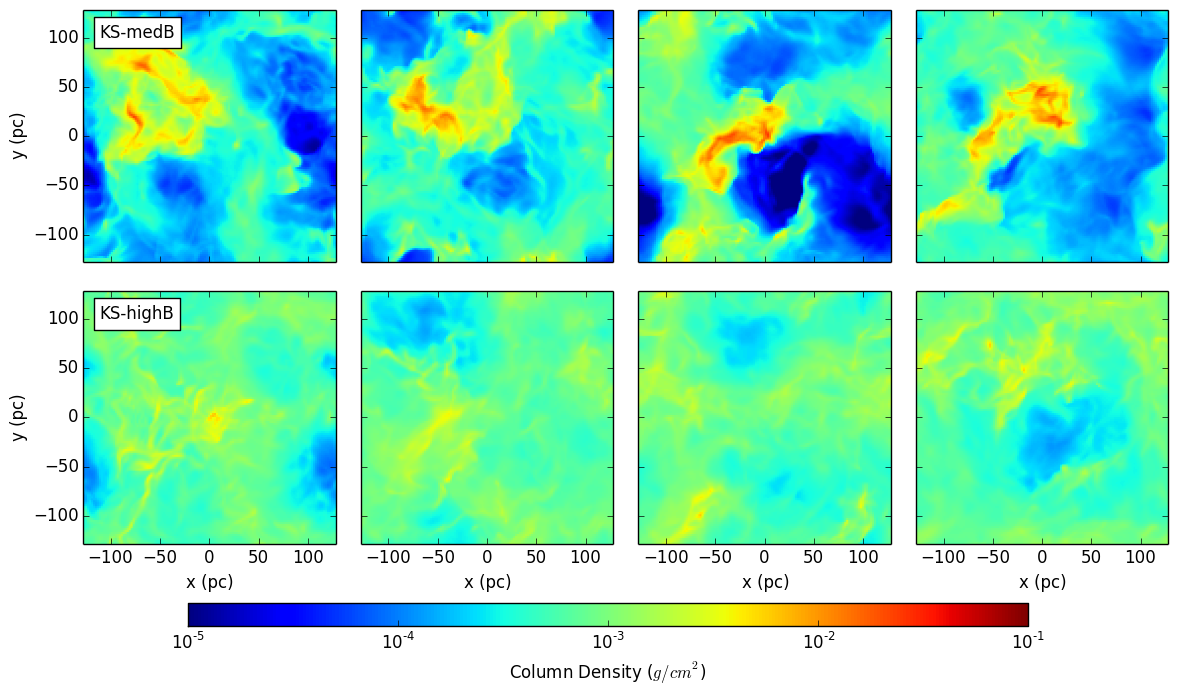}
\caption{Time evolution of the column density for runs KS ({\it top}),
  KS-lowB ({\it second row}), KS-medB ({\it third row}) and KS-highB
  ({\it bottom row}) at 50, 100, 150 and 200 Myr (from left to
  right). The densest clumps develop in the KS-lowB simulation after
  100 Myr while KS-highB has the most homogeneous ISM structure. The
  structural differences between KS and KS-lowB are discussed in
  Section~\ref{Chem}.} 
\label{ColDens}
\end{minipage}
\end{figure*}

\begin{figure*}
\centering
\begin{minipage}{175mm}
\includegraphics[width=175mm]{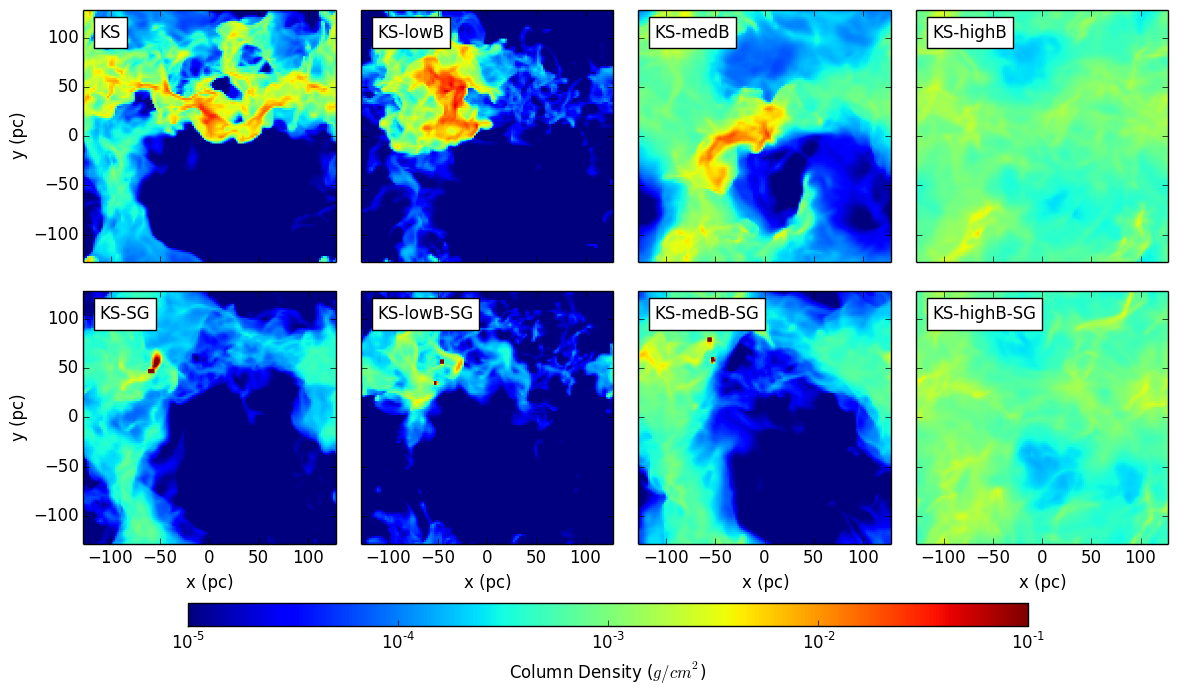}\\
\caption{Column density comparison of simulations KS, KS-lowB, KS-medB
  and KS-highB ({\it top row}) and self-gravity simulations KS-SG,
  KS-lowB-SG, KS-medB-SG and KS-highB-SG ({\it bottom row}), all at
  150 Myr.  With self-gravity the dense structures are much
  more compact. KS-highB-SG is an exception: the ISM is much more
  homogeneous and resembles the ISM in KS-highB.} 
\label{ColDensSG}
\includegraphics[width=175mm]{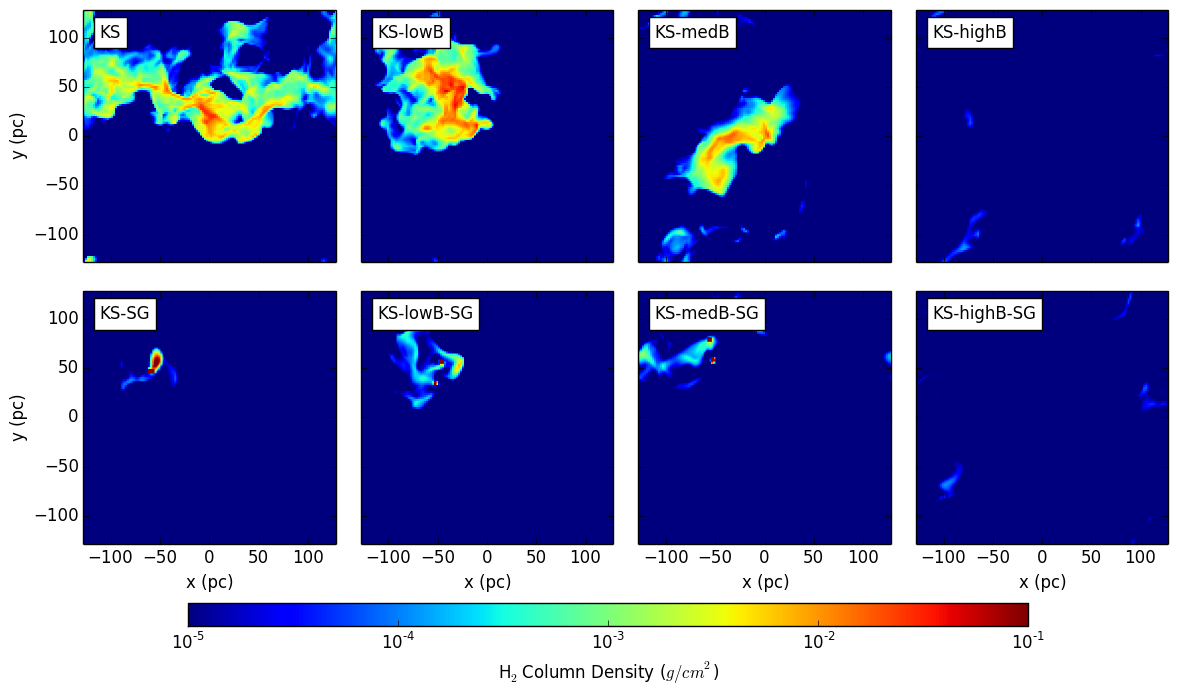}
\caption{Same as Fig. \ref{ColDensSG} but for the H$_{2}$ column
  density, all at
  150 Myr. Without self-gravity, the H$_{2}$ is contained in relatively extended clouds while with self-gravity almost all the
  H$_{2}$ is concentrated in a few very compact clumps. However, in runs
  KS-highB and KS-highB-SG very little H$_{2}$ is formed.}  
\label{ColH2}
\end{minipage}
\end{figure*}

\begin{figure*}
\centering
\begin{minipage}{175mm}
\includegraphics[width=165mm]{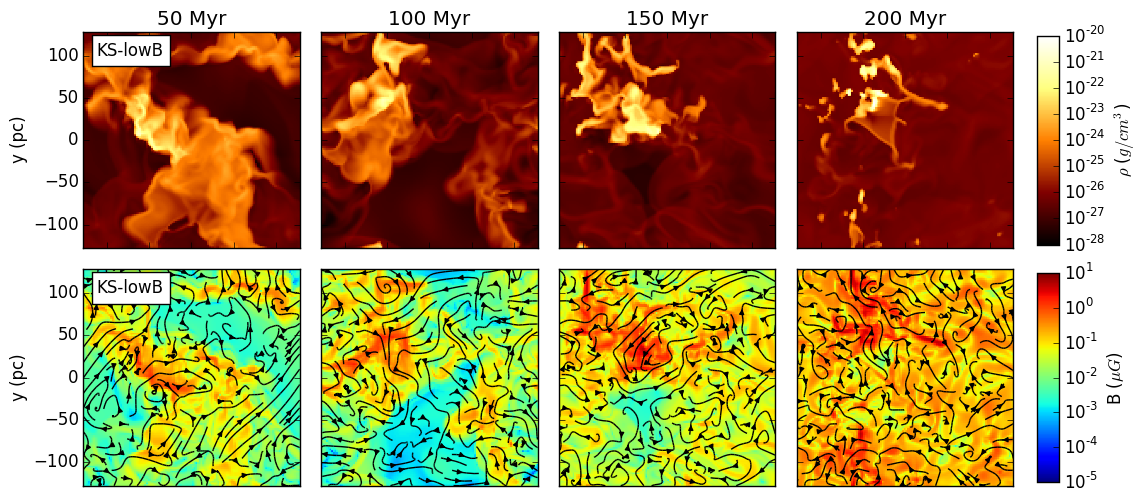}
\includegraphics[width=165mm]{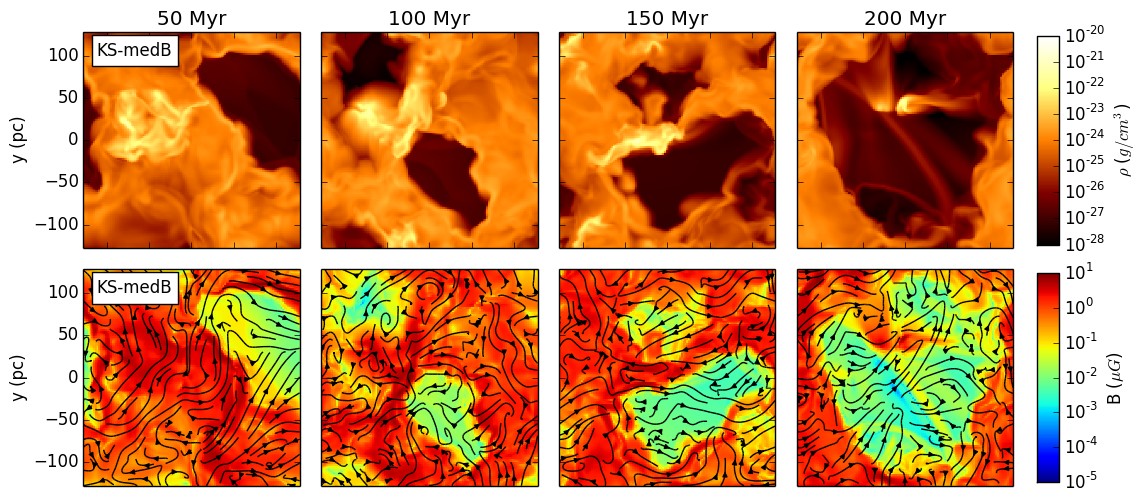}
\includegraphics[width=165mm]{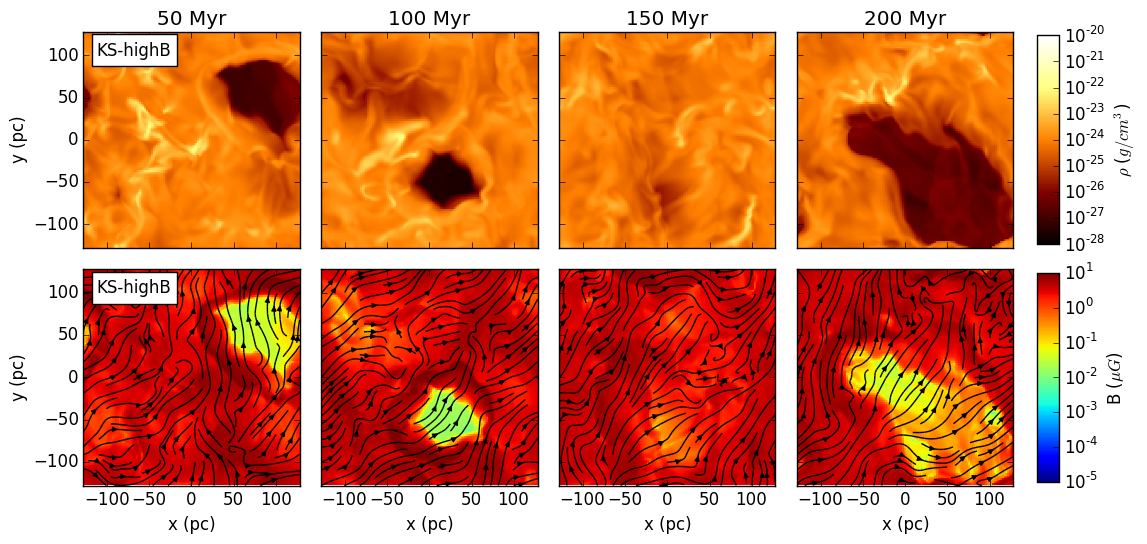}
\caption{Time sequence showing density slices passing through the
densest region in the simulation and oriented parallel to the $z$ axis
(rows 1, 3, 5) and the corresponding magnetic field strengths (rows
2, 4, 6), with magnetic field lines over-plotted. Results are shown for
output times of 50, 100, 150 and 200 Myr and for runs KS-lowB (top
two rows), KS-medB (middle two rows) and KS-highB (bottom two
rows). There is a strong correlation between the density and
magnetic field strength. The magnetic field in KS-lowB is highly tangled
  due to the turbulent motion of the gas.}
\label{DensMag}
\end{minipage}
\end{figure*}

The initial homogeneous distribution of atomic hydrogen is rapidly
stirred by expanding SN shells. The explosions disperse the gas 
locally and compress the regions between the shells that will form
high density structures and molecular hydrogen. Randomly placed SNe
result in a highly structured medium with large variations in column
density (from 10$^{-5}$ to 10$^{-1}$ g/cm$^{2}$) and a significant
fraction of the simulation volume filled with hot gas. This is in
agreement with previous findings
\citep[see][]{Hen14,Gatto,Walch15,Gir15}.   

Figure \ref{ColDens} depicts the morphological differences of the ISM
between KS (top row), KS-lowB (second row), KS-medB (third row) and
KS-highB (bottom row). We compare column density projections after 50,
100, 150 and 200 Myr of evolution. There are no significant
structural differences between KS and KS-lowB but the medium and high
initial field strength simulations show fewer clumpy structures (see the
discussion in Section \ref{Chem}). While the column density in the
first three simulations varies by at least four orders of magnitude, in
KS-highB it varies by only two orders of magnitude.

%


In Fig.~\ref{ColDensSG} we highlight the structural differences
between the runs with and without self-gravity, all at 150 Myr. We
show the column densities for KS and KS-SG (left column), KS-lowB and
KS-lowB-SG (second column), KS-med and KS-med-SG (third column) and
KS-highB and KS-highB-SG (right column). The simulations  including
self-gravity form very dense structures confined to a small volume
while in the runs without self-gravity the structures are less dense
and occupy a larger volume. The same can be seen in Fig. \ref{ColH2}
where the H$_{2}$ column density is shown. However, KS-highB and KS-highB-SG
are clearly different. Because of the strong magnetic fields the gas
cannot be compressed and collapse to high densities. As a results only
traces of H$_{2}$ can be found (see Section \ref{Chem}).

In Fig.~\ref{DensMag} we present the time evolution in density slices
(through the densest region in the box) and the magnetic
field strength with the field lines over-plotted for KS-lowB (top two rows), 
KS-medB (third and forth row) and KS-highB (bottom two
rows). For KS-lowB, even after a very short time (50 Myr), the ISM
becomes very inhomogeneous with densities ranging over 5 orders of
magnitude (from 10$^{-27}$ to 10$^{-22}$ g/cm$^{3}$) and the magnetic
field varying by 3 orders of magnitude (from 10$^{-3}$ to 1 $\mu$G).  
Overall, the highest magnetic field strength is found in dense gas. In
runs KS-lowB and KS-medB, the combination of the strong turbulent
motions and the weakness of the magnetic field lead to the field becoming
highly tangled. This is an expected result, consistent with the ideal MHD
assumption where the magnetic field is dragged along with the gas.
In simulation KS-highB (Fig. \ref{DensMag}
bottom row) the
magnetic field strength remains high during the entire simulation. The magnetic
field lines are less tangled and due to the  strength of the field,
compression is less efficient and the ISM is more homogeneous.


\section[]{Chemistry and magnetic field evolution}
\label{Chem}

\begin{figure}
\includegraphics[width=80mm]{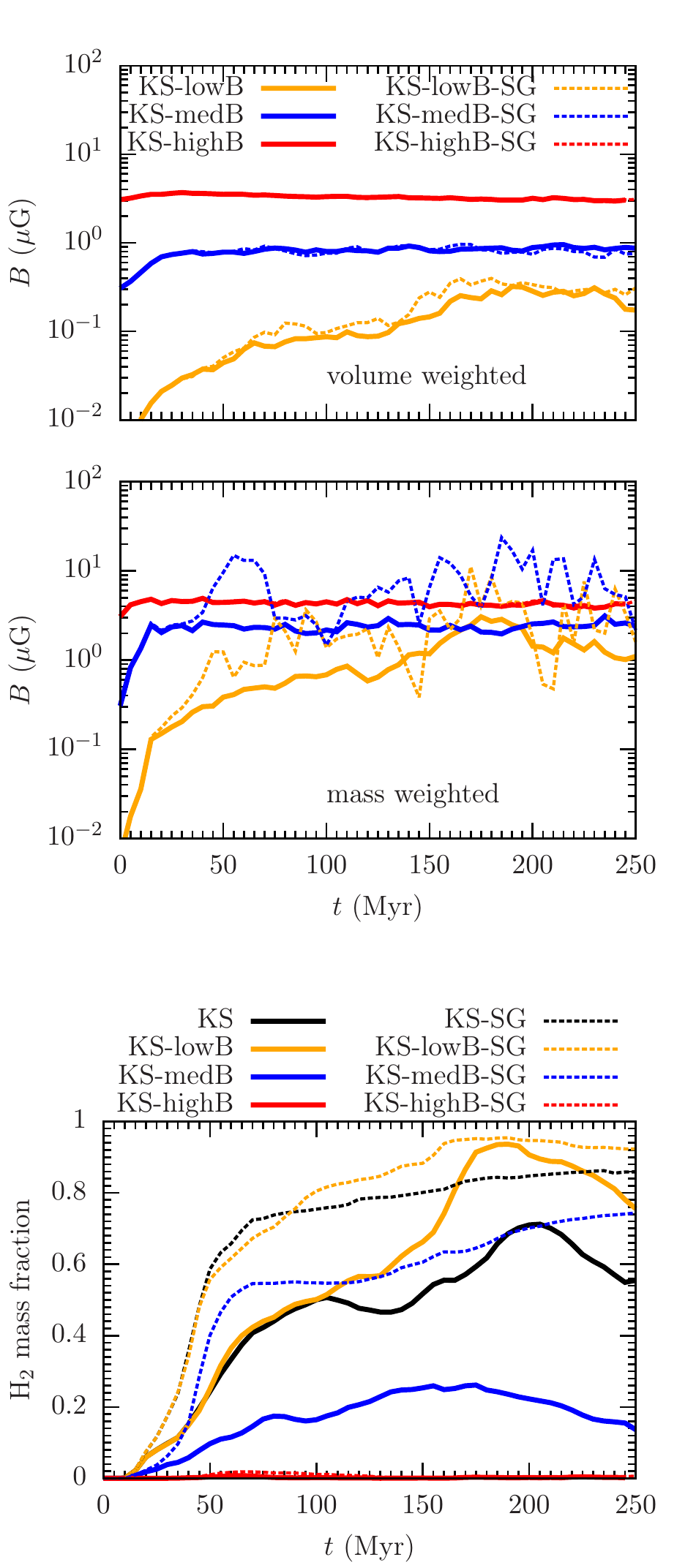}
\caption{Magnetic field evolution over time (volume weighted {\it top panel} and mass weighted {\it middle panel})
and H$_2$ mass fraction evolution over 250 Myr ({\it bottom panel}). 
We show the runs: KS (solid black line), KS-lowB (solid orange line), KS-medB (solid blue line), KS-highB (solid red line), KS-SG (dashed black line), KS-lowB-SG (dashed orange line), 
 KS-medB-SG (dashed blue line) and KS-highB-SG (dashed red line). The average magnetic field of KS-highB is almost
 flat at 3 $\mu$G while the field in KS-medB and KS-lowB are being amplified. Although including a very weak initial magnetic field actually aids the
 formation of H$_{2}$, increasing the field strength to values closer to the observations leads to a significant suppression of H$_2$ formation,
 to the point where there is barely any H$_{2}$ formed in runs KS-highB or KS-highB-SG.}
\label{B_sat_H2}
\end{figure}

\begin{figure*}
 \centering
\begin{minipage}{175mm}
\includegraphics[width=82mm]{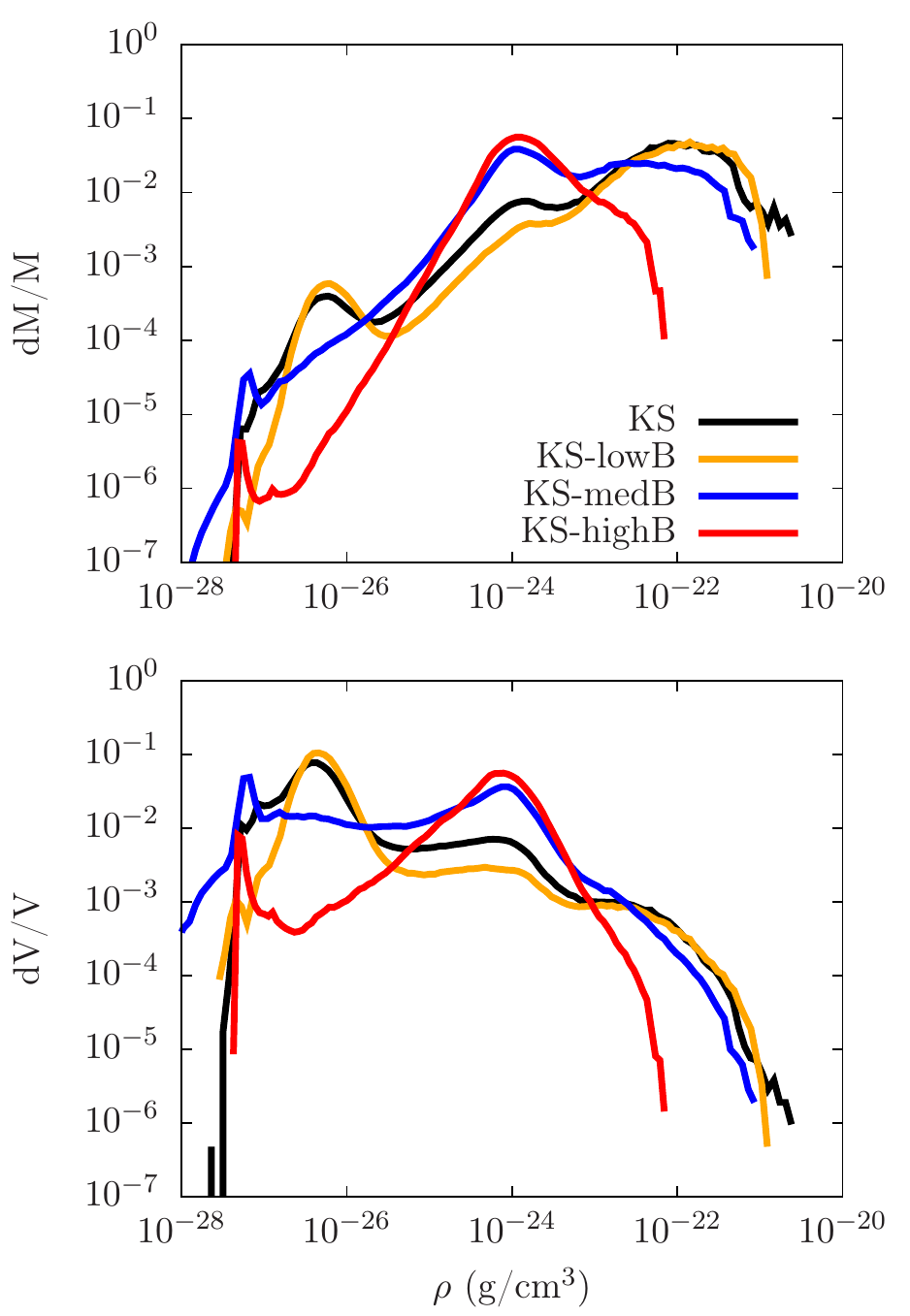}\quad
\includegraphics[width=82mm]{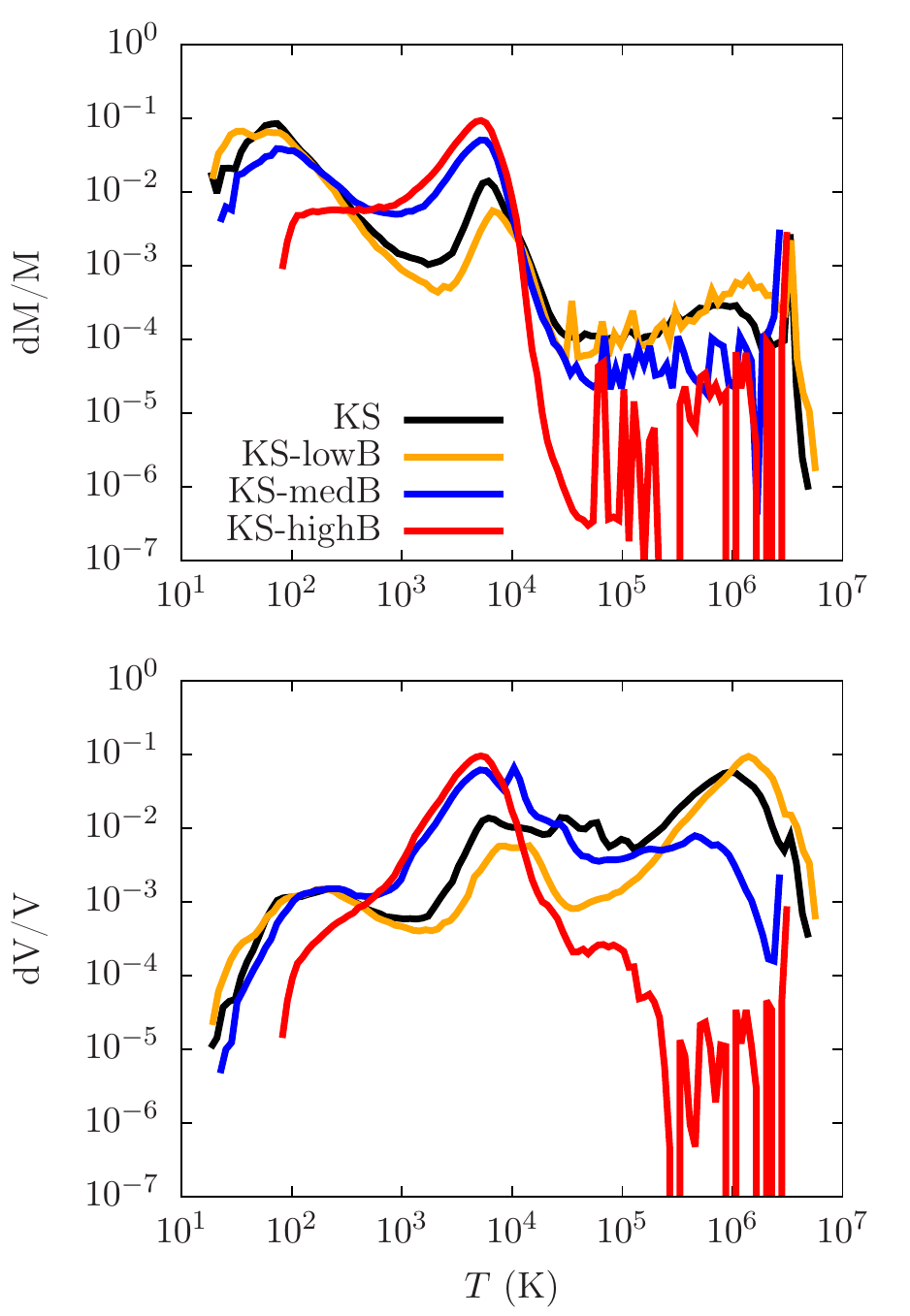}
\caption{Mass weighted ({\it top left panel}) and volume weighted ({\it bottom left panel}) density PDFs and mass ({\it top right panel}) 
and volume weighted ({\it bottom right panel}) temperature PDFs of KS (black), KS-lowB (yellow line), KS-medB (blue line) and KS-highB (red line), all at 150 Myr. 
In both cases, the PDF of KS-highB is narrower at high densities and low temperatures while there is very little hot and low density gas
in this simulation.}
\label{PDF}
\end{minipage}
\end{figure*}

\begin{figure}
\includegraphics[width=80mm]{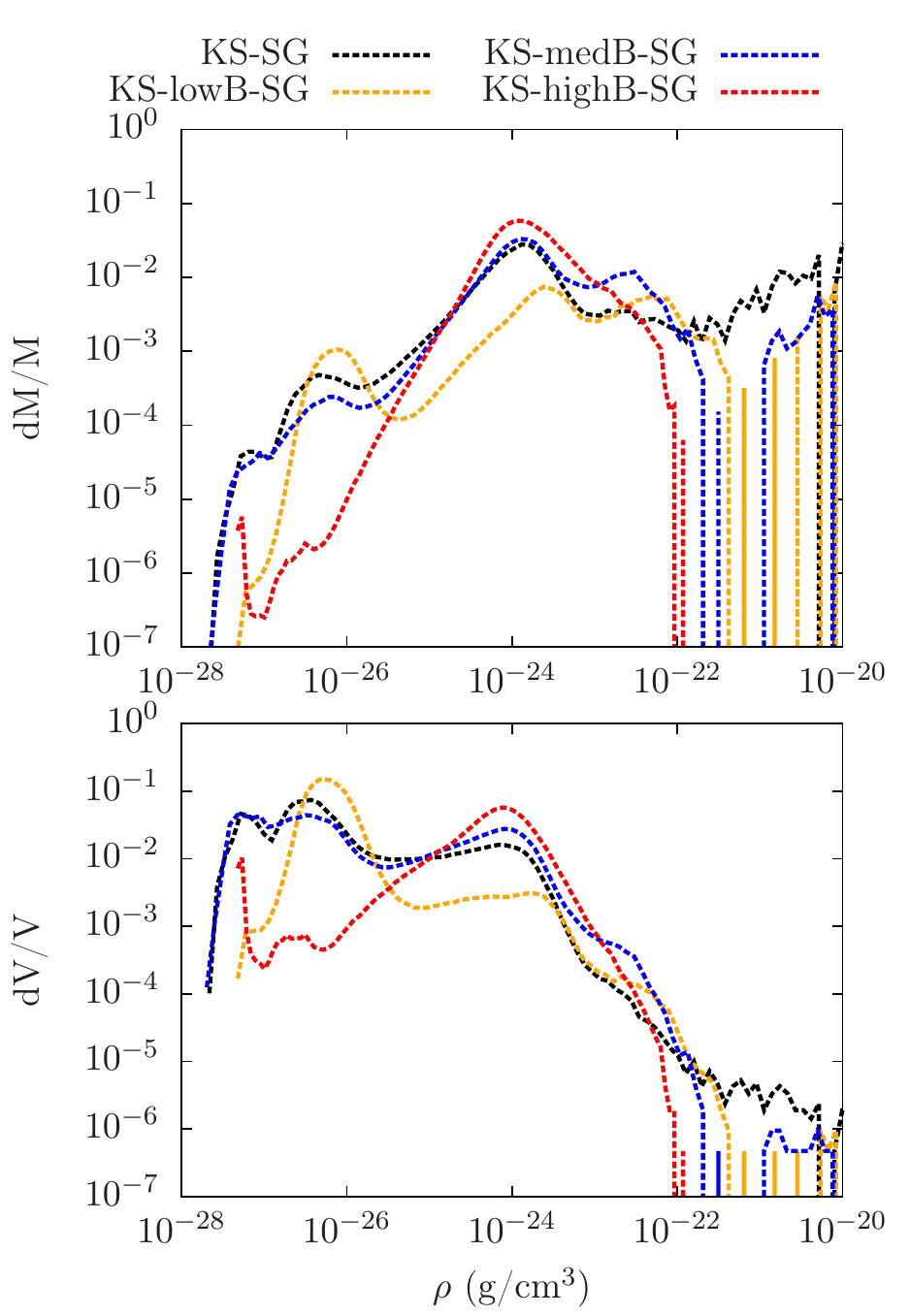}
\caption{Mass weighted ({\it top panel}) and volume weighted ({\it bottom panel}) density PDFs of KS-SG (black dotted line), KS-lowB-SG (yellow dotted line), KS-medB-SG (blue dotted line) and 
KS-highB-SG (red dotted line), all at 150 Myr. The PDF of KS-highB-SG is much narrower and unlike the other PDFs does not show a peak at 10$^{-26}$ g/cm$^{3}$.}
\label{PDFdensSG}
\end{figure}

\begin{figure}
\includegraphics[width=80mm]{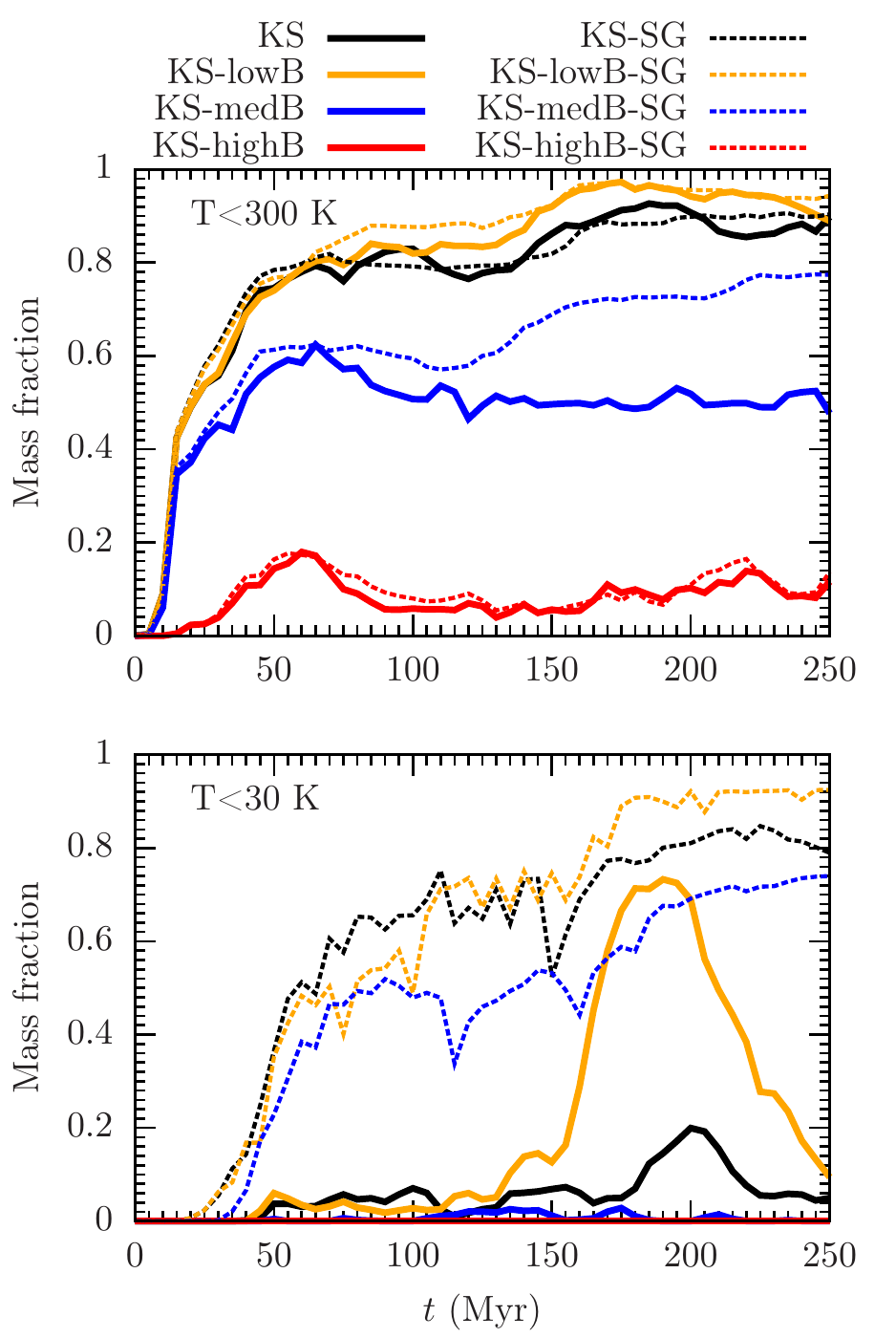}
\caption{Mass fraction evolution of the cold gas (T$< 300$ K, {\it top panel}) and very cold gas
(T$<30$ K, {\it bottom panel}) for all the simulations (same colour-coding as in Fig. \ref{B_sat_H2}). While in the cold gas regime the mass fractions in KS and KS-lowB
are similar, the very cold gas mass fraction starts diverging soon after 100 Myr and in KS-lowB, the very cold gas reaches a fraction almost four times higher than in KS.}
\label{Mfcold} 
\end{figure}

\begin{figure}
\includegraphics[width=80mm]{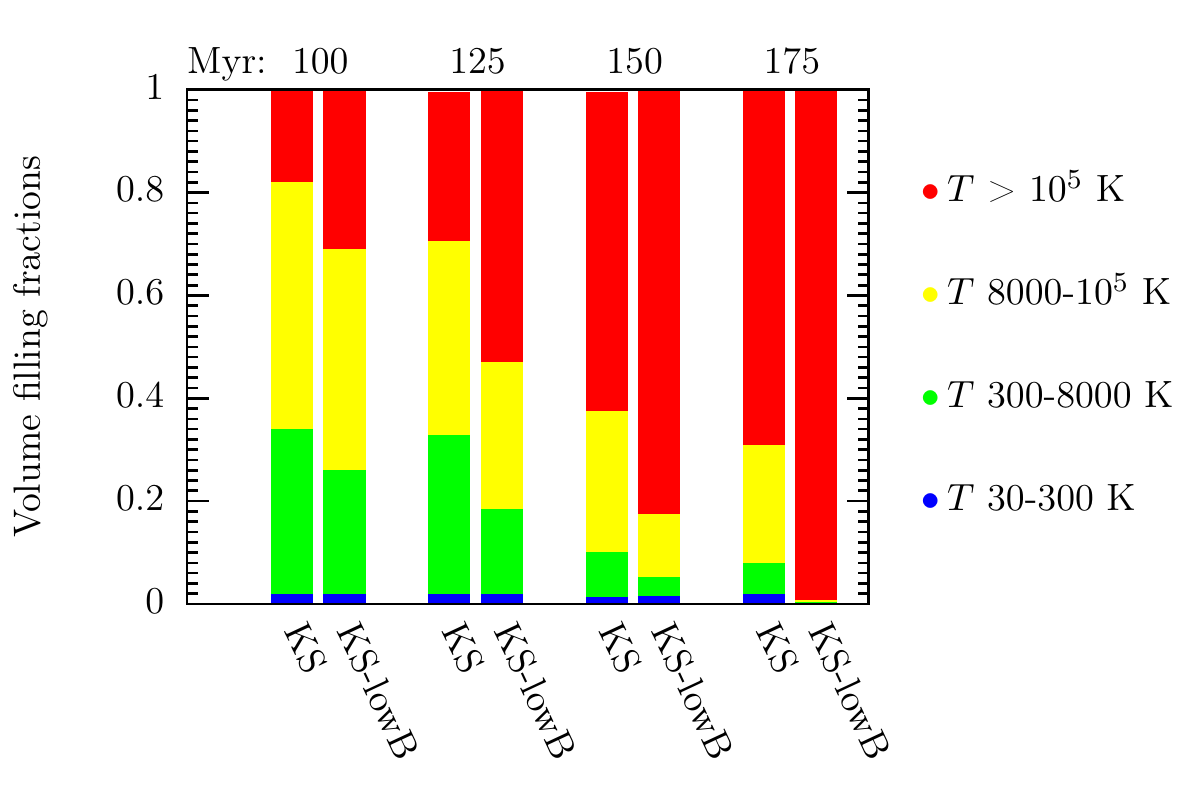}
\caption{Volume filling fractions at
100, 125, 150 and 175 Myr (averaged over 4 Myr). Colour-coding: blue - cold gas ($T \in [30;300)$ K), green - cool gas ($T \in [300;8000)$ K, yellow - warm gas 
($T \in [8000;10^{5})$ K) and red - hot ionised gas ($T \geq10^{5}$ K) for KS and KS-lowB. The cool and warm gas VFFs are higher for KS while the hot gas VFF comes to completely dominate in KS-lowB by 175 Myr. }
\label{VffBlowB} 
\end{figure}

\begin{figure}
\includegraphics[width=80mm]{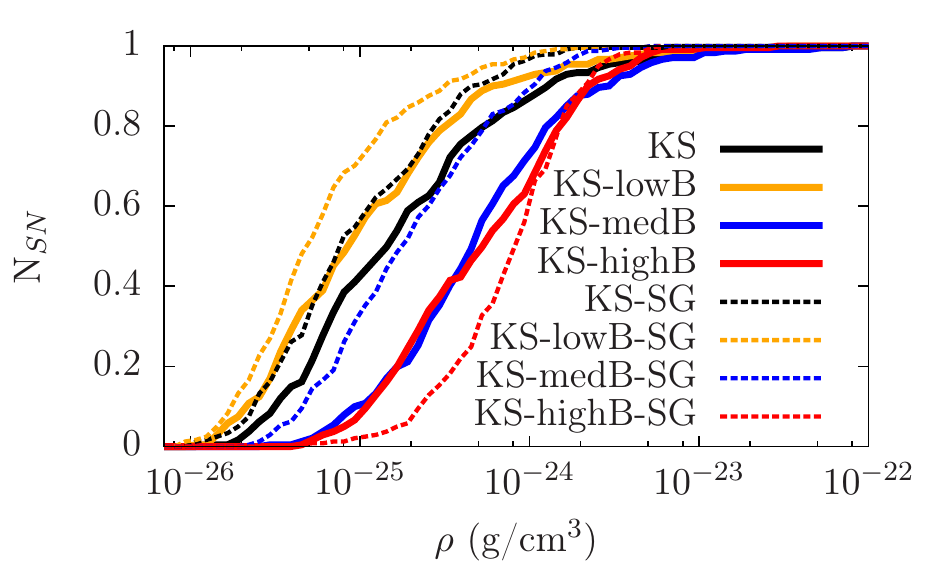}
\caption{Cumulative histogram of the number of SNe as a function of the environment they explode in. Same colour-coding as in Fig. \ref{B_sat_H2} and Fig. \ref{Mfcold}.
The SNe in KS-lowB (and KS-lowB-SG) explode in a more diffuse ISM than the ones in KS (and KS-SG).}
\label{Hist} 
\end{figure}

\begin{table*}
 \centering
 \begin{minipage}{160mm}
 
  \begin{tabular}{@{}llrrrrrrrrrrrrr@{}}
  \hline

 Sim. 		& f$_{\rm H_{2}}$          & f$_{\rm H}$              &  $B$                 & $B$            &  $B$ (H$^{+}$)       &  $B$ (H)        & $B$ (H$_{2}$)       & $B$ ($T_{1}$)            & $B$ ($T_{2}$)  & $B$ ($T_{3}$) & f$_{\beta < 1}$ \\
 name		&                    &                  &     [$\mu$G]         &  [$\mu$G]      &  [$\mu$G]            &  [$\mu$G]       & [$\mu$G]            &  [$\mu$G]                & [$\mu$G]     &  [$\mu$G]   &            \\
weighing        &                    &                  & mass                 & volume         & mass                 & mass            & mass                & mass                     & mass         & mass        & \\
\hline
        KS      		& 0.5                 & 0.4    &                            &                         &                   &                   &&      &                       &      &            & \\
\hline
	KS-lowB    	&0.6   		&0.3 & 1.2 	          & 0.5   	        &  0.5 & 1 & 1.2&1.5& 1.2& 0.9 & 0.07 \\
\hline
	KS-lowB-L4    	&0.6   		&0.4 & 0.4 	          & 0.07   	        &  0.3 & 0.5 & 0.3&0.3& 0.4& 0.4 & 0.002 \\
\hline
	KS-medB  	& 0.2 		 & 0.7 & 2.2  		 & 0.8   	          & 1.8 &  2 & 2.3& 1.9&2.7&2 & 0.3 \\
\hline
	KS-highB 	& 10$^{-3}$	  & 0.9 & 4  		 & 3.2              &  4 & 4  & 4.4& 4.4&3.9& 5.5&  0.7\\
\hline
	KS-SG 		 & 0.8  		 	 &0.2&			   	     	        & & & &&&& \\
\hline
 	KS-lowB-SG & 0.9  		& 0.1& 3	& 0.3 	        & 0.9 & 1.7 & 3&4&2.5&2.5& 0.01 \\
\hline
 	KS-medB-SG& 0.6  		& 0.4& 6.5		 & 0.8 	        &  2.4& 3 & 8.8&22&11.6&2.7 & 0.16\\
\hline
	KS-highB-SG& 0.8$\cdot$10$^{-3}$    & 0.97& 4.04		& 3.2	       & 4.3 &4 &5.2& 4.7 &4.2&5.3&0.7\\
\hline

\end{tabular}
\caption{
List of simulations with the H$_{2}$ and H mass fractions, mass and
volume weighted magnetic field intensity, mass weighted magnetic
field in H$^+$, H and H$_{2}$, mass weighted magnetic field in $T_{1}<$30
K, $T_{2} \in [30;300)$ K and $T_{3} \in [300;8000)$ K and the mass fractions for which $\beta <1$ for all the simulations at 150 Myr. } 
\label{table1}
\end{minipage}
\end{table*}

In a turbulent environment, with the magnetic field ``frozen into'' the gas, the fluid motion drags along the magnetic field, twisting and folding it.
Turbulent compression\footnote{Note that as our simulations do not account for
the effects of Galactic rotation, they are unable to probe any further amplification
that may occur due to the $\alpha$-$\Omega$ dynamo.} quickly amplifies the magnetic field at small scales \citep{Kaz68, Subr} until the saturation scale is reached \citep{Egan}, 
when there 
is a statistical balance between the kinetic and magnetic
energy. This effect is seen in the weak and medium strength simulations. In order to reach this
saturation limit, the system needs several dynamical
times \citep{Pion07,Hen11,Fed11a,Hill12,Schober15}.

We show the average volume weighted and mass weighted magnetic field
evolution in Fig. \ref{B_sat_H2} (top and middle
panels) for all magnetic runs. The volume weighted magnetic field
in KS-highB is constant over time. In KS-medB the field is
amplified over the first 20 Myr and then remains constant. On the other hand, the seed field in KS-lowB is amplified over a 
much longer time, more than 150 Myr. Self-gravity does not
significantly influence the saturation field strength in any of the
simulations. The volume weighted mean magnetic field saturates at about 0.2
$\mu$G for KS-lowB,  at 1 $\mu$G for KS-medB and at 3 $\mu$G for
KS-highB. The mass weighted mean magnetic field (Fig. \ref{B_sat_H2}
middle panel) shows a similar behaviour although the saturated values
are higher: $\sim$ 1 $\mu$G for KS-lowB, 2 $\mu$G for KS-medB and 4
$\mu$G for KS-highB. Self-gravity in KS-medB and KS-lowB induces significant fluctuations in the mass weighted field strength, almost an order of
magnitude, which emphasises the scaling of the field strength with gas
density (shown also in Fig. \ref{B_sc}).  The gas can reach higher densities and magnetic field strengths in the self-gravitating runs 
(see Section \ref{Thmagpress} and \ref{Scal}). However, in all the simulations, the mean magnetic field strength is weaker than in the solar neighbourhood \citep[$6 \pm 2$ $\mu$G, see][]{Beck01}.


We analyse the chemical state of the ISM by focusing on the time evolution
of the H$_2$ mass fractions. As there is only a few percent of mass in H$^+$, the
remaining gas mass is in atomic form. 
Our simulations' resolution limits prevent meaningful conclusions about CO formation (more details are given in Appendix \ref{A}, Fig. \ref{CO}).
The evolution and the
strength of the magnetic field is tightly linked to the evolution of
H$_{2}$. In Fig. \ref{B_sat_H2} (bottom panel), the H$_{2}$ formed in
KS-highB and KS-highB-SG is less than 2\% of the total mass throughout
the simulations (seen also in Fig. \ref{ColH2}). In our setup, high magnetic 
fields can suppress the formation of H$_{2}$. 

Figure \ref{PDF} shows the mass and volume weighed density and temperature probability distribution functions (PDFs) of
the simulations without self-gravity. The density PDFs of the unmagnetised, initially low and medium field strength runs reach a density higher than 
10$^{-21}$ g/cm$^{3}$ while the PDF of KS-highB is much narrower ($\rho_\mathrm{max}\approx10^{-22}$ g/cm$^{3}$). As a result, the
minimum temperature of the gas in KS-highB is much higher: around 100~K, in comparison to 20-30~K in the other simulations. The
lack of high density gas in the KS-highB run also has a pronounced 
impact on the formation of H$_{2}$, explaining the small fraction of $\sim$ 2\% that we see for the KS-highB and KS-highB-SG runs.
The much more homogeneous ISM we see in Fig. \ref{ColDens} is a result of most of the mass and volume of the gas in KS-highB being around
10$^{-24}$ g/cm$^{3}$ (discussed also in Section \ref{Thmagpress}) and temperature of 10$^{4}$ K. In comparison, the other simulations show distinct
density PDF peaks at 10$^{-26}$, 10$^{-24}$ and 10$^{-22}$ g/cm$^{3}$, indicative of a three-phase ISM. In KS and KS-lowB, most of the simulation box is filled with 
10$^{6}$ K gas while most of the mass is at 10$^{2}$ K. 

The fact that the PDF of KS-highB falls off so steeply at the high density and low temperature end can have several reasons.
The random distribution of the SNe in a 3 $\mu$G magnetised environment is apparently not enough to stir the gas efficiently in the simulation.
The compression of the gas is too low to form high density, long-lived structures that would permit the substantial formation of H$_{2}$.
The current simulation setup fails to create a hot phase as well. A more efficient way of driving turbulence would be clustered
SNe generating stronger coherent motions. Extending the simulation box to a stratified box with open boundaries and an external background potential would 
both allow for coherent acceleration towards the centre of the potential as well as allowing the SNe to expand further without the spatial restrictions of the
periodic box (see also the discussion in Section \ref{Thmagpress}).

In Fig.~\ref{PDFdensSG} we show the mass and volume weighted density PDFs for the simulations with self-gravity. The main features are the same as for the simulations without self-gravity. The main systematic difference is the high-density tail that corresponds to gravitationally dominated regions.

Returning to Fig. \ref{B_sat_H2}, in the runs without
self-gravity, higher magnetic fields result in lower
H$_{2}$ mass fractions (at 150 Myr, 60\% in KS-lowB, 25\% in
KS-medB and 1\% in KS-highB). The H$_{2}$ mass fractions in the
simulations with self-gravity reach 50\% - 70\% after  
only 50 Myr and much higher values after 150 Myr, compared to the
runs without self-gravity (90\% in KS-lowB-SG, 60\% in KS-medB-SG and
about 1\% in KS-highB-SG). The evolution of the H$_{2}$ mass fractions
in KS and KS-lowB is very similar until about 100 Myr. After this
point, the evolutionary trend is very similar but the H$_{2}$ mass
fraction is lower in KS than in KS-lowB. The same holds for KS-SG and KS-lowB-SG which diverge at 90 Myr.
To illustrate why we see this counter-intuitive difference between the hydro and KS-lowB simulations in 
H$_{2}$ we show in Fig. \ref{Mfcold} (top panel) the mass fraction of the cold (T $<$ 300 K) and the very cold (T $<$ 30 K, bottom panel) gas. The cold gas in both 
KS and KS-lowB (as well as in KS-SG and KS-lowB-SG) has similar mass fractions ($\sim$ 90\%) while in the T $<$ 30 K regime KS-lowB increases and 
peaks at around 200 Myr at a four times higher value than KS.

This difference occurs due to a systematic effect (tested with simulations using different random positioning for the SNe). Even though an initial field 
of 6 $\times10^{-6} \mu$G is too weak to make a considerable difference in the structure and chemical evolution of the ISM at the beginning of the simulation,
as the field is being amplified, dense magnetised regions are being formed through compression. More compact dense structures are formed in KS-lowB in
comparison to the filamentary structures of KS, seen also in Fig. \ref{ColH2}, 
upper row, first two panels (the aspect of the dense clumps in KS-lowB and KS are consistent in the y-z direction as well). The magnetic tension
in these clumps prevents the dense gas from being dispersed as easily as in the KS simulation,
keeping the molecular hydrogen more confined and hence better shielded.

The cold, dense molecular gas is embedded in the cool  ($T \in [300;8000)$ K) and warm ($T \in [8000;10^{5})$ K) surrounding environment (see Fig. \ref{ColH2}) and the
fact that the H$_{2}$ molecules reside in smaller clumps in KS is shown also by the higher cool and warm gas volume filling fraction (VFF) compared to KS-lowB.
In Fig. \ref{VffBlowB} we present a series of averaged (over 4 Myr) VFFs for KS and KS-lowB at 100, 125, 150 and 175 Myr. Significant differences in cool (yellow
column) and warm (green column) gas VFFs appear around 100 Myr, shortly before the H$_{2}$ mass fractions starts to diverge. The difference in the structure of the
dense gas up to this point is enough to make the self-shielding of the molecular gas more efficient in contrast to the smaller clumps in KS. This effect can
be seen in the self-gravity runs as well.

Because the sequence of SNe is fixed in space and time, they typically explode in higher density regions in run KS.
In Fig. \ref{Hist} we present the cumulative histogram of the number of SNe as a function 
of the density of the environment they explode in. There is a clear difference between KS and KS-lowB (also KS-SG and KS-lowB-SG) showing that the SNe 
in KS-lowB (and KS-lowB-SG) explode in 
a more diffuse ISM than in KS (and KS-SG), leading to much denser clumps containing much more molecular gas (see Fig. \ref{ColDens} second row, 150 and 200 Myr) and an increasingly high hot gas VFF over time (see Fig. \ref{VffBlowB} at 150 and 175 Myr). The structure of the ISM in the KS-lowB simulation, at later stages, points to the onset of thermal runaway (described in detail by \citet{Gatto}).

According to \citet{Hon95} in the Milky Way, the H$_{2}$ mass fraction
varies from $\simeq$ 50\% around the galactic centre to $\simeq$ 2\% at 9.5 kpc distance away from the centre. In a more recent study, \citet{Pin13} show an average H$_{2}$ mass fraction
of $\sim 50$\% when taking into account the CO-dark H$_{2}$ gas component.
By varying the initial magnetic field between 0 and 3 $\mu$G we obtain H$_{2}$ mass fractions ranging from 0.1\% (KS-highB and KS-highB-SG) to 90\% (KS-lowB-SG).
The H$_{2}$ mass fraction we obtain after 150 Myr are in the observed range for KS-medB (20\%). 
If we consider a H$_{2}$:H surface density
ratio we obtain a 0.29 average value for KS-medB at 150 Myr, consistent with the observed values of $0.1-1$ \citep{Schru11}.

We test the dependence of the amplification of the magnetic field and the H$_{2}$ mass fraction evolution on the 
resolution in Appendix \ref{A}. The magnetic filed is slightly stronger for higher resolution, whereas the amount of H$_2$ is converged at our
fiducial resolution. We also investigate the differences between a simulation in which self-gravity acts during the entire evolution and a run 
in which we first allow SNe to generate a structured ISM before switching on self-gravity at t$=150$ Myr. We find that self-gravity switched on at a later time
in the simulation leads to isolated dense clumps embedded in a lower density gas than the MCs formed with self-gravity from the beginning of the simulation (see Appendix \ref{B}).

Figure \ref{B_M_temp} shows the evolution of the mass weighted magnetic field over time (left column) and the mass fractions
(right column) in different temperature regimes.  From the top to the bottom we show the low, medium and high initial field simulations. 
In the top left panel, the magnetic field in KS-lowB is stronger for colder gas (at 150 Myr the field is about 1.5 $\mu$G in $T<30$ K, 1.2 $\mu$G in $T \in [30;300)$ K and 0.9 $\mu$G in
$T \in [300;8000)$ K). In the self-gravity simulation the variation of the magnetic field in the cold ($T \in [30;300)$ K) and very cold ($T < 30$ K) regime is about one order of magnitude. In the top right panel,
the mass fraction in the very cold gas becomes high after the first 20 Myr (40\%) in the self-gravity run while it takes $\sim$ 200 Myr for the same mass fraction 
to be reached without self-gravity (see also Fig. \ref{B_sat_H2}). 

\begin{figure*}
\centering
\begin{minipage}{175mm}
\includegraphics[width=80mm]{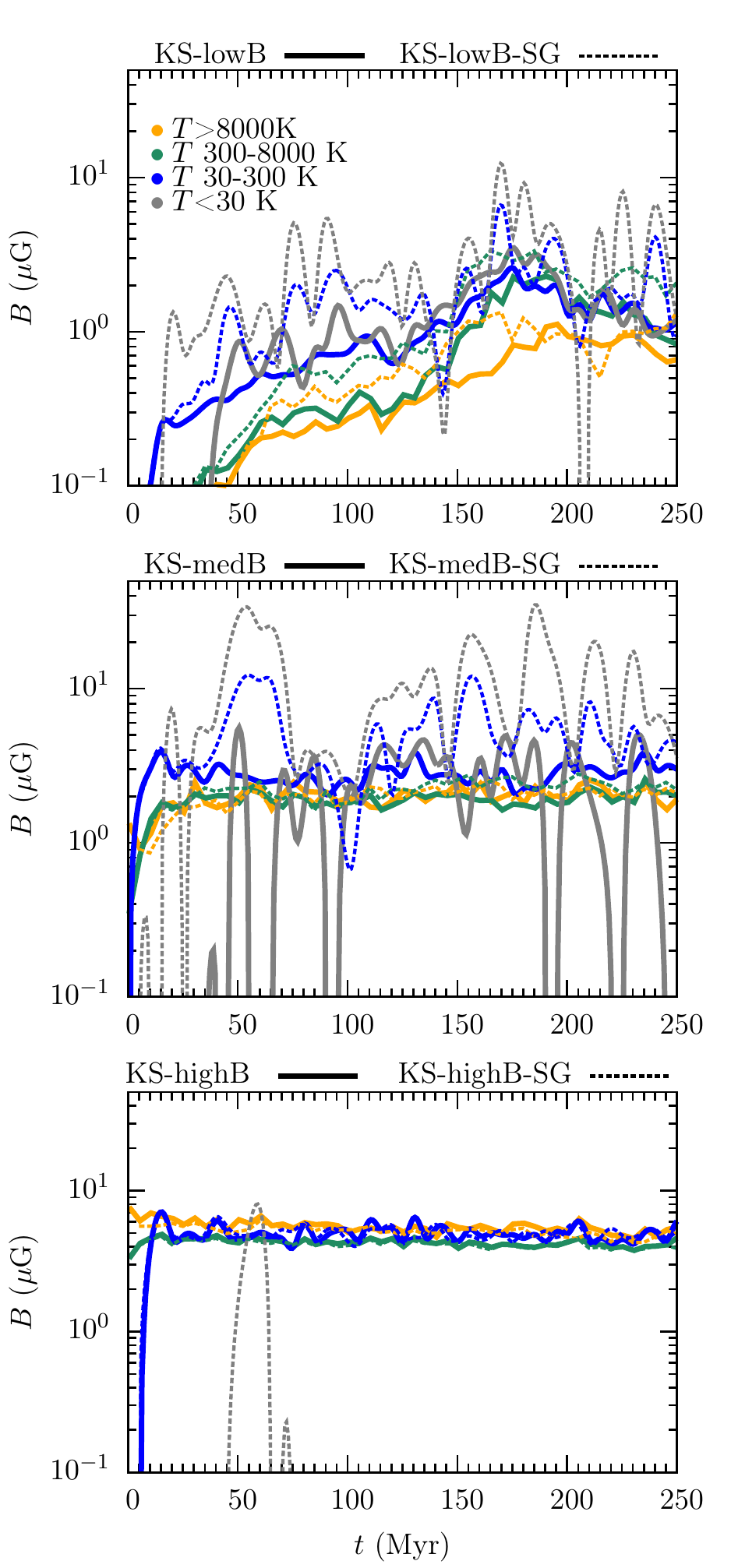}\quad
\includegraphics[width=80mm]{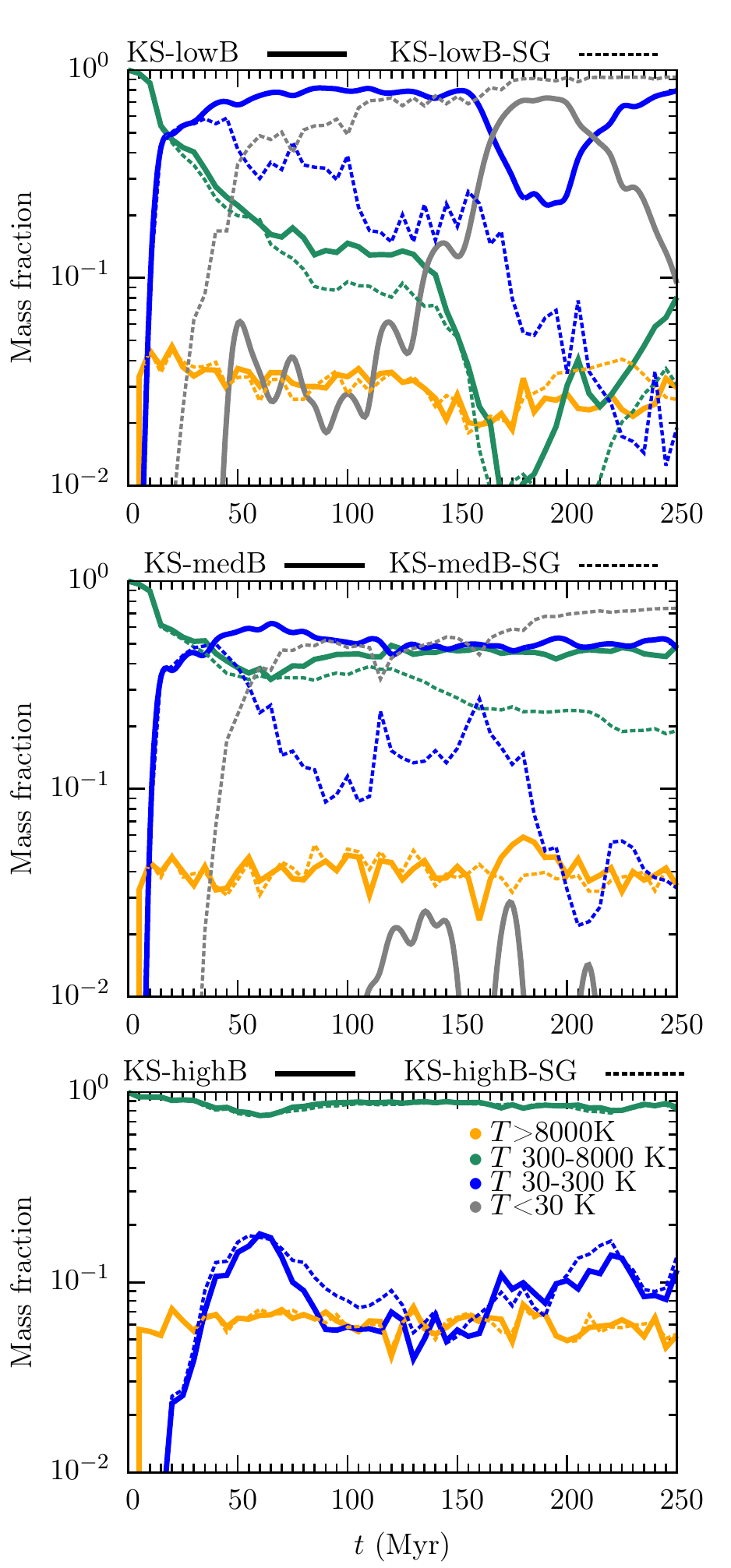}
\caption{Time evolution of the mass weighted magnetic field strength for KS-highB, KS-medB and KS-lowB (solid lines) and KS-highB-SG, KS-medB-SG and KS-lowB-SG (dashed lines) in different temperature regimes
(grey for $T<30$ K, blue for $T \in [30;300)$ K,
green for $T \in [300;8000)$ K and orange for $T \geq 8000$ K)- {\it left column}. The corresponding mass fraction evolution in the {\it right column} with the same line styles and colour-coding. The lower the temperature of the medium, the higher the average magnetic field in these regions. We interpret this as
a consequence of flux freezing, since the mean density of the cold phase is much higher than that of the intermediate or warm phases.}
\label{B_M_temp}
\end{minipage}
\end{figure*}

In Fig. \ref{B_M_temp} left middle panel (for KS-medB and KS-medB-SG) strong magnetic field variations in the very cold regime can be noticed for KS-medB 
(from below 0.1 to about 6 $\mu$G). This can be related to the fact that the mass fraction of the very cold gas is very low (Fig. \ref{B_M_temp} right middle panel) 
remaining less than 2\% over the time of the simulation. Also the magnetic field in the cold and very cold gas in KS-medB-SG is fluctuating by almost an order of 
magnitude above the field strength in the corresponding temperature regime in KS-medB. This is consistent with the mass fraction in $T<30$ K 
temperature regime that can reach 40\% in KS-medB-SG due to the strong compression of the gas. 

In KS-highB and KS-highB-SG, the magnetic field has approximately the same strength in all the temperature regimes (Fig. \ref{B_M_temp}, bottom panels). Here self-gravity 
does not have an impact on the strength of the field. There is less than 1\% gas with  $T < 30$ K in the simulations, hence the magnetic field in this temperature regime 
in KS-highB-SG is present only for a short time around 50 Myr. Most of the gas is at $T \in [300;8000)$ K with a mass fraction close to unity.

The magnetic field is stronger in colder temperature regimes and in denser regions. For our simulations, 
in the cold phase ($T\leq300$ K, $\rho\simeq 10^{-21}$ g/cm$^{3}$) the
magnetic field is around 4 $\mu$G, consistent with the observations of \citet{Beck13}. In the warm diffuse medium the observations show
a field strength of a few $\mu$G, similar to our results in KS-medB (2 $\mu$G) and KS-highB (4 $\mu$G). 

\begin{figure}
\includegraphics[width=82mm]{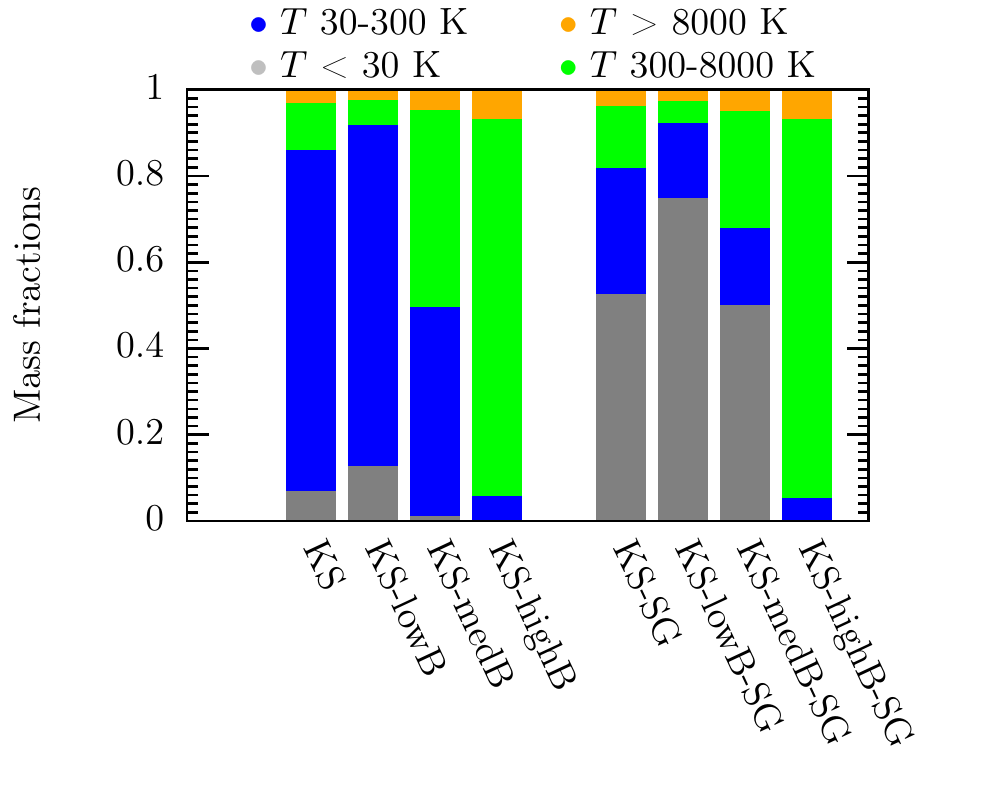}\\
\includegraphics[width=82mm]{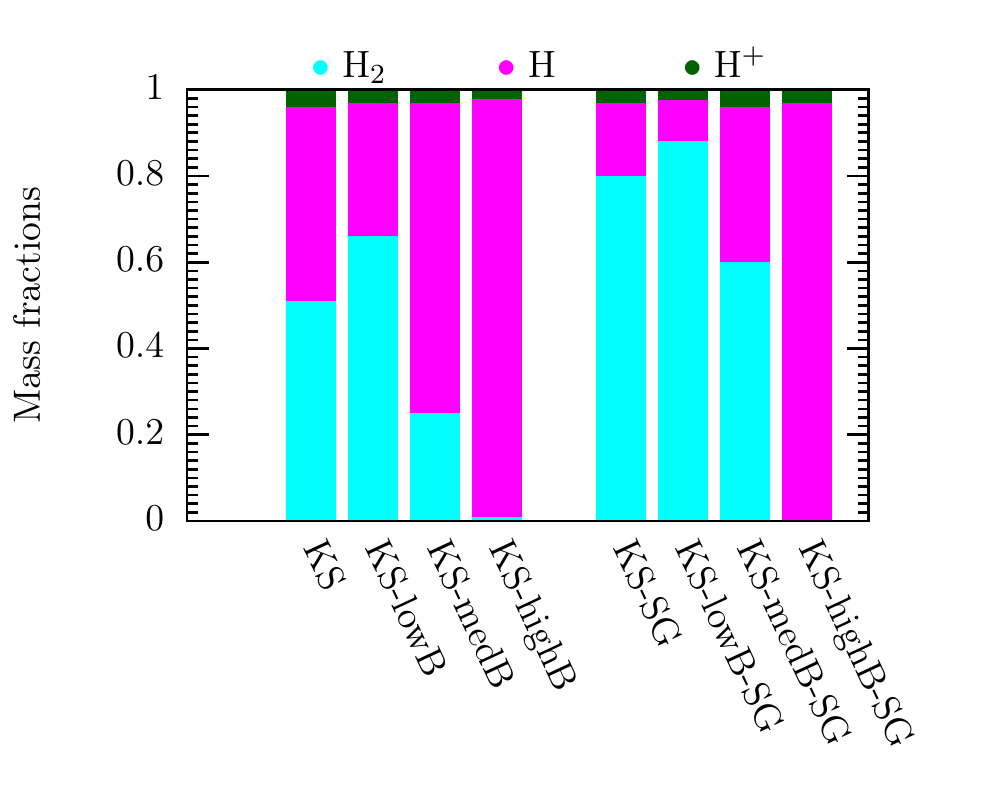}\\
\includegraphics[width=82mm]{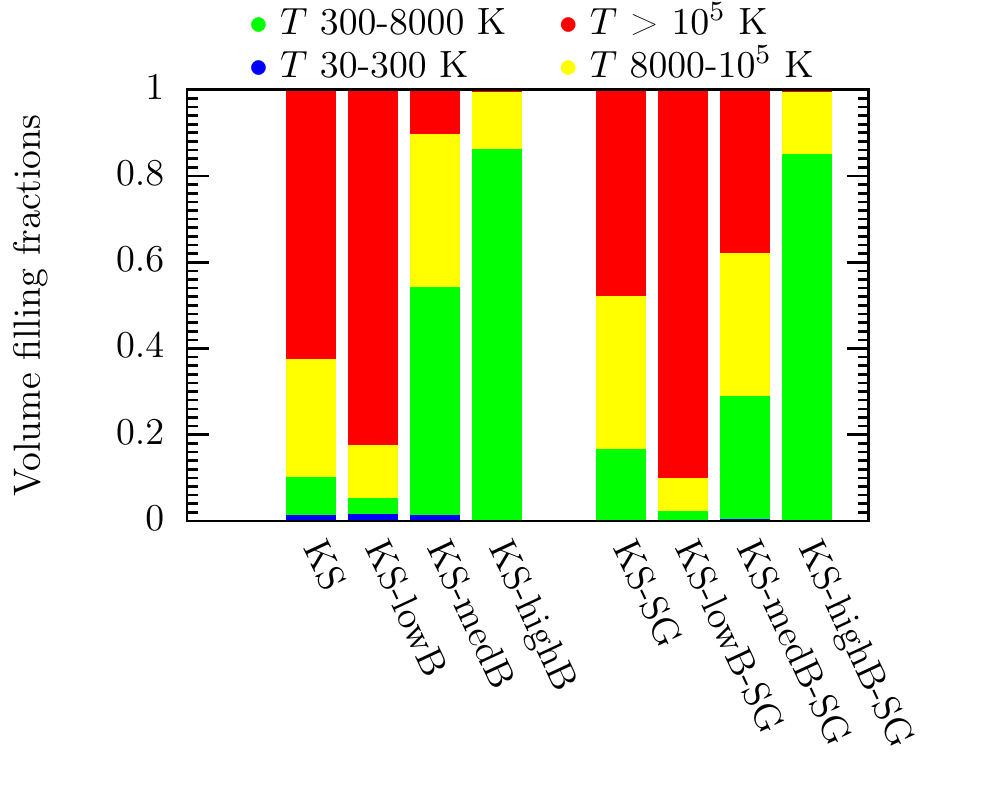}
\caption{Temperature regime mass fractions ({\it top panel}), H$_{2}$, H and H$^{+}$ mass fractions ({\it middle panel}) and volume filling fractions ({\it bottom panel}) at
150 Myr (averaged over 4 Myr) Colour-coding: grey - very cold gas ($T<30$ K), blue - cold gas ($T \in [30;300)$ K), green - cool gas ($T \in [300;8000)$ K, yellow - warm gas 
($T \in [8000;10^{5})$ K) and red - hot ionised gas ($T \geq10^{5}$ K), the orange mass fractions correspond to temperatures above 8000 K, light blue - H$_{2}$, magenta - H
and dark green - H$^{+}$.
High magnetic fields result in low cold gas, low H$_{2}$ mass fractions, high VFFs of warm gas and low VFF of hot gas. 
Self gravity increases the mass fraction of the colder gas component and slightly increases the hot gas VFF. There is no tight correlation between the very cold gas
and H$_{2}$ mass fraction.}
\label{Vff}
\end{figure}

In Fig. \ref{Vff}, we show the mass fractions (in the same temperature regimes as shown in Fig. \ref{B_M_temp}), the H$_{2}$, H and H$^{+}$ mass fractions 
and the volume filling fractions at 150 Myr, averaged over $\pm$ 2 Myr. 
Stronger magnetic fields significantly decrease the mass fractions of cold gas and H$_{2}$ and significantly increase the mass fraction and VFF of warm gas.
We notice this trend independent 
of self-gravity. The main effect of self-gravity is to increase the mass fraction of the coldest (and densest) gas component (grey bars in Fig. \ref{Vff}) and the H$^{+}$ 
mass fractions (light blue bars).
There is also a trend towards higher VFFs of the hot gas in runs with self gravity (red bars on the bottom panel of Fig. \ref{Vff}). This is a result of SNe
going off in low density regions \citep[as also seen in][]{Gatto, Li15}. According to \citet{Snow98} the hot gas VFF is $\simeq$ 60\% while \citet{Hei03} 
give $\simeq$ 50\% for the warm neutral gas and $\simeq$1.5\% for the cold gas.
In all the simulations the hot gas VFF varies between 0.1 \% (KS-highB) and 90\% (KS-lowB-SG) while the KS-medB simulations reproduces best the observations (the same 
simulation that obtains the best match to the H$_{2}$ mass fraction). 
When comparing the top and middle panels of Fig.\ref{Vff} we notice that there is no one-to-one correlation between the very cold gas and H$_{2}$ mass fractions
\citep[also shown by][]{Walch15}. Even though 
the trend with the initial magnetic field is similar, for the runs without self-gravity the ratio of mass fractions of the very cold gas and H$_{2}$ is 13\% for KS, 19\% for 
KS-lowB and 4\% for KS-medB. In comparison to the self-gravity runs where this ratio is much higher: 65\% for KS-SG, 85\% for KS-lowB-SG and 83\% for KS-medB-SG.

\begin{figure}
\includegraphics[width=80mm]{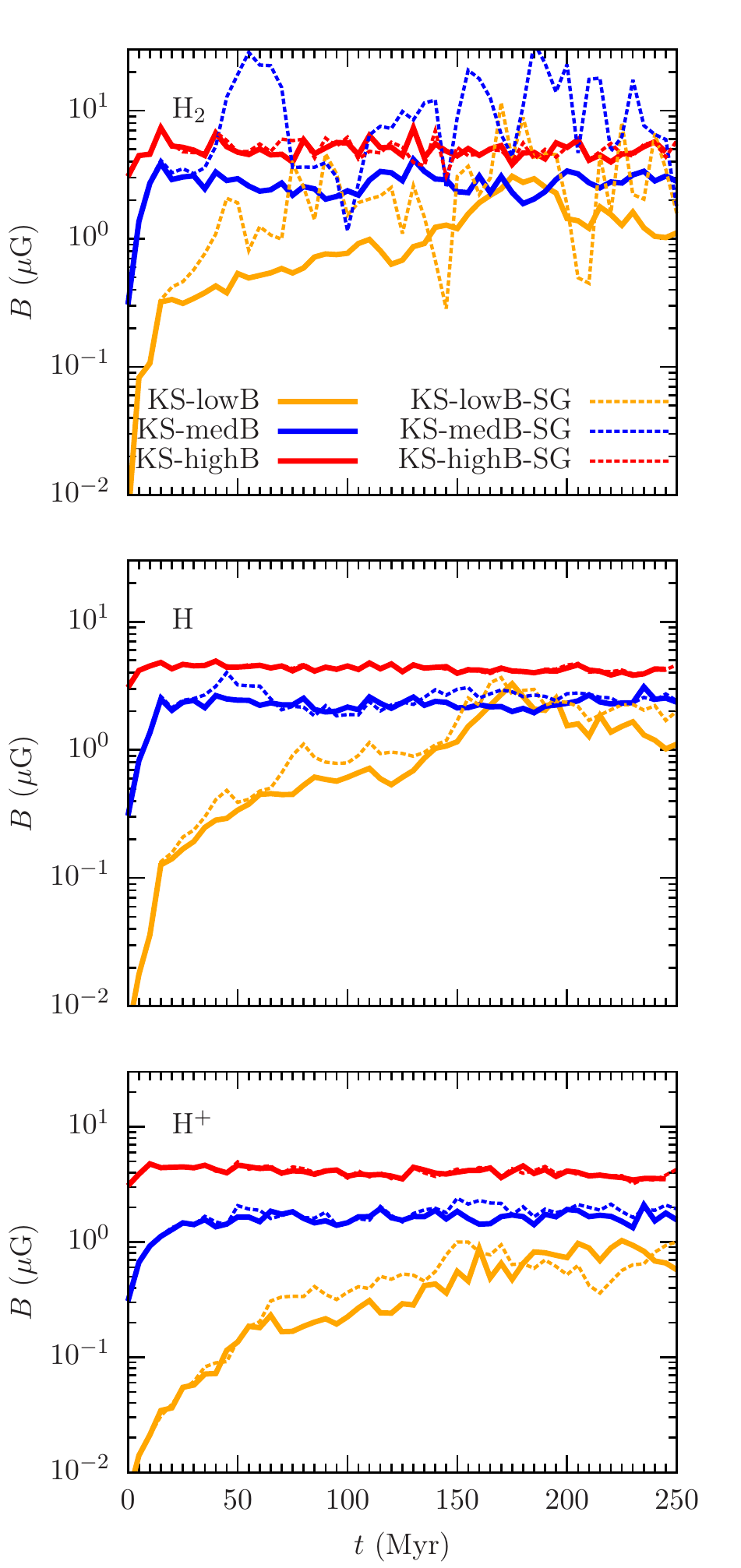}
\caption{Mass weighted magnetic field evolution in H$_{2}$ ({\it top panel}), atomic H ({\it middle panel}) and H$^{+}$ ({\it bottom panel}). 
KS-lowB (solid orange line), KS-medB (solid blue line), KS-highB (solid red line), KS-lowB-SG (dashed orange line), KS-medB-SG (dashed blue line) and KS-highB-SG (dashed red line).
The magnetic field reaches up to 10 $\mu$G in  H$_{2}$ with self-gravity while for KS-medB the field is around 3 $\mu$G in H and 2 $\mu$G in  H$^{+}$.}
\label{B_Chem}
\end{figure}


In Fig. \ref{B_Chem} we present the average H$_{2}$, H and H$^{+}$ mass weighted magnetic field (from top to bottom).
In H$_{2}$ the magnetic field is the strongest: after 150 Myr the field is 1.2 $\mu$G in KS-lowB, 2 $\mu$G in KS-medB and 4.5 $\mu$G in KS-highB. Again self-gravity results in fluctuations in the mass weighted field (also seen in Fig. \ref{B_sat_H2},  middle panel). Due to the efficient formation of the very high density gas, the magnetic field can be locally amplified to over 10 $\mu$G. In atomic and ionised hydrogen, self-gravity does not impact the magnetic field significantly. In atomic hydrogen the magnetic field is about 1 $\mu$G in KS-lowB, 2 $\mu$G in KS-medB and 4.5 $\mu$G in KS-highB. The hot ionised gas is formed when the 
atomic gas is heated by SN blasts. Characterised by low densities and high temperatures, in this regime the field is the weakest: 0.5 $\mu$G in KS-lowB, 1.8 $\mu$G in KS-medB. In the case of KS-highB, the magnetic field strength is very similar 
in all temperature regimes with  4 $\mu$G in the hot gas at 150 Myr. 


For the further characterisation of the physical state of our model ISM, we show the volume and mass weighted root mean square (RMS) of the Mach number in Appendix \ref{C}, Fig. \ref{RMS}.

\subsection{Pressure components}
\label{Thmagpress}

\begin{figure*}
\centering
 \begin{minipage}{\textwidth}
\includegraphics[width=122mm]{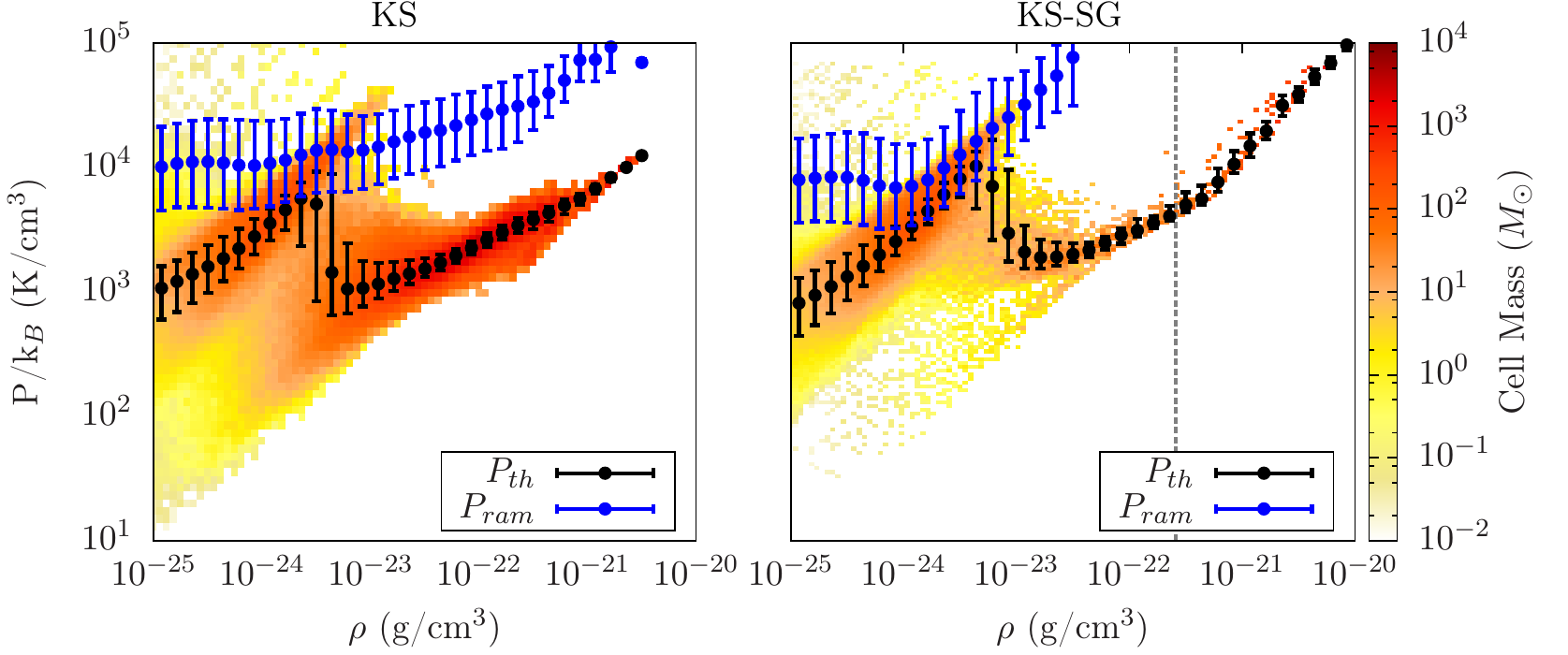}\\
\includegraphics[width=182mm]{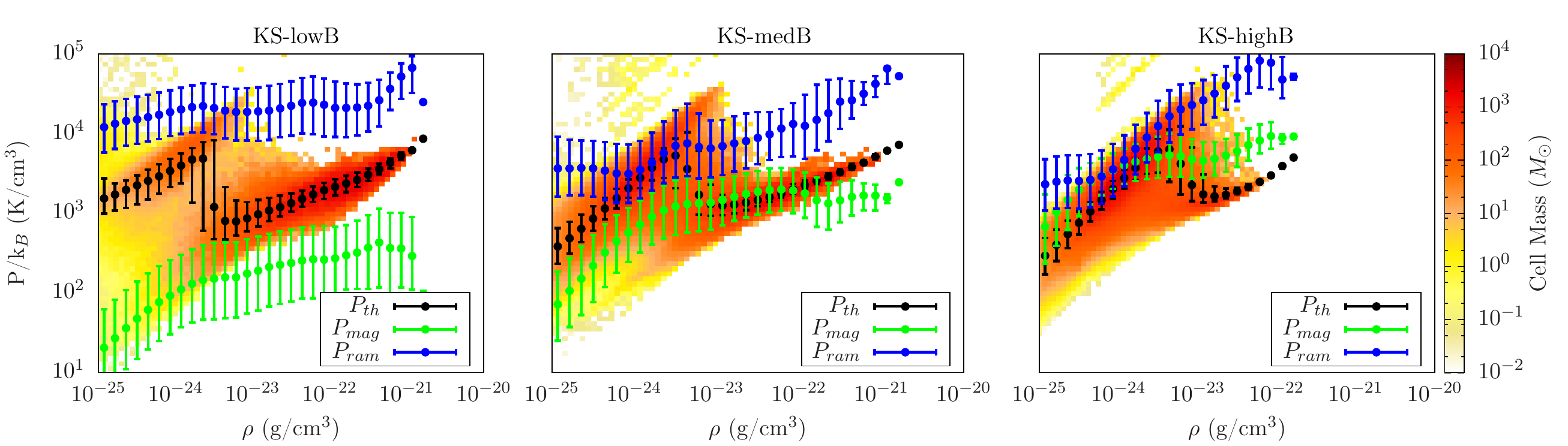}\\
\includegraphics[width=182mm]{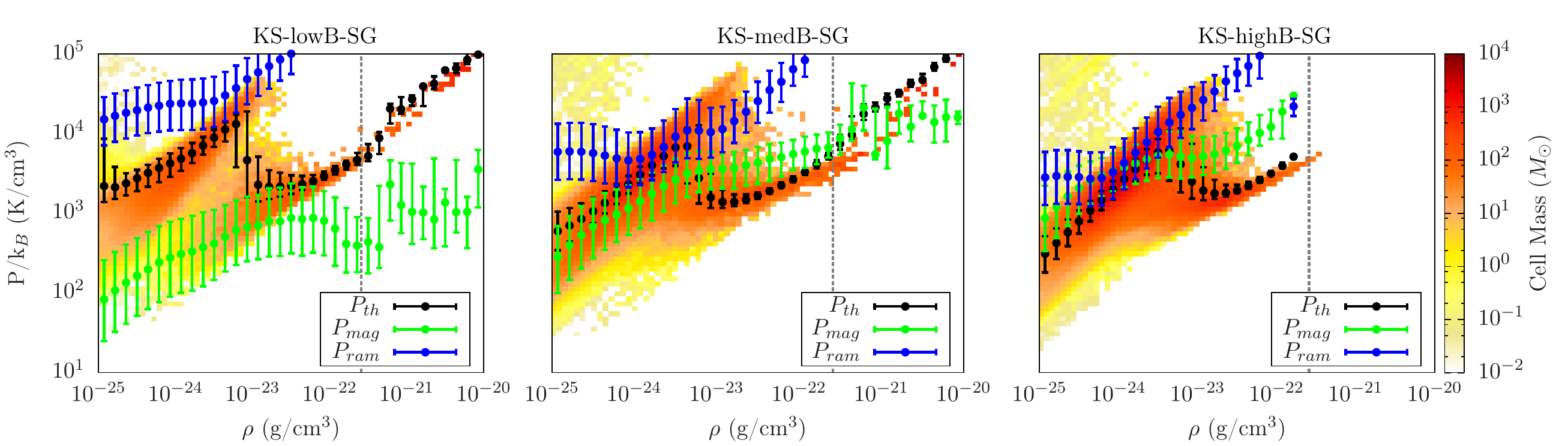}\
\caption{Pressure-density phase plots with mass colour-coding at 150 Myr. {\it Top row}: KS (left) and KS-SG (right), KS-lowB (left), {\it middle row}: KS-medB (middle) and KS-highB (right) and {\it bottom row}: KS-lowB-SG (left), 
KS-medB-SG(middle) and KS-highB-SG (right). The black points stand for the median thermal pressure, green for the median magnetic pressure and blue for the ram pressure averaged over
148-152 Myr, with the bar limits representing
the 25th and 75th percentile. The grey dotted line represents the maximum density we can resolve according to the Truelove criterion, computed assuming a temperature of 100~K.
The stronger the initial field, the lower the maximum density that can be reached.
Ram pressure dominates over the thermal and the magnetic pressure in all the simulations, over the entire density regime.
Self-gravity allows the gas to reach a much higher maximum density.}
\label{phase}
\end{minipage}
\end{figure*}

We present a diagram of thermal, magnetic and ram pressure as a function of density on top of the mass weighted pressure-density distribution
(Fig. \ref{phase}) for the hydro runs (top row), the magnetic simulations 
without self-gravity (middle row) and with self-gravity (bottom row). 
The initial field strength increases from left to right. 
We take the median pressure for each density bin (equal in log space and a number of 35 between $10^{-25}$ and $10^{-20}$ g/cm$^{3}$). The upper limit of
the bars represents the 75th percentile and the lower one the 25th percentile. We average over 4 Myr around 150 Myr. 
The colour map shows the mass of the gas, which indicates how 
the gas is distributed between the thermally unstable regime, the stable warm and the stable cold branch. 
The hydro simulations reach a higher maximum density (see Fig. \ref{PDF} left panels) and contain less thermally unstable gas than the magnetised boxes.
This thermally unstable gas represents between 10\% (low initial field) and 90\% (high initial field) of the total gas mass. According to \citet{Hei03}, 30\% of the mass fraction
of the solar neighbourhood is thermally unstable (from the Millennium Arecibo 21 cm survey) which is very close to the thermally unstable gas mass fraction we 
obtain in the KS-medB and KS-medB-SG runs.

The grey dotted line represents the approximate density we can resolve. Applying the Truelove criterion \citep{True} in the simulations with self-gravity, the Jeans length should 
be resolved with at least 4 cells in order to avoid spurious fragmentation in numerical collapse calculations. For a fixed 8 pc Jeans length the critical density scales linearly 
with the temperature. From Fig. \ref{PDF} right panel we see that there is a considerable amount of gas at T$=100$ K in all the simulations and for this temperature the 
maximum density we can resolve is about 2.6 $\times 10^{-22}$ g/cm$^{3}$.

The highest density reached in the simulations depends on the magnetisation of the environment: $1.65\times 10^{-21}$g/cm$^{3}$ in KS-lowB
(same as KS-medB) and $1.64\times 10^{-22}$ g/cm$^{3}$ in KS-highB (shown also in Fig. \ref{ColDens}).

As discussed earlier, the KS-highB-SG model does not develop as high densities as its counterparts 
at lower field strengths. There could be several reasons for this behaviour. First, we notice that
the density structure in model KS-highB-SG is much more uniform than for lower field strengths. This
suggests that the magnetic field ``cushions" the effect of the supernova explosions, preventing 
efficient compressions and sweep-up of gas to large (column) densities. Second, the accumulation
length along the (strong) field may be too short in our box to allow the build-up of a column
density large enough for self-gravity to dominate. This column density is determined
by the critical mass-to-flux ratio, which for a sheet of infinite extent results in $N_c = B/(2\pi m G)$,
with the critical column density $N_c$, the magnetic field strength $B$, and the particle mass $m$
\citep{Cr12}. In numbers, this corresponds to 
\begin{equation}
  N_c=3.7\times 10^{20} \left(\frac{B}{\mu G}\right)\,{\rm cm}^{-2},
\end{equation}
at a mean molecular weight of $\mu=1$. For our strongest field of $3\mu$G, this corresponds to 
$N_c=1.1\times10^{21}$~cm$^{-2}$, yet the mean column density in our simulation box 
is $N=3.9\times10^{20}$~cm$^{-2}$. In other words, due to the strong magnetic field, the only
way to build up high column densities is by flows {\em along} the field lines, resulting in the
mean column density dictating the onset of gravitational instability as well as H$_{2}$ formation \citep[e.g.][]{Berg}
. Conversely, in the models without, or with sufficiently weak magnetic
fields, motions perpendicular to the field lines can contribute to the build-up of (local)
column densities, enhancing the shielding, and allowing for gravitational collapse \citep[see also discussion in][]{Hei08}.


For the low initial field simulations, up to a density of 10$^{-24}$ g/cm$^{3}$, the thermal pressure
is more than two orders of magnitude higher than the magnetic pressure and one order of magnitude lower than the ram pressure, $P_\mathrm{ram}=\rho v^2$.  
This is the region where the gas
has high temperatures and it is frequently heated by SN blasts. Keep in mind, however, that the mean magnetic field strength is resolution dependent (see Appendix \ref{B}).
Between 10$^{-24}$ and 10$^{-23}$ g/cm$^{3}$, in the
thermally unstable regime, the thermal pressure drops significantly while the magnetic pressure keeps rising. However, 
the magnetic pressure remains always lower than the thermal pressure in KS-lowB and KS-lowB-SG (see Fig. \ref{DensMag}). In the KS-medB case the two pressures are comparable in the cold 
regime while still dominated by turbulence. In KS-highB, the magnetic pressure is high enough in all density regimes to dominate over the thermal pressure but remains 
below the ram pressure.   

Self-gravity allows the formation of denser structures, hence a more than one order of magnitude difference in the maximum density reached between KS-lowB,
KS-medB and KS-highB (top row) and the self-gravity runs (bottom row) develops. Above $\rho\sim10^{-23}\,\mathrm{g/cm^3}$ the ram pressure grows
approximately linearly with density in these simulations indicating that $v^2$ does not vary as a function of $\rho$. In all simulations, $P_\mathrm{ram}$ is
the dominant energy component.
For KS-lowB-SG, the magnetic pressure grows with density up to $\sim10^{-22}$ g/cm$^{3}$ and then flattens while the thermal
pressure grows linearly. In KS-medB-SG we notice similar general trends in the scaling of magnetic pressure. However, $P_\mathrm{mag}$ is overall higher 
and exceeds $P_\mathrm{ram}$ for $10^{-23}\,\mathrm{g/cm^3}\lesssim\rho\lesssim10^{-22}\,\mathrm{g/cm^3}$. In KS-highB-SG the magnetic pressure exceeds the 
thermal one over the entire density range.


\begin{figure*}
\centering
\begin{minipage}{175mm}
\includegraphics[width=168mm]{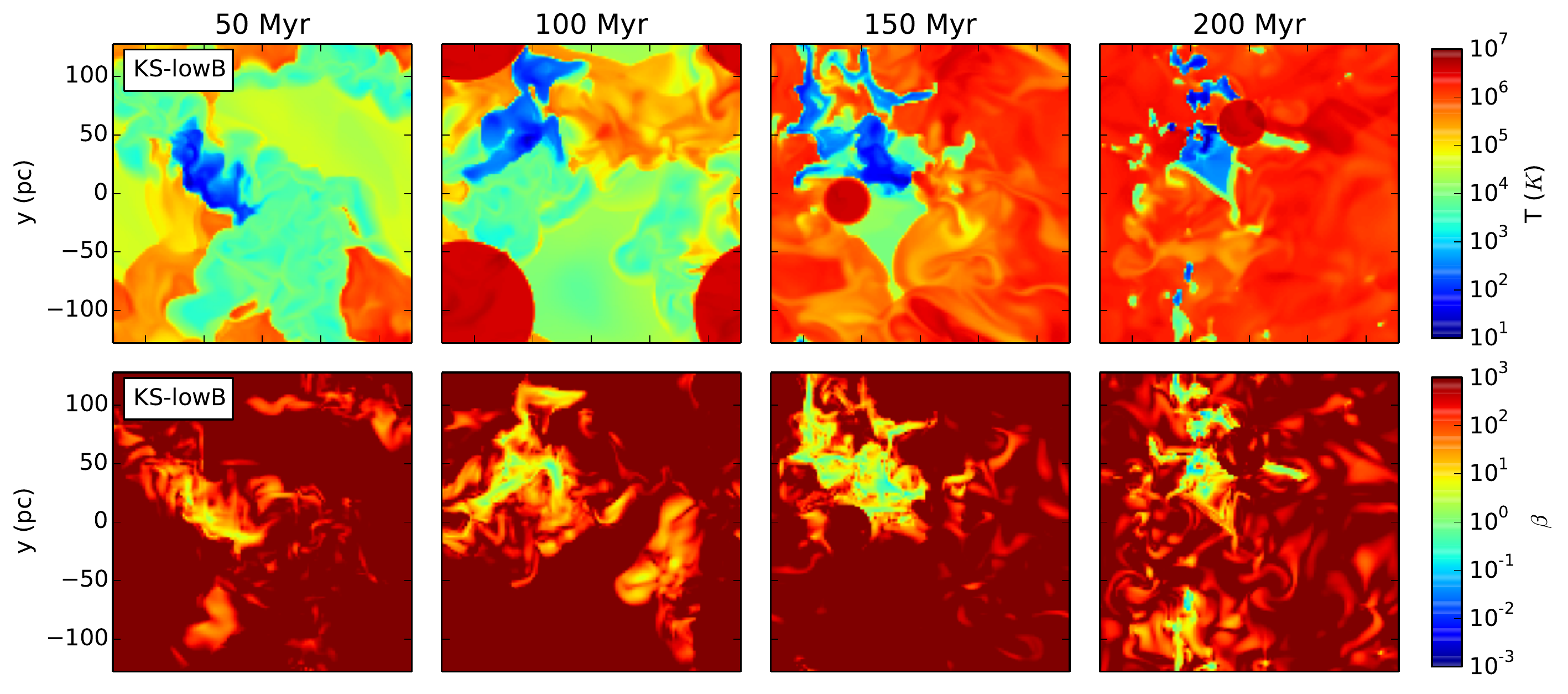}
\includegraphics[width=168mm]{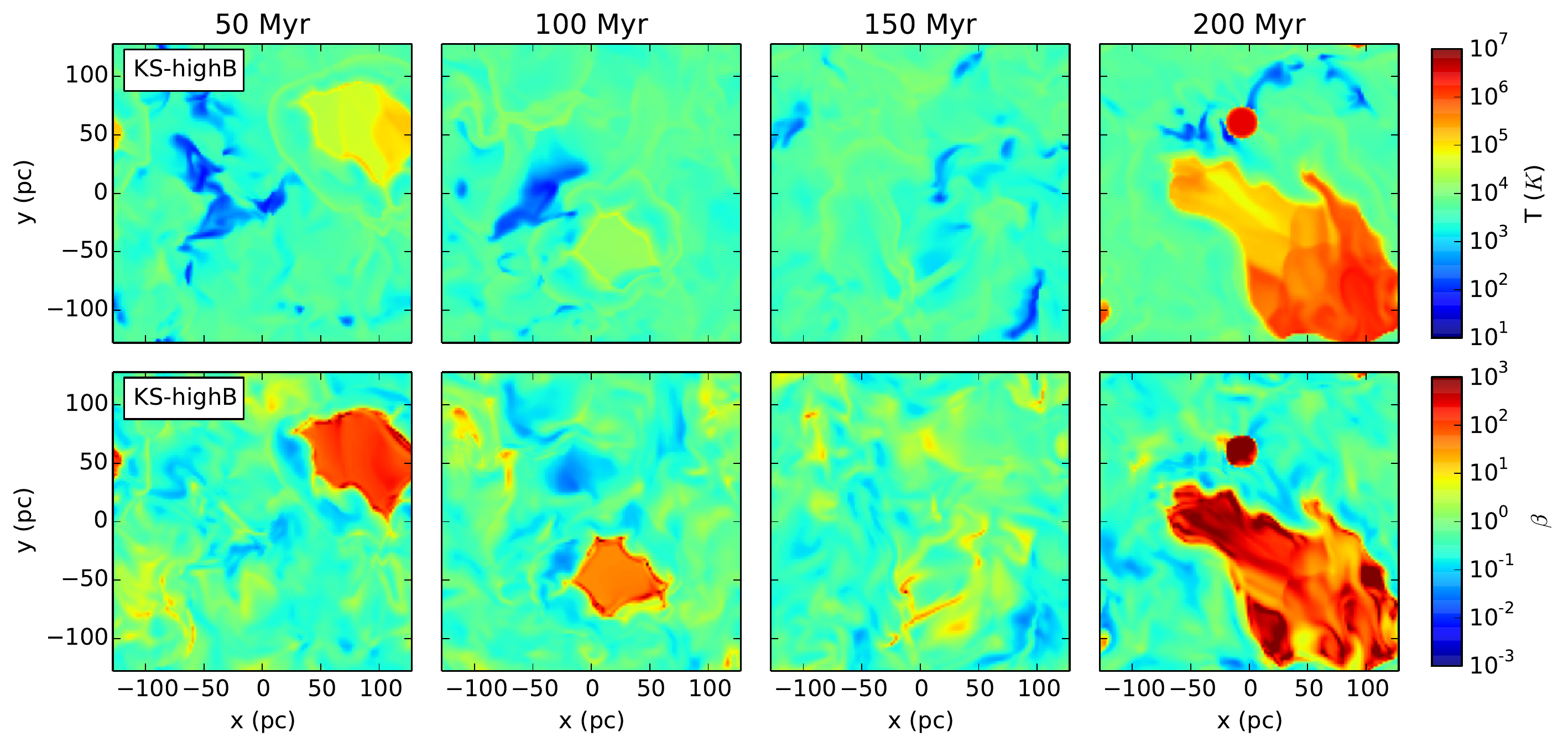}
\caption{Time sequence of temperature slices through the plane defined by the densest region along the z axis ({\it top row}) at 50, 100, 150 and 200 Myr and the plasma beta parameter ({\it second row}) 
for KS-lowB, same for KS-medB ({\it third} and {\it fourth row}) and KS-highB ({\it fifth} and {\it bottom row}). 
High temperature and high plasma beta (low temperature and low plasma beta) are well correlated only locally, for extreme temperatures.}
\label{TempMag}
\end{minipage}
\end{figure*}

\begin{figure}
\includegraphics[width=85mm]{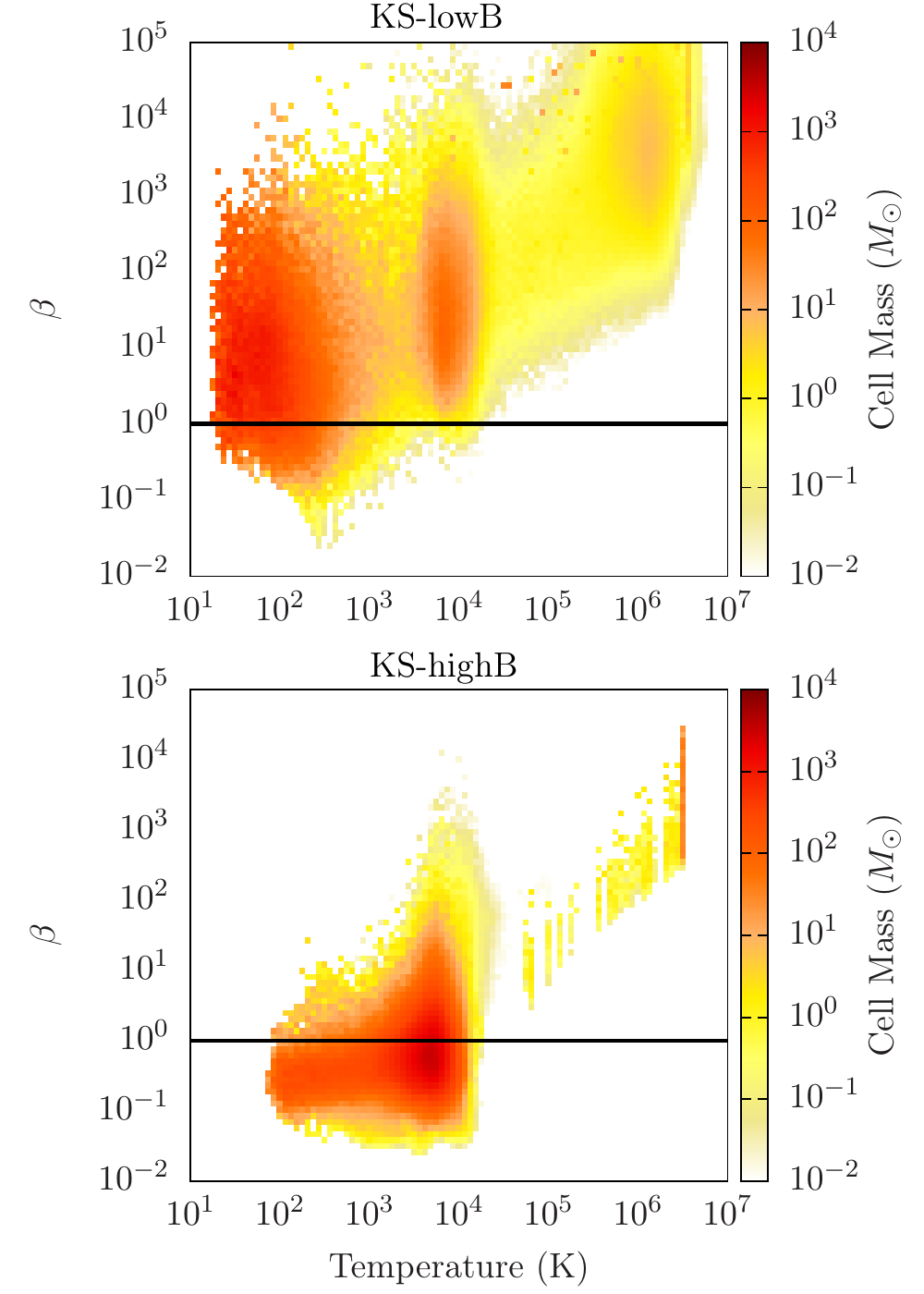}
\caption{Plasma beta parameter as a function of temperature at 150 Myr for KS-lowB ({\it top panel}) and KS-highB ({\it bottom panel}). While only a small mass fraction lies below $\beta = 1$  in KS-lowB, most of the gas in KS-highB is magnetically dominated.}
\label{BetaTemp}
\end{figure}

The plasma beta parameter (the ratio of the thermal to magnetic pressure $\beta= P_{\mathrm{th}}/P_{\mathrm{mag}}=8 \pi P_{\mathrm{th}}/B^{2}$) is shown in Fig. \ref{TempMag}. We present slices of the temperature and the corresponding plasma beta through the densest region in the simulation box for different times. The top two rows are for KS-lowB and the bottom two rows for KS-highB. 
Locally, in the extreme cases, regions with very low temperatures correspond to low plasma beta and also extremely high temperature regions correlate with very high plasma beta (see the upper tails in Fig. \ref{BetaTemp}). However, overall this correlation is not obvious for intermediate temperatures (e.g. in the upper two rows, the regions with temperatures above $10^{4}$ K correspond to high plasma beta, but in the bottom rows temperatures lower than 10$^{4}$ K do not have a one-to-one correspondence to low plasma beta values).

We illustrate in Fig. \ref{BetaTemp} a two-dimensional histogram of the plasma beta parameter as function of temperature, 
with the black line indicating the $\beta = 1$ limit. For KS-lowB, the scatter in $\beta$ is very large (4 to 5 orders of magnitude) in any temperature regime,
showing that the magnetic field is still being amplified at this time, even in the cold, dense clumps. In contrast, in KS-highB, most of the gas is magnetically
dominated. However, the same large scatter in $\beta$ can be seen for temperatures between $10^{3}$ and $10^{4})$ K. This region corresponds to the peak in the volume and mass weighted
density PDFs (see Fig. \ref{PDF}).



\begin{figure*}
\centering
 \begin{minipage}{\textwidth}
\includegraphics[width=182mm]{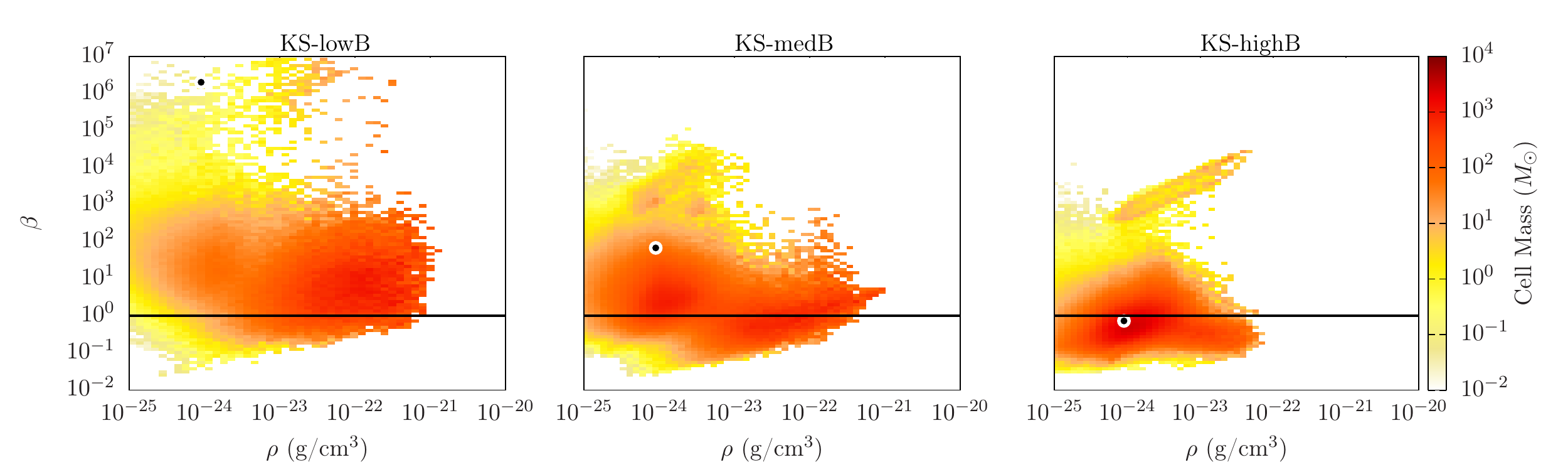}\\
\includegraphics[width=182mm]{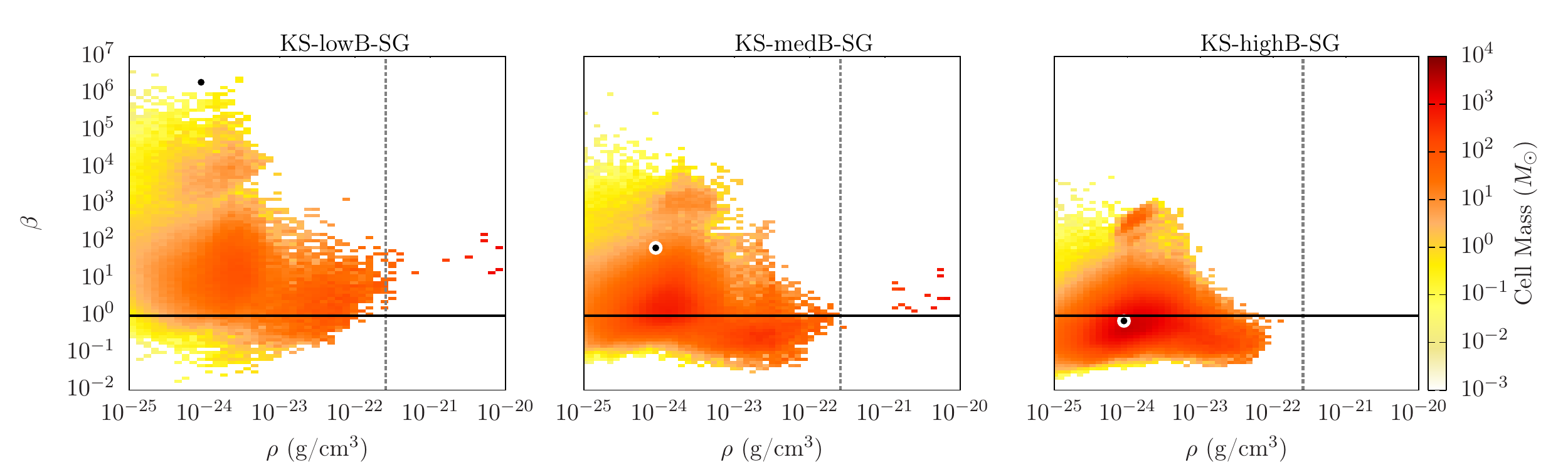}
\caption{Plasma beta as a function of density at 150 Myr. In the {\it top row}  we show KS-lowB (left), KS-medB (middle) and KS-highB (right). In the {\it bottom row}: 
KS-lowB-SG (left), KS-medB-SG (middle) and KS-highB-SG (right). The black line marks the limit $\beta$ = 1. The black point at $10^{-24}$ g/cm$^{3}$ represents 
the initial state of the simulation. The grey dotted line represents the maximum density we can resolve according to the Truelove criterion. 
The stronger the initial field the more the magnetic pressure dominates over the thermal pressure in the dense gas. With self-gravity we obtain very 
dense structures with high thermal pressures. There is less mass below  $\beta$ = 1 with self-gravity in comparison to the simulations without 
self-gravity. }
\label{Beta}
\end{minipage}
\end{figure*}

 Figure \ref{Beta} is a two-dimensional histogram of the plasma beta parameter as a function of density.  In the top row we plot $\beta$ for the simulations without self-gravity with increasing magnetic field strength from the left to the right and on the bottom row, the corresponding simulations with self-gravity, all at 150 Myr. The black points at $10^{-24}$ g/cm$^{3}$ represent the initial states of the simulations.
Because of the very low initial seed field, the initial plasma beta of KS-lowB and KS-lowB-SG is about $\beta = 2\times10^{6}$.
Most of the mass in the intermediate and low initial field simulations shifts from a large beta towards the $\beta = 1$ limit and below (for KS-medB and KS-medB-SG) 
which reflects the amplification of the field via the dynamo effect and compression of the gas in combination with flux-freezing. In KS-highB and KS-highB-SG 
the gas is initially completely magnetically dominated and over time due to the SNe that inject 
thermal energy into the ISM, some of the gas reached higher plasma beta values. However, during the time of the simulations, 70\% of the gas mass is magnetically dominated. 

For the simulations without self-gravity the magnetically dominated gas mass is larger with stronger magnetic field (7\% in KS-lowB, 30\% in KS-medB and 71\%in KS-highB). 
For the self-gravity simulations, the KS-highB-SG mass fraction with $\beta < 1 $ is very similar to KS-highB. In KS-lowB-SG and KS-medB-SG the very dense clumps formed early in the simulations (see Fig. \ref{ColDens}) have high thermal pressures leading to a lower mass fraction of the magnetically dominated gas (1\% in KS-lowB-SG and 16\% in KS-medB). 


\subsection[]{Magnetic field scaling with density}
\label{Scal}

The energy balance between kinetic and magnetic energy in simulations depends on two factors: the magnetic field amplification and dissipation.
Recent studies describe the amplification of a magnetic seed field due to gravitational collapse through the small-scale turbulent dynamo
\citep{Sr10,Fed11,Peters,Sch12b}. 
Unfortunately, in this paper, we cannot discuss the mechanisms that lead to the amplification of the magnetic field in detail
because of limited resolution. 
The magnetic field evolution has been studied in detail in isothermal
environments with 1024$^{3}$ and 2048$^{3}$ cells \citep[see][]{Krit09,Jones11, Fed14b} but our demanding thermodynamics makes a resolution of 512$^{3}$ cells already
hard to achieve (see Appendix B). Another reason is the wide temperature range and the chemistry of the ISM model
which makes it difficult to recover key parameters for describing the dynamo (Magnetic Prandtl number $P_{\rm m}$, Reynolds number $R_{\rm e}$, Magnetic
Reynolds number $R_{\rm m}$, for details see \citet{Krit,Sch12b}). In addition, we face numerical dissipation of the magnetic field
without specifically including diffusive terms in the MHD equation.
Tests of decaying supersonic MHD turbulence have been conducted by \citet{Krit} on state-of-the-art MHD
codes. The study shows that the magnetic energy in all codes
drops by 6\% to 30\% over 0.2 turnover times, after the turbulent driving stops. As the initial conditions are identical, this difference is only due to the intrinsic
numerical dissipation of the MHD solvers. However, \citet{Laz} concluded the numerical diffusivity can account for fast magnetic reconnection which removes 
the magnetic
field from the turbulent plasma during star formation. In this sense, numerical dissipation could provide a good representation of reconnection diffusion.

In our case, the situation is more complex. Our simulations are not isothermal and  we introduce motions by SN explosions. Hence, 
the dynamical driving scales as well as the scales on which magnetic dissipation and amplification occur are not as well defined as in 
idealised turbulence simulations with, e.g. Ornstein-Uhlenbeck driving \citep[see, e.g.][]{Pope}. The magnetic field amplification and dissipation 
occurs in gas with very different densities and temperatures because SNe go off in an already highly structured, magnetised, multi-phase ISM.

\begin{figure}
\includegraphics[width=85mm]{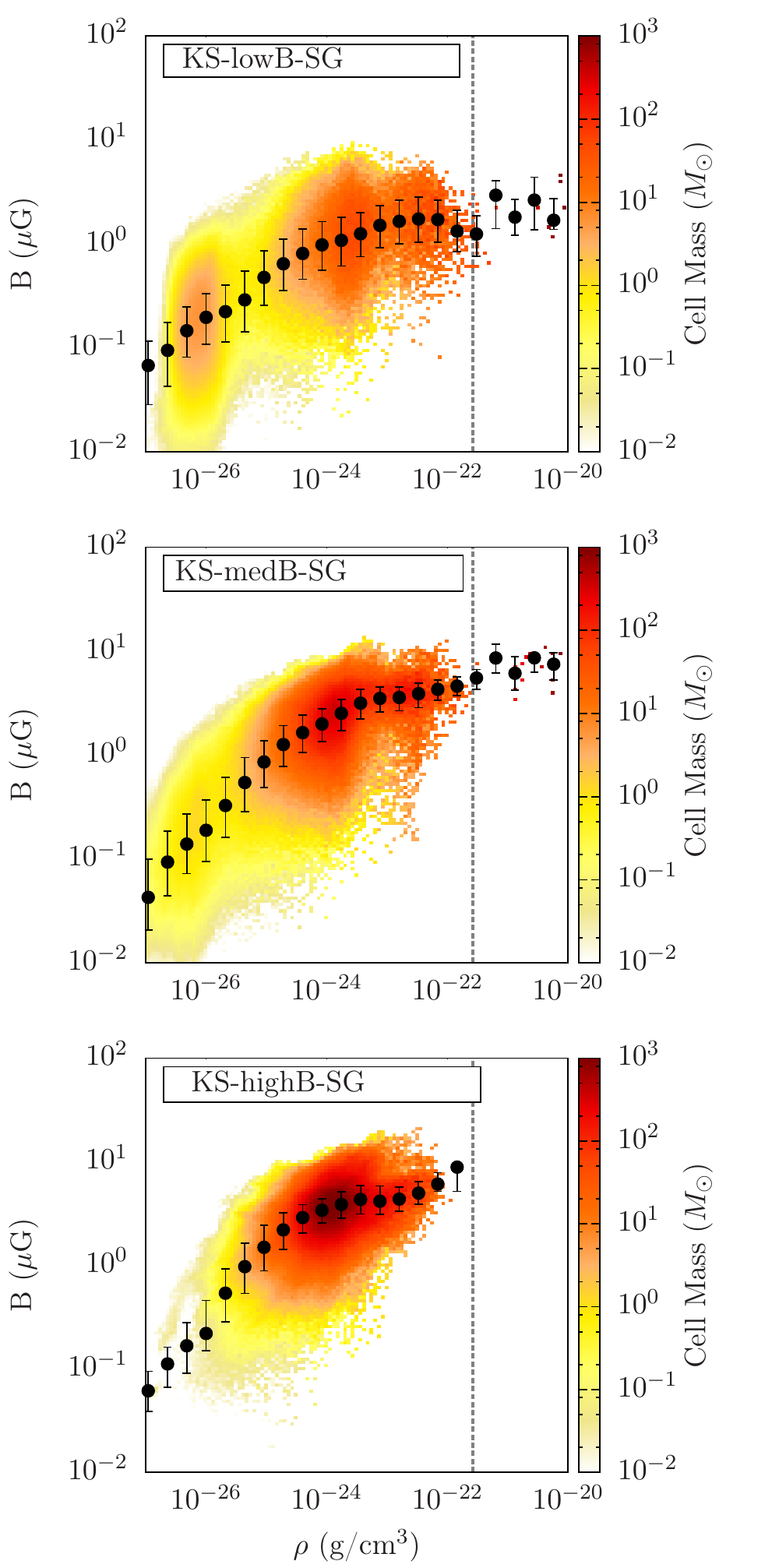}
\caption{Scaling of the magnetic field with density for KS-lowB-SG around 150 Myr
(black dots)- {\it top panel}, KS-medB-SG ({\it second panel}) and KS-highB-SG ({\it bottom panel}). The grey dotted line represents the maximum
den
sity we can resolve according to the Truelove criterion. There is a large scatter in the magnetic field strength with density indicating that the gas in the box is dominated 
by the kinetic energy density.}
\label{B_sc}
\end{figure}

In Fig. \ref{B_sc} we show a 2D profile of the magnetic field as a
function of density, at 150 Myr, with the mass color-coded.
We over-plot the median volume weighted magnetic field with the bars representing the 75th and 25th percentile and 
average the field over 4 Myr (between 148-152 Myr). The scaling in the runs without self-gravity is very similar to 
the one in the simulations with self-gravity, which is shown here. 

There is a large scatter in the magnetic field values for a given density suggesting that the gas is driven by turbulence rather than channeled by the magnetic field. This is the case in all our simulations, as
seen in Fig. \ref{phase}; the ram pressure dominates over the magnetic and thermal pressure \citep[see also][]{Avill05}. 
At low densities, below 10$^{-24}$ - 10$^{-25}$ g/cm$^{3}$, the gas is frequently heated by the SN. Every time the SNe blasts blow away 
the gas they swipe along the field lines too, so the low density tail of the scaling reflects the way we introduce turbulence 
into the simulations.

The magnetic field strength does not increase with $\rho$ over the entire density range. Above $\sim10^{-24}\,\mathrm{g/cm^3}$ the scaling flattens. This
flattening could be a numerical artefact due to the limited resolution and a missing dynamo effect in dense regions, which are confined to small spatial scales 
and thus unresolved at higher densities.

Due to the numerical limitations, it is difficult to compare our scaling to other previous studies \citep{Li,Collins}. Furthermore, we have only a small number of molecular clouds formed in our simulations.
However, we compare the magnetic field strength in the high density gas to the observations of \citet{Cr12} and even though the magnetic field in our densest resolved structures
is lower than the upper limit of the line-of-sight magnetic field, we are within the observed range. 

\section{Conclusions}
\label{Conc}

We use the 3D magnetohydrodynamic code FLASH in version 4 to study the structure of the magnetised, multiphase interstellar medium and focus on the connection between magnetic
fields and the chemical composition. We use periodic boxes with a volume of $(256\,\mathrm{pc})^3$ with different initial magnetic field strengths of 
6$\times10^{-3}$, 0.3 and $3\,\mu\mathrm{G}$. SNe are injected as individual explosions at a fixed rate at random locations. We focus on the chemical 
evolution following mainly ionised, atomic and molecular hydrogen over a total simulation time of 250 Myr. Our results can be summarised as follows:

\begin{itemize}

\item The mass weighted magnetic field in simulations with low and
medium initial field strengths saturate at 1-3 $\mu$G. In simulations including self-gravity the field strength reaches up to $\sim$10 $\mu$G.
The time to reach saturation is $\sim$ 20 Myr for a medium initial field and about five to eight times
larger for very weak initial fields. Our strong initial field of $3\,\mu\mathrm{G}$ is already above the dynamically reached saturation limit and 
remains unchanged. Increasing the resolution indicates that the field strength in our simulations is not converged.
\item The dense molecular clouds and structures show magnetisations of the order of $1-10\,\mu\mathrm{G}$, 
consistent with the range of of field strengths observed in molecular clouds.
\item For the fixed supernova rate and identical positions for all simulations the initial field strength has a strong impact on 
the resulting structure of the interstellar medium. Starting from a homogenous warm phase,
the runs with low and medium initial fields evolve into a multi-phase
interstellar medium with tangled 
magnetic fields and well defined cold, warm and hot phases. The simulations with the
lowest initial field strength develop the highest hot gas volume filling fractions
($\sim$ 80\%) and have the highest mass fraction in cold and molecular gas.

\item The simulations with high initial fields do not develop into a distinct multi-phase ISM. Instead, they
  remain dominated by warm gas at densities close to the mean density, permeated by a coherent large-scale magnetic field.
  In those simulations very little molecular hydrogen is formed. The injected SNe do not provide sufficient dynamical driving in order to
  form long-lived dense structures that are able to resist the opposing magnetic pressure long enough to exceed the cooling time and the time for H$_2$ formation.

\item Self gravity increases the fraction in molecular gas for low and intermediate magnetic fields but has
only a moderate impact on the overall multi-phase structure.

\item The mass fraction of magnetically dominated gas ($\beta\equiv P_{\rm th}/P_{\rm mag}<1$) increases with the initial strength of 
the magnetic field ranging from 7\% for $B_\mathrm{init}=6\,\mathrm{nG}$ to 70\% for $B_\mathrm{init}=3\,\mu\mathrm{G}$.

\end{itemize}


\section*{Acknowledgements}

All simulations have been performed on the Odin and Hydra clusters hosted by the Max Planck Computing \& Data
Facility (http://www.mpcdf.mpg.de/). We thank the anonymous referee for his/her useful
comments, which helped to improve the paper. We also thank M. Hanasz for stimulating
discussions. AG, SW, TN, PG, SCOG, RSK, and TP acknowledge the Deutsche Forschungsgemeinschaft (DFG)
for funding through the SPP 1573 ``The Physics of the Interstellar
Medium''. SW acknowledges funding by the Bonn-
Cologne-Graduate School, by SFB 956 ``The conditions and
impact of star formation'', and from the European Research
Council under the European Community's Framework Programme
FP8 via the ERC Starting Grant RADFEEDBACK
(project number 679852). TN acknowledges support
by the DFG cluster of excellence 'Origin and structure of
the Universe'. RW acknowledges support by the Czech Science
Foundation project 15-06012S and by the institutional
project RVO: 67985815. RSK and SCOG acknowledge support
from the DFG via SFB 881 ``The Milky Way System''
(sub-projects B1, B2 and B8). RSK acknowledges support
from the European Research Council under the European
Community's Seventh Framework Programme (FP7/2007-
2013) via the ERC Advanced Grant STARLIGHT (project
number 339177). The software used in this work was in part
developed by the DOE NNSA-ASC OASCR Flash Center
at the University of Chicago. We thank C. Karch for the
program package FY and M. Turk and the yt community for
the yt project \citep{Turk}.




\appendix


\section{Resolution study}
\label{A}

\subsection{CO mass fractions}

\begin{figure}
 \includegraphics[width=80mm]{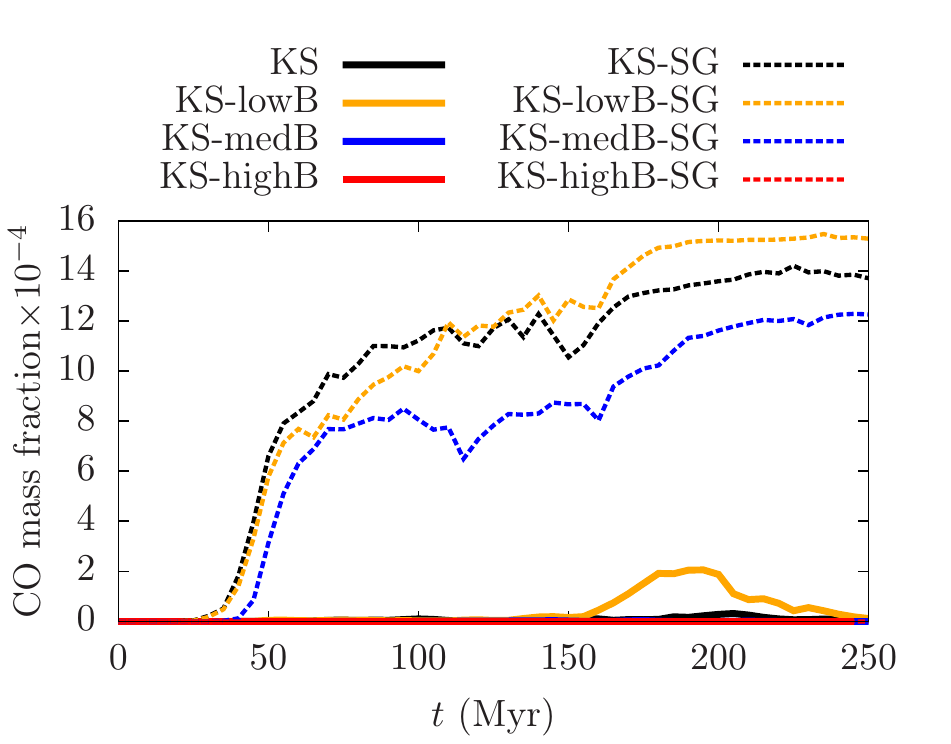}
 \caption{CO mass fraction evolution over 250 Myr for the simulations: KS (solid black line), KS-lowB (solid orange line), KS-medB (solid blue line), KS-highB (solid red line), KS-SG (dashed black line), KS-lowB-SG (dashed orange line), 
 KS-medB-SG (dashed blue line) and KS-highB-SG (dashed red line). The maximum CO mass fraction is reached in KS-lowB-SG with a maximum of 1.5$\times 10^{-3}$.}
 \label{CO}
 \end{figure}

Figure \ref{CO} shows the CO mass fraction\footnote{Our simplified chemical network does not track the abundance of atomic carbon.}evolution over 250 Myr.
 The self-gravity runs form the most CO with fractions over 0.01\% for KS-lowB-SG and KS-SG. The only simulation without self-gravity that forms mass fractions over $10^{-5}$ is KS-lowB.

``Zoom-in'' simulations of molecular clouds carried out by Seifried et al. (in prep.) show that the CO fraction only converges once the spatial resolution is better than 0.25
pc. This requirement is consistent with the finding by \citet{Gl10} and \citet{GlCl12} that the CO abundance is resolved in their simulations of turbulent clouds, which have effective
resolutions in the CO-rich gas of 0.3 pc or better. In addition, a new study also shows that  CO-bright regions have a narrow distribution of temperatures between 10 and 30 K \citep{Gl16}. In our simulations, 
there are only a few cells with gas within this temperature regime, namely 214 for KS, 572 for KS-lowB, 39 for KS-medB, 1 for KS-SG, 5 for KS-lowB-SG, 6 for KS-medB-SG and
none for KS-highB and KS-highB-SG (at 150 Myr). Considering that these cells are deeply embedded in cold, dense molecular hydrogen clumps (that represent less than 3\% of the total volume at most) and there are only a 
small number of cells temperatures relevant for CO, we conclude that a spatial resolution of 2 pc/cell is not sufficient for resolving the evolution of the CO.

\subsection{Magnetic field amplification and H$_{2}$ mass fractions}

MHD simulations are affected by
numerical dissipation that acts on scales much larger than the scale on which real physical dissipation occurs.
For example, in typical ISM conditions, after repeating the calculations by \citet{Bal96} and \citet{Zwe97}, \citet{ML04} obtained an ambipolar diffusion dissipation scale larger
than 0.04 pc while the Kolmogorov dissipation scale is of the order of $10^{-5}$ pc ($10^{14}$ cm), calculated by \citet{Krit}. Both scales are orders
of magnitude smaller than the spatial resolution in our simulations. The highest magnetic Prandtl number that can be resolved
in numerical simulations is around unity while in a typical MC it is about an order of magnitude higher than that \citep[see][]{Fed14}.
This means that the simulations themselves introduce artificial diffusivity, with effects that depend on the numerical resolution
of the simulation \citep{Bal,Waa11}. 

\begin{figure}
\includegraphics[width=80mm]{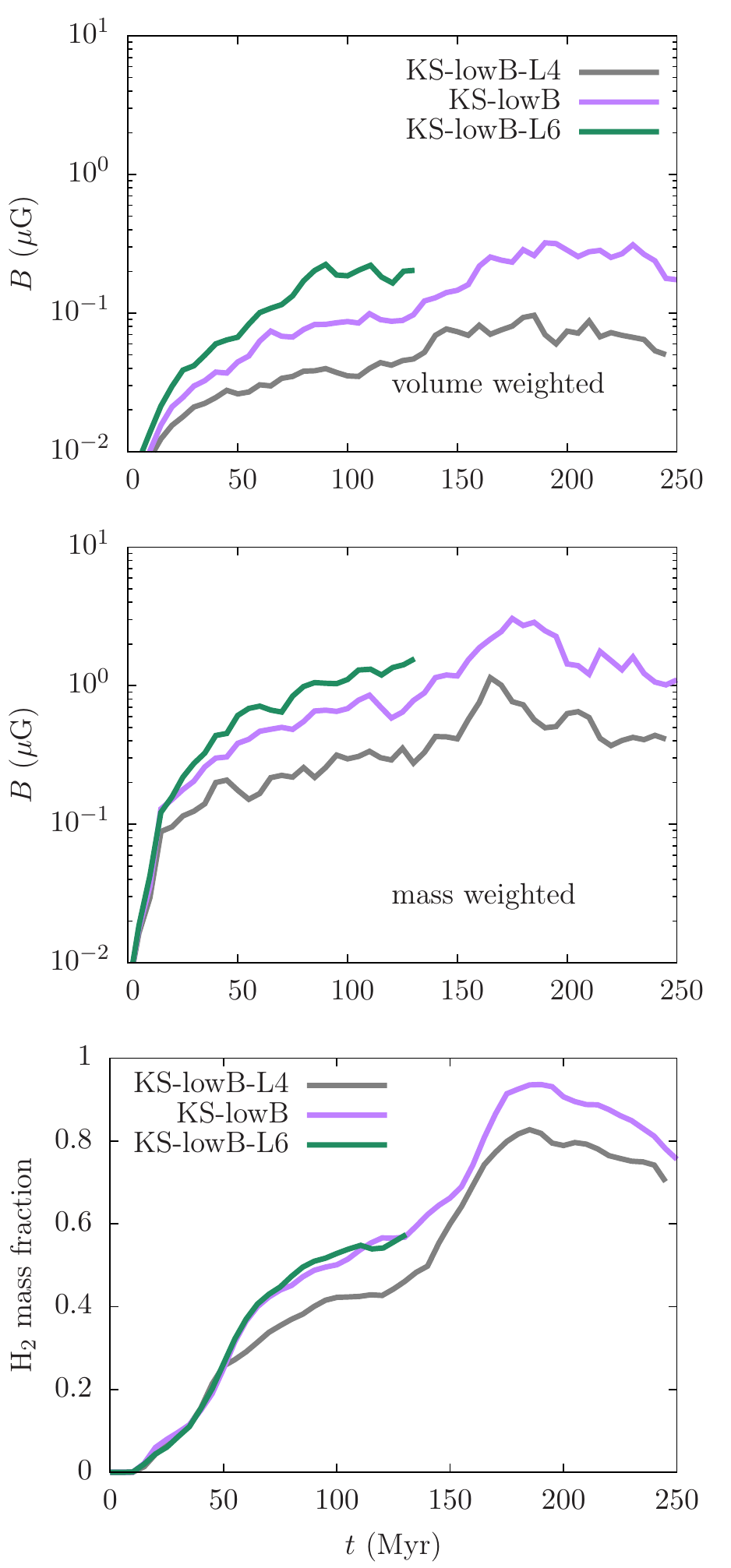}
\caption{Time evolution of the mean magnetic field strength and H$_2$ mass fraction for 
	three different resolutions (grey line: 4 pc per cell, run KS-lowB-L4; purple line:
	2 pc per cell, run KS-lowB-L5; green line: 1 pc per cell, run KS-lowB-L6). {\it Top panel}: volume weighted mean magnetic field strength. {\it Middle panel}: mass weighted mean magnetic field strength. {\it Bottom panel}: H$_{2}$ mass fraction. The magnetic field is stronger with higher resolution and the H$_2$ mass fraction as well, but not significantly.  }
\label{L4L5L6}
\end{figure}

We have tested the resolution dependence of the amplification of the magnetic seed field in our simulations (Fig. \ref{L4L5L6},
first and second panels) by performing simulations with 4, 5 and 6 refinement levels (runs KS-lowB-L4, KS-lowB, and KS-lowB-L6),
corresponding to physical resolutions of 4~pc, 2~pc and 1~pc per cell, respectively. The early stages of the simulations ($t < 20$~Myr)
capture the growth of the seed field and we find that during this period the difference between the three simulations is small.
However, the field amplification occurs more rapidly with increasing numerical resolution.
 We follow the evolution of the mean magnetic field in runs KS-lowB-L4 and KS-lowB for a period of $200$~Myr
after saturation and find that the difference in the mass weighted field strengths is 63\% between runs KS-lowB and KS-lowB-L6 and 29\% between runs KS-lowB and KS-lowB-L4. 
The higher computational demands of
run KS-lowB-L6 mean that we were only able to follow it until shortly after saturation, but it is nevertheless clear that the field
strength in this run saturates at slightly higher value in comparison to the lower resolution runs. We therefore conclude that even 
though we do not recover a convergent behaviour for the field
strength, 
a resolution of 2~pc per
cell is sufficient to model the global behaviour of the magnetic field in our simulations.
We note that the resolution necessary to fully
capture the behaviour of the small-scale turbulent dynamo can be estimated to be at least two orders of magnitude higher 
than this value
\citep{Sr10}, putting it far out of reach of current simulations. 

The other important quantity that is potentially sensitive to resolution is the H$_{2}$ mass fraction.
Work by \citet{GlMc07b} and \citet{Micic12} has shown that small-scale transient density enhancements produced by turbulence 
can significantly
enhance the H$_{2}$ formation rate in interstellar gas, provided that the turbulent mixing time is shorter than the H$_{2}$
formation time in the over-dense regions. As we do not resolve the turbulent cascade, we potentially miss
many of these small over-densities and hence underestimate the H$_{2}$ formation rate. In practice, however, this does not appear
to be an important effect for the size scales and timescales considered here. In the last panel of Fig.~\ref{L4L5L6}, we show how
the H$_{2}$ mass
fraction evolves with time in runs KS-lowB-L4, KS-lowB, and KS-lowB-L6. We see that although there are minor differences, 
there is no significant increase when we increase our resolution from 2 pc per
cell (run KS-lowB) to 1 pc per cell (run KS-lowB-L6). This suggests that our fiducial resolution is high enough to yield
converged results for the evolution of the H2 mass fraction on these timescales. Note, however, that we do find a small but systematic
difference between the H2 mass fractions in runs KS-lowB-L4 and KS-lowB, suggesting that a resolution of 4 pc per cell is not
sufficient to fully resolve H2 formation.


%

\section{Self-gravity switched on at a later time}
\label{B}

\begin{figure}
\includegraphics[width=80mm]{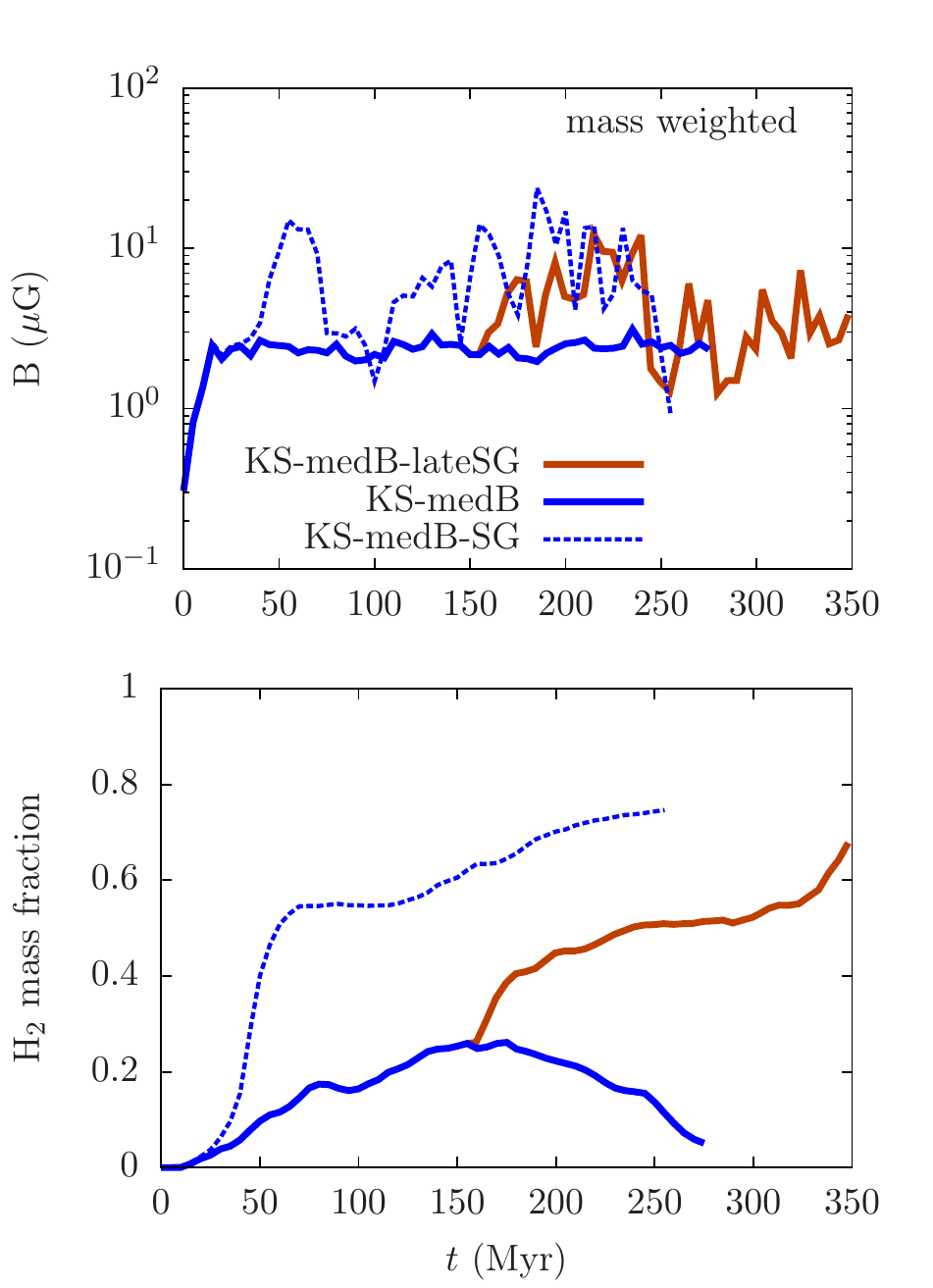}
\caption{Top panel: time evolution of the mean mass weighted magnetic field strength. Bottom panel: time evolution of the H$_{2}$ mass fraction. 
We compare runs KS-medB-lateSG (dark-orange line), KS-medB (blue solid line) and KS-medB-SG (blue dotted line). We see that the H$_{2}$ mass
fraction increases dramatically once self-gravity is switched on at 150 Myr in run KS-med-lateSG. We also see an increase in the mass weighted mean magnetic field 
strength at the same time.}
\label{Chem-lateSG}
\end{figure}

In Fig. \ref{Chem-lateSG} (top panel) we compare the H$_{2}$ mass fraction for KS-medB-lateSG, KS-medB-SG and KS-medB. Once we switch on self-gravity at 150 Myr, the H$_{2}$ mass fraction
increases rapidly. However, the impact of self-gravity on the total volume weighted field strength is negligible as gravity acts on the small dense clumps which have a low volume filling fraction
\citep[see e.g.][]{Gatto}. The effect on the mass weighed field strength is stronger as the compressed dense regions become denser and their magnetic field increases. 

\begin{figure*}
 \begin{minipage}{\textwidth}
  \centering
\includegraphics[width=170mm]{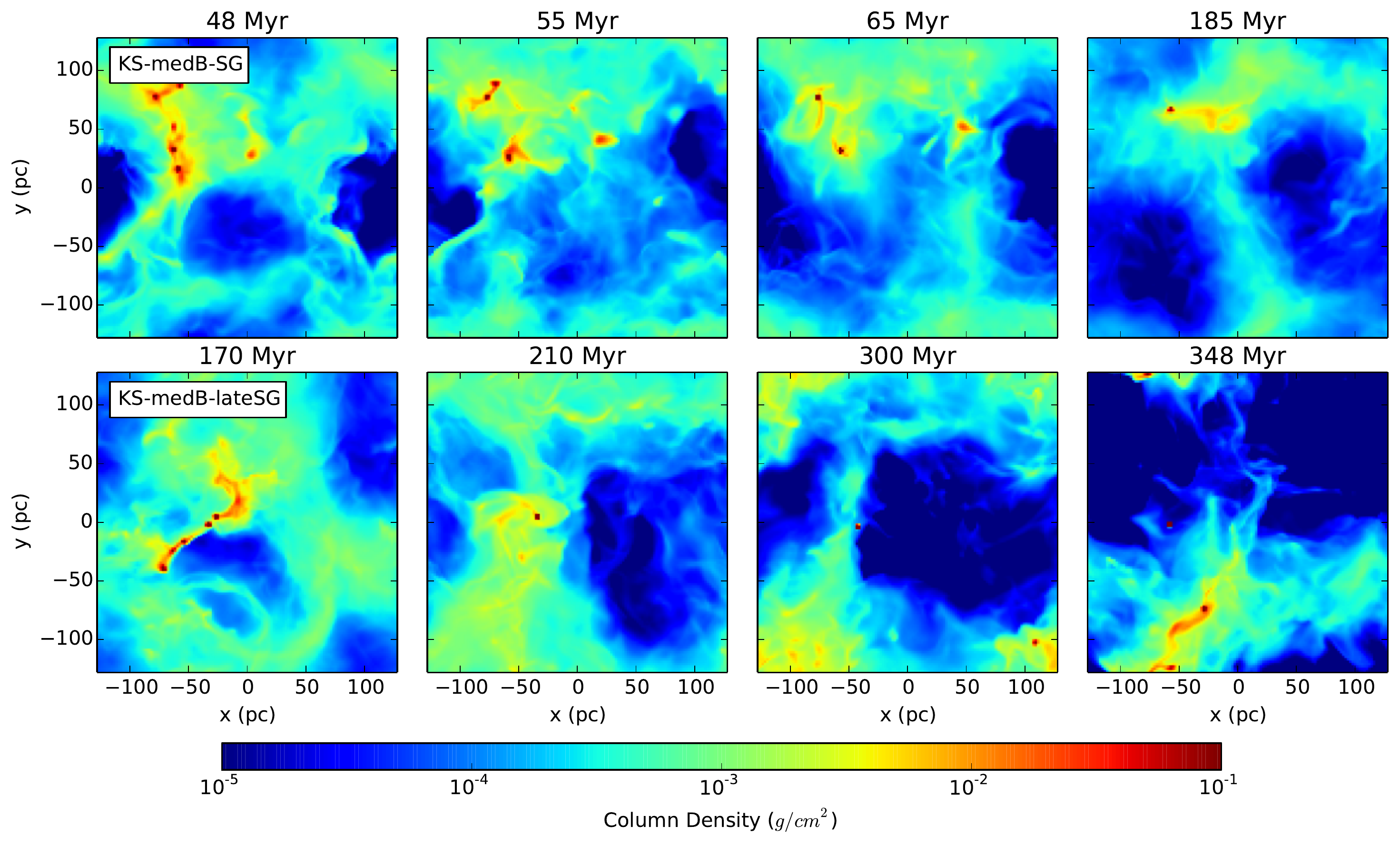}
\caption{Column density projection for KS-medB-SG ({\it first row}) and KS-medB-lateSG ({\it second row} ). In each column, the snapshots compared have the same H$_{2}$ mass fraction. Note that as H$_{2}$ forms at a different rate in the two simulations, this means that we have to compare them at different output times. We see 
that the dense blobs formed in run KS-medB-SG are embedded in higher density gas than those in run KS-medB-lateSG. This occurs because in run KS-medB-lateSG, we switch on self-gravity after dense structures have already formed.
Consequently, once they start to collapse, they very efficiently capture all of the dense gas surrounding them.
This leads to more isolated dense blobs and a higher volume of the box dominated by hot, low density gas.}
\label{lateSG}
\end{minipage}
\end{figure*}

We have selected four values of the H$_{2}$ mass fraction (0.35, 0.45, 0.52 and 0.67) and noted the time it takes for KS-medB-SG and KS-medB-lateSG to reach these values. 
In Fig. \ref{lateSG} we show column density snapshots for these two simulations at the times they reach the same H$_{2}$ mass fraction. In each column the H$_{2}$ mass fraction is equal in both
simulations. What we see is that when self-gravity is acting on the dense regions from the beginning of the simulation the cold, dense clumps are embedded in dense gas. If we let the
structures evolve before switching on self-gravity, the already formed structures will rapidly incorporate all the gas around them 
leaving the newly-formed dense clumps isolated. At 20 Myr after the time at which self-gravity was switched on, the clumps are still accreting the gas around them (Fig. \ref{lateSG} left bottom row)
while after 60 Myr they are already isolated.

\begin{figure}
\includegraphics[width=40mm]{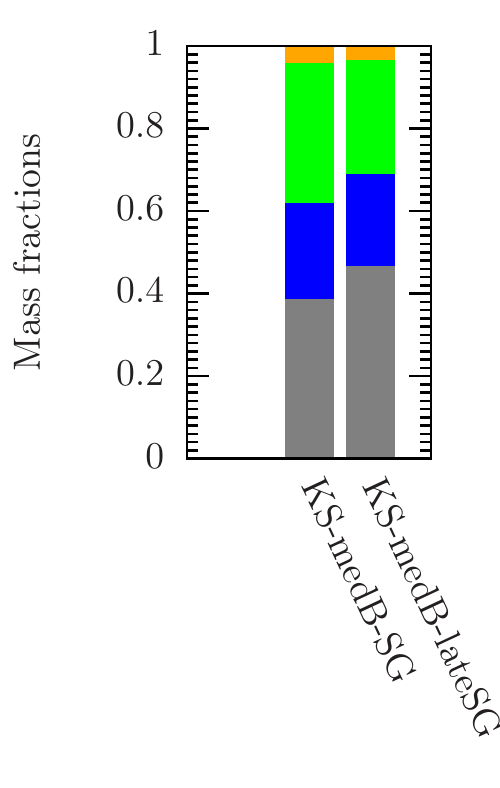}\quad
\includegraphics[width=40mm]{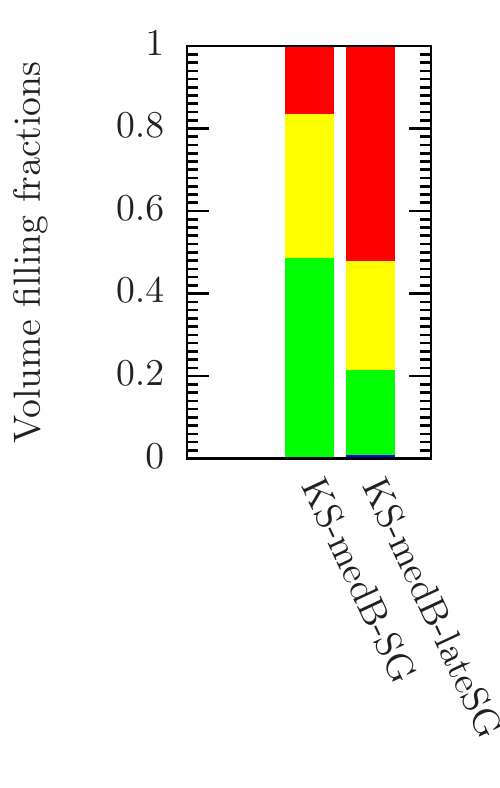}
\caption{Mass fractions ({\it left panel}) and VFFs ({\it right panel}) for runs KS-medB-SG and KS-medB-lateSG. We average over 5 Myr in both cases (63 to 68 Myr for KS-medB-SG and  297 to 302 Myr for KS-medB-lateSG: at these times, the H$_{2}$ mass fraction is the same in both simulations) to avoid discreteness effects due to the explosion of a single SN. The plots use the same colour-coding as in Fig.~\ref{Vff}. The hot gas VFF is twice as high in KS-medB-lateSG as in KS-medB-SG.}
\label{VffMf-lateSG}
\end{figure}

Switching on self-gravity at a later time (150 Myr) yields a larger cold mass fraction (Fig. \ref{VffMf-lateSG}) for the same amount of H$_{2}$ but not significantly 
in comparison to switching on self-gravity from the beginning 
of the simulation. It also yields a larger hot gas VFF in KS-medB-lateSG in comparison to KS-medB-SG. Because the already formed dense structures 
contract very suddenly, they incorporate a lot of the neighbouring gas. The cold clumps end up surrounded by low density gas that is easily heated by 
the exploding SNe, leading to a VFF of hot gas that is almost twice as high as in the run with self-gravity switched on from the beginning.

\begin{figure*}
 \begin{minipage}{\textwidth}
  \centering
\includegraphics[width=140mm]{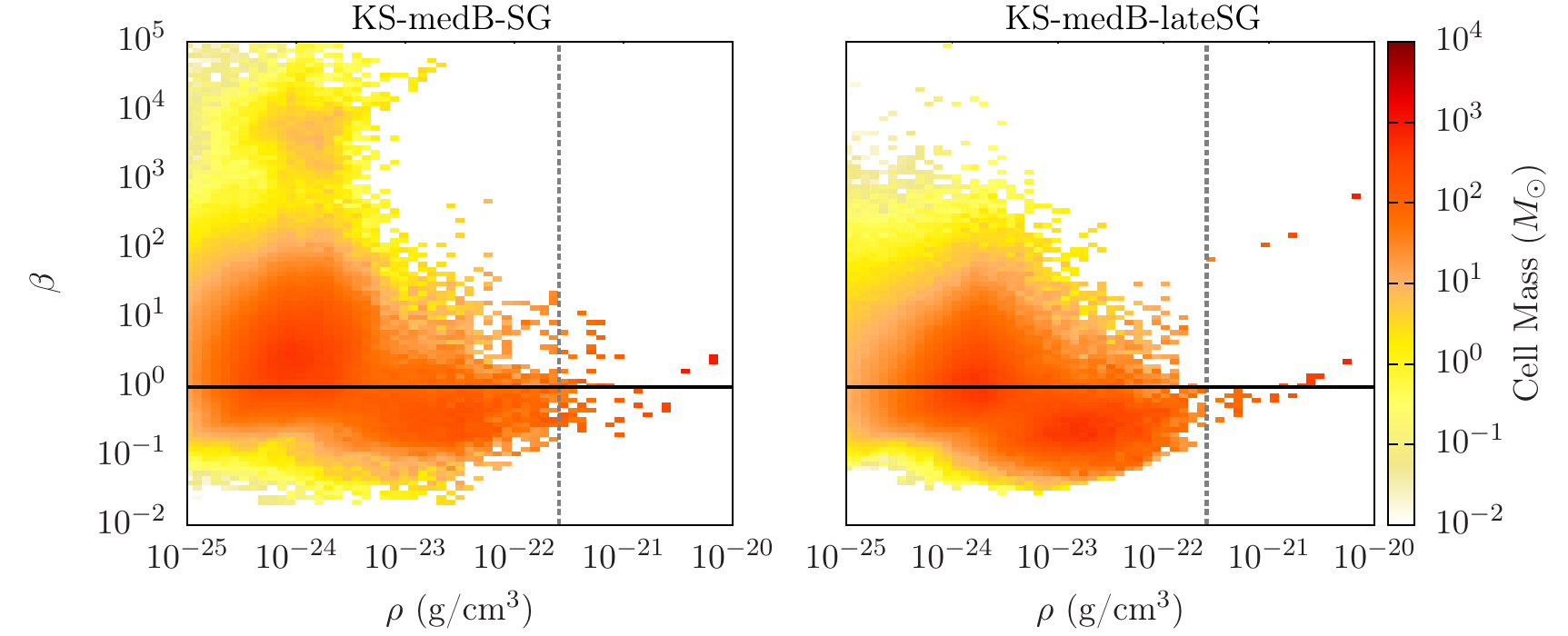}
\caption{Plasma beta as a function of density with mass colour-coding for KS-medB-SG (left panel) at 65 Myr and KS-medB-lateSG (right panel) at 300 Myr
(when the H$_{2}$ mass fraction is the same in both simulations). The grey dotted line represents the maximum
density we can resolve according to the Truelove criterion. In KS-medB-lateSG the gas mass that is magnetically dominated is almost double that in KS-med-SG. }
\label{Beta-lateSG}
\end{minipage}
\end{figure*}

In Fig. \ref{Beta-lateSG} we show the plasma beta parameter as a function of density for KS-medB-SG and KS-medB-lateSG at 65 Myr and 300 Myr when the H$_{2}$ mass fraction
is 0.52 in both cases. In KS-medB-lateSG 31.6\% of the gas mass is magnetically dominated in comparison to only 18\% in KS-medB-SG. This suggests that switching 
on self-gravity at a later time, after dense structures have already formed, leads to a higher cold ($T<30$ K) mass fraction that will be strongly magnetised 
(due to flux-freezing). This is an additional factor that decreases the H$_{2}$ mass fraction.  

\section{RMS Mach number}
\label{C}

\begin{figure}
\includegraphics[width=80mm]{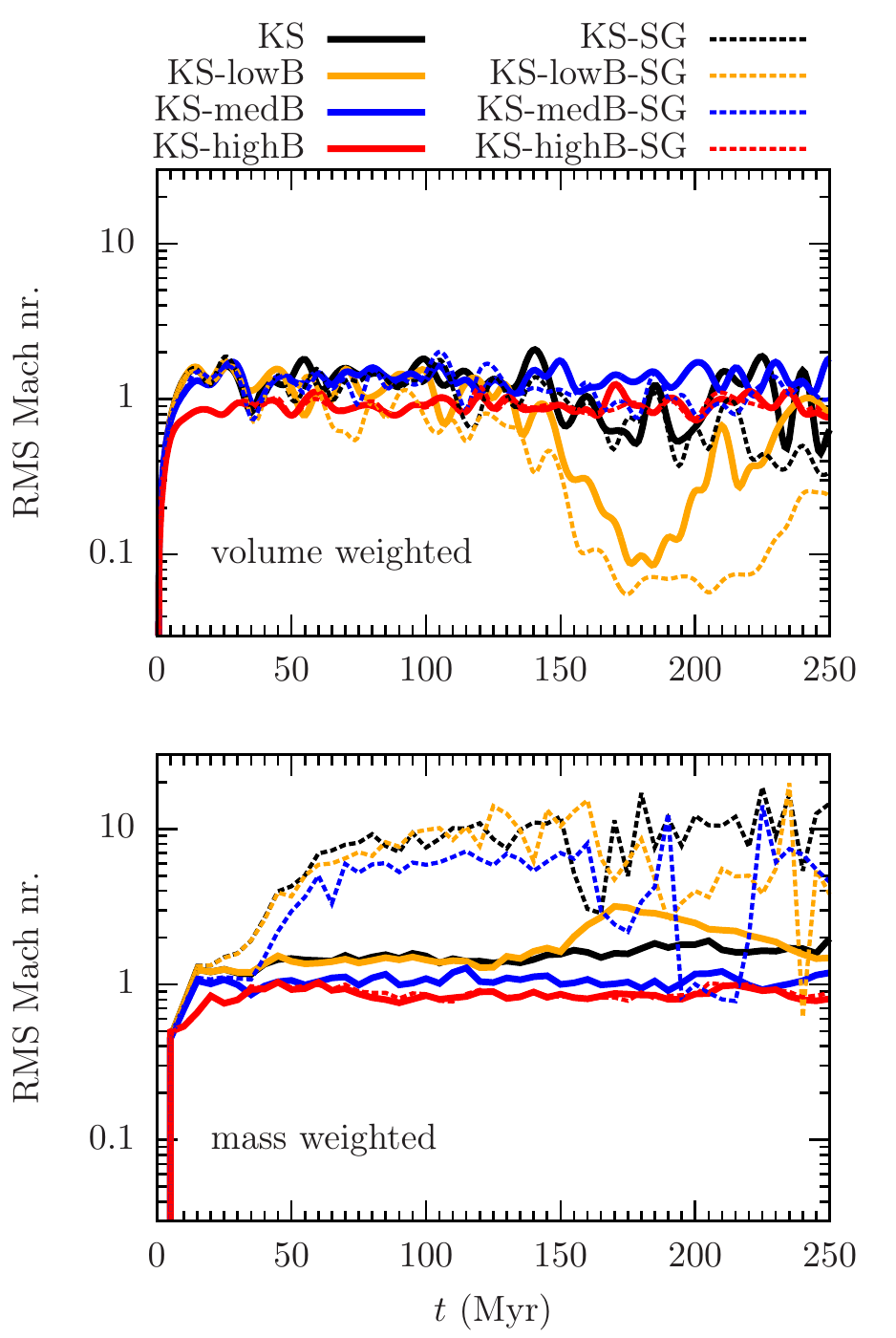}
\caption{Time evolution of the root mean square Mach number, volume weighted ({\it top panel}) and mass weighed ({\it bottom panel}) 
for KS (solid black line), KS-lowB (solid orange line), KS-medB (solid blue line), KS-highB (solid red line), KS-SG (dashed black line), 
KS-lowB-SG (dashed orange line), KS-medB-SG (dashed blue line) and KS-highB-SG (dashed red line). The volume weighted
RMS Mach numbers are close to unity except for KS-lowB, KS-lowSG, KS-highB and KS-highB-SG. The mass weighted values are higher for the self-gravity
runs than the rest, reaching values up to 10.}
\label{RMS}
\end{figure}

We present, in Fig. \ref{RMS}, the volume and mass weighted root mean square of the Mach numbers of all the simulations. We calculate the volume/mass weighted RMS of the local flow divided by the sound speed in every cell, with the adiabatic index equal to 1.4, if the cell is dominated by diatomic molecules, or 1.66, otherwise. 
In both cases the RMS Mach number is lower than unity for KS-highB and KS-highB-SG. The volume weighted values for KS-lowB and KS-lowB-SG decrease and 
reach a minimum
between 150 and 200 Myr. At this point the very hot gas fills more than 80\% of the volume of the simulation (see Fig. \ref{Vff}) and the high sound
speed of this environment is affecting the overall Mach number of these simulations. The mass wighted values are about unity for KS-medB and between
1 and 10 for the rest of the simulations. The average sound speeds at 150 Myr are 100 km/s (KS), 150 km/s (KS-lowB), 21 km/s (KS-medB), 7 km/S (KS-highB), 68 km/s (KS-SG), 172 km/s (KS-lowB-SG), 63 km/s 
(KS-medB-SG) and 8 km/s (KS-highB-SG).

\begin{figure}
\includegraphics[width=80mm]{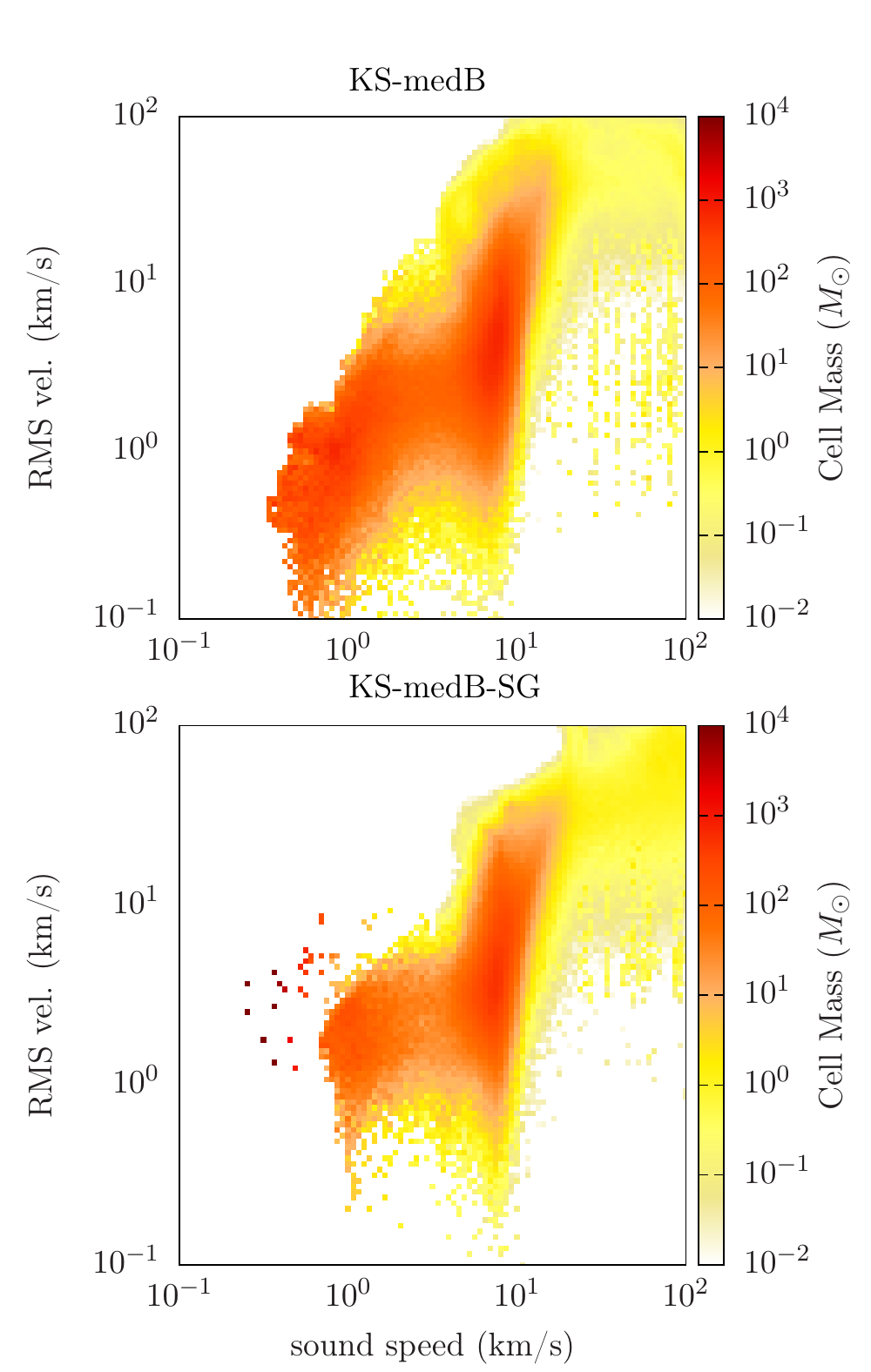}
\caption{RMS velocity as function of sound speed for KS-medB ({\it top panel}) and KS-medB-SG ({\it top panel}) at 150 Myr. Most of the mass in KS-medB has the RMS velocity 
increasing proportionally  
with the sound speed while in KS-medB-SG most of the mass is concentrated in a few cells having one order of magnitude higher RMS velocities for low sound speeds.}
\label{VelCO}
\end{figure}

The highest RMS Mach numbers are obtained in the self-gravity runs where most of the mass is concentrated
in small and very dense clumps. To illustrate better the difference between the runs with and without self gravity, we plot a 2D histogram of the RMS velocity and sound speed in Fig. \ref{VelCO}
 for KS-medB (top panel) and KS-medB-SG (bottom panel), both at 150 Myr. The main difference is that while for KS-medB the RMS velocity is tightly correlated with the sound speed, explaining 
 the close to unity RMS Mach number, in KS-medB-SG most of the mass resides in a few cell which have a high RMS velocity (between 1 and 10 km/s) but have low sound speeds (below 1 km/s).

\label{lastpage}


\begin{thebibliography}{99}

\bibitem[\protect\citeauthoryear{Alves et al.}{2014}]{Alv14} Alves, F.~O., Frau, P., Girart, J.~M., Franco, G.~A.~P., Santos, F.~P., \& Wiesemeyer, H., 2014, A\&A, 569, L1
\bibitem[\protect\citeauthoryear{Ballesteros-Paredes \& Hartmann}{2007}]{BalHar07} Ballesteros-Paredes, J. \& Hartmann, L., 2007, RMAA, 43, 123
\bibitem[\protect\citeauthoryear{Balsara}{1996}]{Bal96} Balsara, D.~S., 1996, ApJ, 465, 775
\bibitem[\protect\citeauthoryear{Balsara et al.}{2001}]{Bal01} Balsara, D.~S., Benjamin, R.~A., \& Cox, D.~P., 2001, ApJ, 563, 800
\bibitem[\protect\citeauthoryear{Balsara et al.}{2004}]{Bal} Balsara, D.~S., Kim, J., {Mac Low}, M.-M., \& Mathews, G.~J., 2004, ApJ, 617, 339
\bibitem[\protect\citeauthoryear{Banerjee et al.}{2009}]{Bane09} Banerjee, R., Vazquez-Semadeni, E., Hennebelle, P.,  \& Klessen R.~S., 2009, MNRAS 398, 1082
\bibitem[\protect\citeauthoryear{Beck}{2001}]{Beck01} Beck, R., 2001, Space Sci.\ Rev., 99, 243
\bibitem[\protect\citeauthoryear{Beck \& Wielebinski}{2013}]{Beck13} Beck, R., \& Wielebinski, R., 2013, in ``Planets, Stars and Stellar Systems'', Volume 5, eds.\ T.~D. Oswalt, G.~Gilmore, (Springer: Dordrecht)
\bibitem[\protect\citeauthoryear{Bergin et~al.}{2004}]{Berg} Bergin, E.~A., Hartmann, L.~W., Raymond, J.~C., \& Ballesteros-Paredes, J., 2004, ApJ, 612, 921
\bibitem[\protect\citeauthoryear{Bouchut et al.}{2007}]{Bou7} Bouchut, F., Klingenberg, C., \& Waagan, K., 2007, Numer. Math., 108, 7
\bibitem[\protect\citeauthoryear{Bouchut et al.}{2010}]{Bou10} Bouchut, F., Klingenberg, C., \& Waagan, K., 2010, Numer. Math., 115, 647
\bibitem[\protect\citeauthoryear{Brandenburg \& Subramanian}{2005}]{Bran05} Brandenburg, A., \& Subramanian K., 2005, Phys.\ Rep.\, 417, 1
\bibitem[\protect\citeauthoryear{Brown}{2010}]{Br10} Brown J.~C., 2010, in ``The Dynamic Interstellar Medium'', ASP Conference Series Vol.\ 438, eds.\ R. Kothes, T.~L. Landecker, A.~G. Willis,  (ASP: San Francisco)
\bibitem[\protect\citeauthoryear{Burkhart et al.}{2009}]{Burkh09} Burkhart, B., Falceta-Goncalves, D., Kowal, D. \& Lazarian, A., 2009, ApJ, 693, 250
\bibitem[\protect\citeauthoryear{Burn}{1966}]{Burn66} Burn B.~J., 1966, MNRAS 133, 67

\bibitem[\protect\citeauthoryear{Chen \& Ostriker}{2014}]{Chen14} Chen, C., \& Ostriker E.~C., 2014, ApJ, 789, 69 
\bibitem[\protect\citeauthoryear{Childress \& Gilbert}{1995}]{Chil95} Childress, S., \& Gilbert, A.~D., 1995, ``Stretch, Twist, Fold: The Fast Dynamo'', (Springer: Berlin)
\bibitem[\protect\citeauthoryear{Clark et al.}{2012}]{Cl12}  Clark, P.~C., Glover, S.~C.~O., Klessen R.~S., \& Bonnell I.~A., 2012, MNRAS, 424, 2599
\bibitem[\protect\citeauthoryear{Clark et al.}{2012}]{Cl122}  Clark, P.~C., Glover, S.~C.~O. \& Klessen R.~S., 2012, MNRAS, 420, 745
\bibitem[\protect\citeauthoryear{Crutcher et al.}{1993}]{Cr93} Crutcher, R.~M., Troland, T.~H., Goodman, A.~A., Heiles, C., Kazes, I., \& Myers, P.~C., 1993, ApJ, 407, 175
\bibitem[\protect\citeauthoryear{Crutcher}{2004}]{Cr03} Crutcher, R.~M., 2004, in ``The Magnetized Interstellar Medium'',  eds.\ B.\ Uyaniker, W.\ Reich, \& R.\ Wielebinski, (Copernicus GmbH: Katlenburg-Lindau), 123
\bibitem[\protect\citeauthoryear{Crutcher et al.}{2004}]{Cr04} Crutcher, R.~M., Nutter, D., Ward-Thompson, D., \& Kirk J.~M., 2004, ApJ, 600, 279
\bibitem[\protect\citeauthoryear{Crutcher, Hakobian \& Troland}{2009}]{Cr09} Crutcher, R.~M., Hakobian, N., \& Troland, T.~H., 2009, ApJ, 692, 844
\bibitem[\protect\citeauthoryear{Crutcher, Hakobian \& Troland}{2010}]{CrHak10} Crutcher, R.~M., Hakobian, N., \& Troland, T.~H., 2010, MNRAS, 402, L64
\bibitem[\protect\citeauthoryear{Crutcher et al.}{2010}]{Cr10} Crutcher, R.~M., Wandelt, B., Heiles, C., Falgarone, E., \& Troland, T.~H., 2010, ApJ, 725, 466
\bibitem[\protect\citeauthoryear{Crutcher}{2012}]{Cr12} Crutcher, R.~M., 2012, ARA\&A, 50, 29
\bibitem[\protect\citeauthoryear{Collins et al.}{2011}]{Collins} Collins, D.~C., Padoan, P., Norman, M.~L., \& Xu, H., 2011, ApJ, 731, 59
\bibitem[\protect\citeauthoryear{Cox \& Smith}{1974}]{Cox74} Cox, D.~P., \& Smith, B.~W., 1974, ApJ, 189, L105

\bibitem[\protect\citeauthoryear{{de Avillez} \& Breitschwerdt}{2005}]{Avill05} {de Avillez}, M.~A., \& Breitschwerdt D., 2005, A\&A, 436, 585
\bibitem[\protect\citeauthoryear{{de Avillez} \& Breitschwerdt}{2007}]{Avill07} {de Avillez}, M.~A., \& Breitschwerdt D., 2007, ApJ, 665, L35
\bibitem[\protect\citeauthoryear{Draine}{2011}]{Dr11} Draine, B.~T., 2011, Physics of the Interstellar and Intergalactic Medium, (Princeton University Press: Princeton)
\bibitem[\protect\citeauthoryear{Draine}{1978}]{Dr78} Draine, B.~T., 1978, ApJS, 36, 595
\bibitem[\protect\citeauthoryear{Dubey et al.}{2008}]{DubeyEtAl2008} Dubey, A., et~al., 2008, in ``Numerical Modeling of Space Plasma Flows: Astronum 2007'', ASP Conference Series Vol.\ 385, 145
\bibitem[\protect\citeauthoryear{Dubey et al.}{2012}]{Dub12} Dubey, A., Daley, C., ZuHone, J., Ricker, P.~M., Weide, K., \& Graziani, C., 2012, ApJS, 201, 27
\bibitem[\protect\citeauthoryear{Dubey et al.}{2013}]{Dub13} Dubey et al., 2013, in ``Proceedings of the 5th International Workshop on Software Engineering for Computational Science and Engineering'', pp. 1-8
\bibitem[\protect\citeauthoryear{Dobbs \& Pettitt}{2014}]{Dobbs14} Dobbs, C., \& Pettitt, A., 2014, to appear in ``Lessons from the Local Group'',  eds.\ K.~C.~Freeman, B.~G.~Elmegreen, D.~L.~Block \& M.~Woolway, (Springer: Dordrecht); arXiv:1407.0250

\bibitem[\protect\citeauthoryear{Egan et al.}{2016}]{Egan} Egan, H., O'Shea, B.~W., Hallman, E., Burns, J., Xu, H., Collins, D., Li, H. \& Norman, M.~L., 2016, arXiv:1601.05083v2 
\bibitem[\protect\citeauthoryear{Elmegreen}{2000}]{Elm} Elmegreen, B.~G., 2000, ApJ, 530, 277
\bibitem[\protect\citeauthoryear{Eswaran \& Pope}{1988}]{Pope} Eswaran, V. \& Pope, S.~B., Computers and Fluids, 1988, 16, 257

\bibitem[\protect\citeauthoryear{Federrath et al.}{2011a}]{Fed11a} Federrath, C., Chabrier, G., Schober, J., Banerjee, R., Klessen, R.~S., \& Schleicher, D.~R.~G., 2011a, Phys.\ Rev.\ Lett., 107, 114504
\bibitem[\protect\citeauthoryear{Federrath et al.}{2011b}]{Fed11} Federrath, C., Sur, S., Schleicher, D.~R.~G., Banerjee, R., \& Klessen, R.~S., 2011b, ApJ, 731, 62
\bibitem[\protect\citeauthoryear{Federrath et al.}{2014a}]{Fed14} Federrath, C., Schober, J., Bovino, S., \& Schleicher, D.~R.~G., 2014a, ApJ, 790, 128
\bibitem[\protect\citeauthoryear{Federrath et al.}{2014b}]{Fed14b} Federrath, C., Schober, J., Bovino, S., \& Schleicher, D.~R.~G., 2014b, ApJ, 797, L19
\bibitem[\protect\citeauthoryear{Fosalba et al.}{2002}]{Fas02} Fosalba, P., Lazarian, A., Prunet, S., \& Tauber J.~A., 2002, ApJ, 564, 762
\bibitem[\protect\citeauthoryear{Frau et al.}{2012}]{Frau12} Frau, P., et~al., 2012, ApJ, 759, 3
\bibitem[\protect\citeauthoryear{Fryxell et al.}{2000}]{Fryx} Fryxell B., et~al., 2000, ApJS, 131, 273

\bibitem[\protect\citeauthoryear{Gatto et al.}{2015}]{Gatto} Gatto, A., et~al., 2015, MNRAS, 449, 1057
\bibitem[\protect\citeauthoryear{Gatto et al.}{2016}]{Gatto16} Gatto, A., et~al., 2016, MNRAS, submitted; arXiv:1606.05346
\bibitem[\protect\citeauthoryear{Gent et al.}{2013}]{Gent13} Gent, F.~A., Shukurov, A., Sarson, G.~R., Fletcher, A., \& Mantere M.~J., 2013, MNRAS 430, L40
\bibitem[\protect\citeauthoryear{Girichidis et al.}{2016}]{Gir15} Girichidis, P., et~al., 2016, MNRAS, 456, 3432
\bibitem[\protect\citeauthoryear{Glover \& {Mac Low}}{2007a}]{GlMc07a} Glover, S.~C.~O., \& {Mac Low}, M.-M., 2007a, ApJS, 169, 239
\bibitem[\protect\citeauthoryear{Glover \& {Mac Low}}{2007b}]{GlMc07b} Glover, S.~C.~O., \& {Mac Low}, M.-M., 2007b, ApJ, 659, 1317
\bibitem[\protect\citeauthoryear{Glover et al.}{2010}]{Gl10} Glover, S.~C.~O., Federrath, C., {Mac Low}, M.-M., \& Klessen R.~S., 2010, MNRAS, 404, 2
\bibitem[\protect\citeauthoryear{Glover \& Clark}{2012}]{GlCl12} Glover, S.~C.~O., \& Clark, P.~C., 2012, MNRAS, 421, 116
\bibitem[\protect\citeauthoryear{Glover \& Smith}{2016}]{Gl16} Glover, S.~C.~O., \& Smith, R.~J., 2016, MNRAS, 462, 3011
\bibitem[\protect\citeauthoryear{Gnat \& Ferland}{2012}]{GF12} Gnat, O., \& Ferland G.~J., 2012, ApJS, 199, 20
\bibitem[\protect\citeauthoryear{Goldreich \& Kylafis}{1981}]{Gol81} Goldreich, P., \& Kylafis, N.~D., 1981, ApJ, 243, L75

\bibitem[\protect\citeauthoryear{Habing}{1968}]{Hb68} Habing, H.~J., 1968, Bull.\ Astron.\ Inst.\ Netherlands, 19, 421
\bibitem[\protect\citeauthoryear{Han}{2004}]{Han03} Han, J.~L., 2004, in ``The Magnetized Interstellar Medium'',  eds.\ B.\ Uyaniker, W.\ Reich, \& R.\ Wielebinski, (Copernicus GmbH: Katlenburg-Lindau), 3
\bibitem[\protect\citeauthoryear{Han}{2009}]{Han09} Han, J., 2009, in ``Cosmic Magnetic Fields: From Planets, to Stars and Galaxies'', Proc.\ IAU Symp.\ 259, eds.\ K.~G.~Strassmeier, A.~G.~Kosovichev, \& J.~E.~Beckman, (CUP: Cambridge), 455

\bibitem[\protect\citeauthoryear{Hennebelle et. al}{2008}]{Hen08} Hennebelle, P., Banerjee, R., Vazquez-Semadeni, E., Klessen, R.~S. \& Audit, E., 2008, A\&A, 486, L43
\bibitem[\protect\citeauthoryear{Hennebelle et. al}{2011}]{Hen11} Hennebelle, P., Commercon, B., Joos, M., Klessen, R.~S., Krumholz, M., Tan, J.~C., \& Teyssier, R., 2011, A\&A, 528, A72
\bibitem[\protect\citeauthoryear{Hennebelle \& Iffrig}{2014}]{Hen14} Hennebelle, P., \& Iffrig, O., 2014, A\&A, 570, 81
\bibitem[\protect\citeauthoryear{Heiles \& Troland}{2003}]{Hei03} Heiles, C., \& Troland T.~H., 2003, ApJ, 586, 1067
\bibitem[\protect\citeauthoryear{Heiles \& Troland}{2005}]{Hei05} Heiles, C., \& Troland T.~H., 2005, ApJ, 624, 773
\bibitem[\protect\citeauthoryear{Heitsch et al.}{2011}]{Hei11}  Heitsch, F., Naab, T., \& Walch S., 2011, MNRAS, 415, 271
\bibitem[\protect\citeauthoryear{Heitsch et al.}{2005}]{Hei205}  Heitsch, F., Burkert, A., Hartmann L.~W., Slyz A.D. \& Devriendt J.~E.~G., 2011, MNRAS, 415, 271
\bibitem[\protect\citeauthoryear{Heitsch \& Hartmann}{2008}]{Hei08}  Heitsch, F. \& Hartmann L.~W., 2008, ApJ, 689, 290
\bibitem[\protect\citeauthoryear{Heitsch \& Hartmann}{2014}]{HeiHar14}  Heitsch, F. \& Hartmann L., 2014, MNRAS, 443, 230
\bibitem[\protect\citeauthoryear{Hill et. al}{2012}]{Hill12} Hill, A.~S, Joung, M.~R., {Mac Low}, M.-M., Benjamin, R.~A., Haffner, L.~M., Klingenberg, C., \& Waagan, K., 2012, ApJ, 750, 104
\bibitem[\protect\citeauthoryear{Hildebrand}{1988}]{Hilde88} Hildebrand, R.~H., 1988, QJRAS, 29, 327
\bibitem[\protect\citeauthoryear{Hoang \& Lazarian}{2008}]{Laz08} Hoang, T., \& Lazarian, A., 2008, MNRAS 388, 117
\bibitem[\protect\citeauthoryear{Honma et al.}{1995}]{Hon95} Honma, M., Sofue, Y., \& Arimoto, N., 1995, A\&A, 304, 1
\bibitem[\protect\citeauthoryear{Hopkins et al.}{2012}]{Hop12} Hopkins, P.~F., Quataert, E., \& Murray, N., 2012, MNRAS, 421, 3522
\bibitem[\protect\citeauthoryear{Hughes et al.}{2010}]{Hug10} Hughes, A., et~al., 2010, MNRAS, 406, 2065

\bibitem[\protect\citeauthoryear{Ib$\rm \acute{a} \rm \tilde{n}$ez-Mej$\rm \acute{i}$a et al.}{2016}]{Huan16}  Ib$\rm \acute{a} \rm \tilde{n}$ez-Mej$\rm \acute{i}$a, J. C.,  et~al., 2016, ApJ, 824, 41

\bibitem[\protect\citeauthoryear{Jansson \& Farrar}{2012}]{Jas12} Jansson, R., \& Farrar, G.~R., 2012, ApJ 757, 14
\bibitem[\protect\citeauthoryear{Jones et~al.}{2011}]{Jones11} Jones, T.~W., Porter, D.~H., Ryu, D., \& Cho, J., 2011; arXiv:1108.1369
\bibitem[\protect\citeauthoryear{Joung \& Mac Low}{2006}]{Jou06} Joung, M.~K.~R., \& {Mac Low}, M.-M., 2006, ApJ, 653, 1266
\bibitem[\protect\citeauthoryear{Joung et al.}{2009}]{Jou09} Joung, M.~K.~R., {Mac Low}, M.-M., \& Bryan, G.~L., 2009, ApJ, 704, 137

\bibitem[\protect\citeauthoryear{Kazantsev}{1968}]{Kaz68} Kazantsev, A. P. 1968, Soviet Journal of Experimental and Theoretical Physics,
26, 1031
\bibitem[\protect\citeauthoryear{Kennicutt}{1998}]{Ken98} Kennicutt, R.~C., 1998, ApJ, 498, 541
\bibitem[\protect\citeauthoryear{Kim et al.}{2013}]{Kim13} Kim, J.-H., Krumholz, M.~R., Wise, J.~H., Turk, M.~J., Goldbaum, N.~J., \& Abel, T., 2013, ApJ, 779, 8
\bibitem[\protect\citeauthoryear{Kim \& Ostriker}{2015}]{Kim14} Kim, J.-H., \& Ostriker, E.~C., 2015, ApJ, 802, 99
\bibitem[\protect\citeauthoryear{Klessen \& Glover}{2016}]{Kle16} Klessen, R.~S. \& Glover, S.~C.~O., 2016, in ``Star Formation in Galaxy Evolution: Connecting Numerical Models to Reality'',  Saas-Fee Advanced Course Vol.\ 43, (Springer-Verlag:Berlin), 85; arXiv:1412.5182
\bibitem[\protect\citeauthoryear{Kritsuk et al.}{2011}]{Krit} Kritsuk, A.~G., et al., 2011, ApJ, 737, 13
\bibitem[\protect\citeauthoryear{Kritsuk et al.}{2009}]{Krit09} Kritsuk, A.~G., Ustyugov, S., Norman, M.~L., \& Padoan P., 2009, J.\ Phys.\ Conf.\ Ser., 180, 012020
\bibitem[\protect\citeauthoryear{Krumholz \& Matzner}{2009}]{Kru09} Krumholz M.~R., \& Matzner, C.~D., 2009, ApJ, 703, 1352
\bibitem[\protect\citeauthoryear{Kudoh \& Basu}{2007}]{KB07} Kudoh T., \& Basu, S., 2007, in ``Triggered Star Formation in a Turbulent ISM", eds.\ B.~G.~Elmegreen, J.~Palous,  
Proc.\ IAU Symp.\ 327, (CUP:Cambridge), 437
\bibitem[\protect\citeauthoryear{Kudoh \& Basu}{2008}]{KB08} Kudoh T., \& Basu, S., 2008, ApJ, 679, L97
\bibitem[\protect\citeauthoryear{Lazarian}{2003}]{Laz03} Lazarian, A., 2003, J.\ Quant.\ Spec.\  Rad.\ Trans., 79, 881
\bibitem[\protect\citeauthoryear{Lazarian}{2013}]{Laz} Lazarian A., 2013, arXiv:1304.3133
\bibitem[\protect\citeauthoryear{Li et al.}{2015}]{Li15} Li, M., Ostriker, J.~P., Cen, R., Bryan, G.~L., \& Naab, T., 2015 ApJ, 814, 4
\bibitem[\protect\citeauthoryear{Li, McKee \& Klein}{2015}]{Li} Li, P.~S., McKee, C.~F., \& Klein, R.~I., 2015, MNRAS, 452, 2500

\bibitem[\protect\citeauthoryear{{Mac Low} \& Klessen}{2004}]{ML04} Mac Low, M.-M., \& Klessen, R.~S., 2004, Rev.\ Mod.\ Phys., 76, 125
\bibitem[\protect\citeauthoryear{{Mac Low} et al.}{2005}]{ML05} {Mac Low}, M.-M., Balsara, D.~S., Kim, J., \& {de Avillez}, M.~A., 2005, ApJ, 626, 864
\bibitem[\protect\citeauthoryear{Mao et al.}{2012}]{Mao12} Mao, S.~A., McClure-Griffiths, N.~M., Gaensler, B.~M., et al., 2012, ApJ, 759, 25
\bibitem[\protect\citeauthoryear{McKee \& Ostriker}{1977}]{McK77} McKee, C.~F., \& Ostriker, J.~P., 1977, ApJ, 218, 148
\bibitem[\protect\citeauthoryear{McKee \& Ostriker}{2007}]{McK07} McKee, C.~F., \& Ostriker, E.~C., 2007, ARA\&A, 45, 565

\bibitem[\protect\citeauthoryear{Micic et al.}{2012}]{Micic12} Micic, M., Glover, S.~C.~O., Federrath, C., \& Klessen, R.~S., 2012, MNRAS, 421, 2531
\bibitem[\protect\citeauthoryear{Mouschovias}{1976}]{Mou75} Mouschovias, T.~C., 1976, ApJ, 206, 753
\bibitem[\protect\citeauthoryear{Murray}{2011}]{Murray} Murray, N., 2011, ApJ, 729, 133

\bibitem[\protect\citeauthoryear{Nelson \& Langer}{1997}]{NL97} Nelson, R.~P., \& Langer, W.~D.\ 1997, ApJ, 482, 796
\bibitem[\protect\citeauthoryear{Nakamura \& Li}{2003}]{Nak} Nakamura, F., \& Li, Z.\ 2003, ApJ, 594, 363
\bibitem[\protect\citeauthoryear{Ostriker et al.}{2010}]{Ost10} Ostriker, E.~C., McKee, C.~F., \& Leroy, A.~K., 2010, ApJ, 721, 975

\bibitem[\protect\citeauthoryear{Pavel et al.}{2012}]{Pav12} Pavel, M.~D., Clemens, D.~P. \& Pinnick, A.~F., 2012, ApJ, 749, 71
\bibitem[\protect\citeauthoryear{Piontek \& Ostriker}{2007}]{Pion07} Piontek, R.~A.,\& Ostriker, E~C., 2007, ApJ, 663, 183
\bibitem[\protect\citeauthoryear{Peters et al.}{2011}]{Pet11} Peters, T., Banerjee, R., Klessen, R.~S., \& {Mac Low}, M.-M., 2011, ApJ, 729, 72
\bibitem[\protect\citeauthoryear{Peters et al.}{2012}]{Peters} Peters, T., Schleicher, D.~R.~G., Klessen, R.~S., Banerjee, R., Federrath,  C., Smith R.~J., \& Sur, S., 2012, ApJ, 760, L28
\bibitem[\protect\citeauthoryear{Peters et al.}{2016a}]{Thom1} Peters, T., Zhukovska, S., Naab, T., Girichidis, P., Walch,  S., Glover, S.~C.~O., Klessen, R.~S., Clark, P.~C. \& Seifried D., 2016, arXiv:1610.06579
\bibitem[\protect\citeauthoryear{Peters et al.}{2016b}]{Thom2} Peters, T., Naab, T., Walch,  S., Glover, S.~C.~O., Girichidis, P., Pellegrini, E., Klessen, R.~S., Wunsch R., Gatto, A., Baczynski C., 2016, arXiv:1610.06569 
\bibitem[\protect\citeauthoryear{Padoan et al.}{2016}]{Pad16} Padoan, P., Pan, L., Haugboelle, T., \& Nordlund, A., 2016, ApJ, 822, 11
\bibitem[\protect\citeauthoryear{Price \& Bate}{2008}]{PB08} Price, D.~J. \& Bate, M.~R., 2008, MNRAS, 385, 1820
\bibitem[\protect\citeauthoryear{Pineda et al.}{2013}]{Pin13} Pineda, J.~L., Langer, W.~D., Velusamy, T. \& Goldsmith, P.~F., 2013, A\&A 554, A103
\bibitem[\protect\citeauthoryear{Salpeter}{1955}]{Sal} Salpeter, E.~E., 1955, ApJ, 121, 161
\bibitem[\protect\citeauthoryear{Schleicher et al.}{2011}]{Sch11} Schleicher, D.~R.~G., Schober, J., Federrath, C., Miniati, F., Banerjee, R., \& Klessen, R.~S., 2011, in ``Magnetic Fields in the Universe III: From Laboratory and Stars to Primordial Structures'', eds.\ M.~Soida, K.~Otmianowska-Mazur, E.~M.~{de Gouveia Dal Pino}, A.~Lazarian; arXiv:1110.2880
\bibitem[\protect\citeauthoryear{Schober et al.}{2012a}]{Sch12} Schober, J., Schleicher, D.~R.~G., Federrath, C., Glover, S., Klessen, R.~S., \& Banerjee R., 2012a, ApJ, 754, 99
\bibitem[\protect\citeauthoryear{Schober et al.}{2012b}]{Sch12b} Schober, J., Schleicher, D.~R.~G., Bovino, S. \& Klessen, R.~S., 2012b, Phys.\ Rev.\ E, 86, 066412
\bibitem[\protect\citeauthoryear{Schober et al.}{2015}]{Schober15} Schober, J., Schleicher, D.~R.~G., Federrath, C., Bovino, S., \& Klessen, R.~S., 2015, Phys.\ Rev.\ E, 92, 023010
\bibitem[\protect\citeauthoryear{Schruba et al.}{2011}]{Schru11} Schruba, A., Leroy, A.~K., Walter, F., et~al., 2011, AJ, 142, 37
\bibitem[\protect\citeauthoryear{Sembach et al.}{2000}]{Sem00} Sembach, K. R., Howk, J.~C., Ryans, R. S. I. \& Keenan, F. R., 2000, ApJ, 528, 310
\bibitem[\protect\citeauthoryear{Slyz et al.}{2005}]{Slyz} Slyz, A.~D., Devriendt, J.~E.~G., Bryan, G.\& Silk, J., 2005, MNRAS, 356, 737
\bibitem[\protect\citeauthoryear{Snowden et al.}{1998}]{Snow98} Snowden, S.~L., Egger, R., Finkbeiner, D.~P., Freyberg, M.~J., \& Plucinsky, P.~P.,  1998, ApJ, 493, 715
\bibitem[\protect\citeauthoryear{Subramanian}{1997}]{Subr} Subramanian, K. 1997, ArXiv e-prints, arXiv:astro-ph/9708216
\bibitem[\protect\citeauthoryear{Sun \& Reich}{2010}]{Sun10} Sun, X.-H., \& Reich, W., 2010, Res.\ Astron.\ Astrophys., 10, 1287
\bibitem[\protect\citeauthoryear{Sur et al.}{2010}]{Sr10} Sur, S., Schleicher, D.~R.~G., Banerjee, R., Federrath, C., \& Klessen, R.~S., 2010, ApJ, 721, L134

\bibitem[\protect\citeauthoryear{Tan et al.}{2014}]{Tan14} Tan, J.~C.,  Beltr\'an, M.~T., Caselli, P., Fontani, F., Fuente, A., Krumholz, M.~R., McKee, C.~F., \& Solte, A., 2014, arXiv:1402.0919
\bibitem[\protect\citeauthoryear{Truelove et al.}{1997}]{True} Truelove, J.~K., Klein, R.~I., McKee, C.~F., Holliman, J.~H., Howell, L.~H., \& Greenough J.~A., 1997, ApJ, 489, L179
\bibitem[\protect\citeauthoryear{Turk et al.}{2011}]{Turk} Turk, M.~J., Smith, B.~D., Oishi, J.~S., Skory, S., Skillman, S.~W., Abel, T., Norman, M.~L., 2011, ApJS, 192, 9
\bibitem[\protect\citeauthoryear{Vazquez-Semadeni et al.}{2005}]{Vaz05} Vazquez-Semadeni, E., Kim, J., Shadmehri, M. \& Ballesteros-Paredes, J., 2005, ApJ 618, 344
\bibitem[\protect\citeauthoryear{Vazquez-Semadeni et al.}{2011}]{Vaz11} Vazquez-Semadeni, E., Banerjee, R., Gomez, G.~C., Hennebelle, P., Duffin, D., \& Klessen, R.~S., 2011, MNRAS 414, 2511

\bibitem[\protect\citeauthoryear{Waagan et al.}{2011}]{Waa11} Waagan, K., Federrath, C., \& Klingenberg, C., 2011, J.\ Comp.\ Phys., 230, 3331
\bibitem[\protect\citeauthoryear{Waagan}{2009}]{Waa9} Waagan, K., 2009,  J.\ Comp.\ Phys., 228, 8609
\bibitem[\protect\citeauthoryear{Walch et al.}{2015}]{Walch15} Walch, S., et al., 2015, MNRAS, 454, 238
\bibitem[\protect\citeauthoryear{Woody at al.}{1989}]{Woody} Woody, D.~P., Scott, S.~L., Scoville, N.~Z., Mundy, L.~G., Sargent, A.~I., Padin, S., Tinney, C.~G., \& Wilson, C.~D., 1989, ApJ, 337, L41
\bibitem[\protect\citeauthoryear{Wu et al.}{2009}]{Wu09} Wu, Q., Kim, J., Ryu, D., Cho, J. \& Alexander, P., 2009, ApJ, 705, L86
\bibitem[\protect\citeauthoryear{Wunsch et al.}{in prep}]{Wu} Wunsch, R., et al., in prep.

\bibitem[\protect\citeauthoryear{Yoast-Hull et al.}{2013}]{Yoas13} Yoast-Hull, T.~M., Everett, J.~E., Gallagher, J.~S., \& Zweibel, E.~G., 2013, ApJ, 768, 53

\bibitem[\protect\citeauthoryear{Zweibel \& Brandenburg}{1997}]{Zwe97} Zweibel, E.~G. \& Brandenburg, A., 1997, ApJ, 478, 563
\bibitem[\protect\citeauthoryear{Zasov \& Khoperskov}{2015}]{Zas15} Zasov, A.~V., \& Khoperskov, S.~A., 2015, MNRAS, 452, 4247
\end{thebibliography}
\end{document}